\documentclass[12pt]{article}
 \usepackage{filecontents}
\usepackage{geometry}
\usepackage{setspace}
\usepackage{tabularx}
\usepackage{float}
\usepackage{xr}
\usepackage{afterpage}
\usepackage{ulem}

\usepackage{amssymb}
\usepackage{xr}
\usepackage{hyperref}
\usepackage{booktabs}
\usepackage{amsmath}
\usepackage{graphicx}
\usepackage{xcolor}
\usepackage{pdflscape}
\usepackage{mfirstuc}
\usepackage{enumitem}
\usepackage{float}
\usepackage{bbm}
\usepackage{caption}

\def\paperName{Credit Information in Earnings Calls}
\def\numobs{13,899} 
\def\numobsig{9830}
\def\numobshy{4069}

\def\corr{\textsf{corr}}
\def\var{\textsf{var}}
\def\one{\mathbf{1}}

\def\dateStrOld{2021-03-11}

\begin{document}

\title{\paperName \thanks{We thank Amit Goyal, Xu Guo, Hai Lin, and Avanidhar Subrahmanyam, as well
    as seminar participants at Balyasny Asset Management, Bank of America, Columbia Business School,
    CUNY, Vanguard, and the Wolfe NLP/ML Conference for helpful suggestions.}}

\author{Harry Mamaysky\thanks{Columbia Business School, hm2646@columbia.edu.}
  \and Yiwen Shen\thanks{HKUST Business School, yiwenshen@ust.hk.}
  \and Hongyu Wu\thanks{Yale School of Management, hw499@yale.edu.}}

\maketitle

\begin{abstract}
  We develop a novel technique to extract credit-relevant information from the text of quarterly
  earnings calls. This information is not spanned by fundamental or market variables and forecasts
  future credit spread changes. One reason for such forecastability is that our text-based measure
  predicts future credit spread risk and firm fundamentals. More firm- and call-level complexity
  increase the forecasting power of our measure for spread changes. Out-of-sample portfolio tests
  show the information in our measure is valuable for investors. Our results suggest that investors
  do not fully internalize the credit-relevant information contained in earnings calls.
\end{abstract}

\vskip 15pt
\noindent
Keywords: Corporate credit, credit default swaps, return forecasting, NLP \\
JEL Codes: G11, G12, G14 \\
Online Appendix: \url{https://sites.google.com/view/hmamaysky/research}

\thispagestyle{empty}

\newpage
\setcounter{page}{1}

\section{Introduction}

The U.S. corporate bond market is large and growing, both in absolute terms and relative to GDP, and
represents one of the key sources of capital for U.S. corporations.  \vskip -10pt
\begin{figure}[H]
  \centering
  \includegraphics[width=0.8\textwidth]{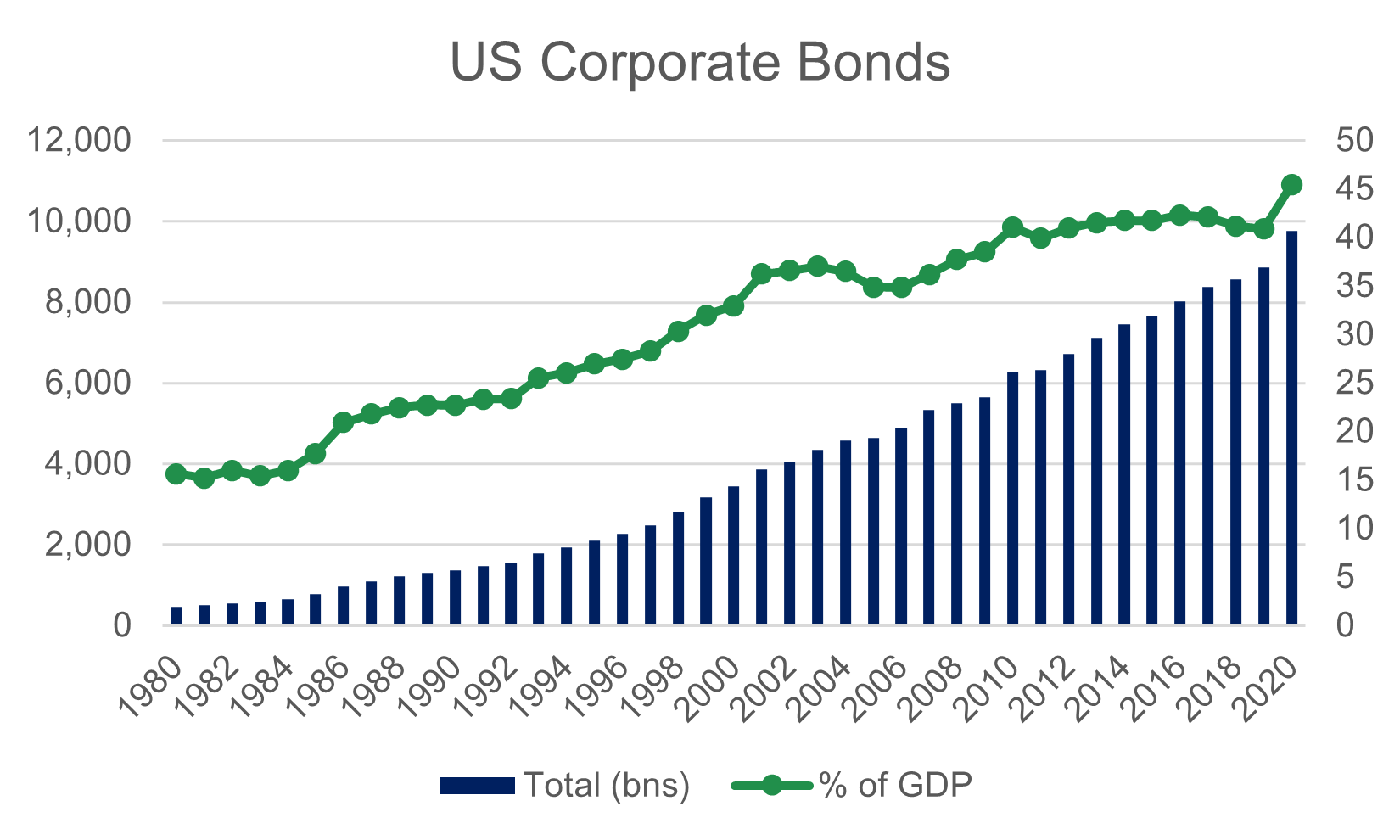}
  \caption{U.S. corporate bonds amount outstanding in billions (from SIFMA) and as a percent of GDP
    (using nominal GDP data from FRED).}
\end{figure}
\vskip -10pt
\noindent
Corporate bond prices reflect investor assessments of a firm's current and {\it future} default
risk. However, many traditional credit measures, such as debt ratios or firm profitability, are more
informative about current rather than future credit risk. Market-based measures of credit risk, like
credit spreads or implied volatilities, are, of course, forward looking, but contain other
confounding influences, like risk or liquidity premia.

An important source of forward-looking credit information for market participants is communication
with management teams. In addition to the publicly disclosed financial metrics that are available
from 10-Ks and 10-Qs, management teams convey to investors their thinking about how a company's
leverage and balance sheet will appear in the future. Earnings calls represent one -- and perhaps
the most -- important channel for regular communication between investors and management
teams. Earnings calls generally occur on a quarterly basis and last between one and one and a half
hours. Given that calls are both infrequent and relatively short, if management teams or investors
choose to discuss information relevant to a company's credit risk, it is likely because such
information is important.  In this paper, we show that quarterly earnings calls contain valuable
information about the future pricing of credit risk, information that is not already reflected in
either credit spreads or in other firm characteristics which have been shown in prior work to
forecast corporate bond returns and risk.

One of the challenges with identifying credit-relevant portions of earnings calls is that call
transcripts often run into the dozens of pages, many of which contain little credit-relevant
information. To address this, we identify a set of credit-related words and phrases, and focus our
analysis on the portions of the earnings call transcripts that are in the vicinity of mentions of
these credit-related terms. Our sample contains all U.S. earnings calls from 2009--2020. Most of
these, and virtually all recent ones, have at least one credit-relevant portion (see Panel A of
Figure~\ref{f:summ-stats}). Using natural language processing (NLP) techniques, we calculate an
implied credit spread for each earnings call. The implied credit spread is obtained by regressing (a
measure related to) firms' credit default swap (CDS) spreads on the count of words, bigrams and
trigrams (collectively, {\it tokens}) that occur in the vicinity of credit mentions in earnings
calls,%
\footnote{We discuss our reasons for using CDS rather than corporate bond data below.}
and then applying the estimated model to the token count of a particular earnings call.

Because of the high dimensionality of this problem -- there are tens of thousands of tokens -- we
use a novel, computationally efficient approach to select the subset of tokens with the highest
explanatory power for CDS spreads.  For each token, we regress the CDS level prevailing immediately
after each call on the number of times that token appears in the credit-relevant portion of the
call. We first select the token whose count leads to the highest regression $R^2$.  We then
recursively regress the residual from the prior step on each of the still-unselected token counts to
find the one with the highest $R^2$.  We implement the above procedure for the panels of investment
grade (IG) or high-yield (HY) CDS spreads separately because the credit-relevant language used in IG
and HY calls is distinct.  Our NLP methodology uses either the full-sample of data or rolling
subsamples. In the latter case, we use no forward-looking information in the construction of the
implied credit spread. The computational efficiency of our methodology -- which results from not
having to run multivariate regressions -- allows us to conduct fully out-of-sample rolling analysis
where new tokens are re-selected in each successive subsample.

We then run a lasso regression of post-call CDS spreads on the selected, high explanatory power
tokens. The lasso, or least absolute shrinkage and selection operator, is a modified regression
which minimizes the mean squared model error while penalizing the sum of the absolute values of the
model coefficients. It is an efficient dimensionality reduction technique; see Hastie, Tibshirani,
and Friedman (2009) for details. We include sector dummies in the lasso, and exclude tokens that are
concentrated in a specific sector (e.g., ``oil wells''). The coefficient estimates from this
regression allow us to associate words and phrases with either better (lower spreads) or worse
(higher spreads) credit news. We do not have to assign an a priori tone to any token as the
algorithm learns this tone endogenously. Furthermore, the rolling (out-of-sample) version of our
analysis allows the credit tone of tokens to vary over time.

Our measure of the information content of earnings calls is called the {\it credit score}, defined
as the difference between a firm's actual credit spread immediately following the earnings call and
the credit spread implied by the lasso model. In the full-sample model, this is just the residual of
the lasso regression; in the rolling model, this is the out-of-sample forecast error. A higher
credit score indicates the market trades at wider spreads than suggested by the language of earnings
calls, and a lower credit score indicates the opposite.%
\footnote{It is common to refer to decreasing (increasing) credit spreads as {\it tightening} ({\it
    widening}) spreads.}
The credit score reflects the degree of disagreement between the market's and management team's
assessment of corporate creditworthiness.

We show that the lagged credit score negatively forecasts 12-month ahead changes in CDS spreads,
even after contemporaneous changes in interest rates, implied volatilities, and firm leverage, as
well as a large set of other variables, which we discuss below, are included as controls. Following
the logic of Collin-Dufresne, Goldstein, and Martin (2001), the contemporaneous regressors in this
specification capture the influences of a Merton (1974)-type model, and thus provide a stringent
test of the forecasting power of lagged credit scores for future spread changes. In a pure
forecasting regression, after the contemporaneous regressors have been dropped from the right-hand
side, the credit score remains a significant and negative forecaster of future CDS changes. Not
surprisingly, without the contemporaneous controls, the credit score effect becomes larger. These
results hold whether credit score is calculated using the full text sample or in rolling
windows. Since the rolling credit score would have been known to investors in real time, this
suggests that credit scores contain valuable out-of-sample credit information, something we
investigate further below. The forecasting results continue to hold for six-month ahead CDS changes,
using both the full-sample and rolling text models.

The evidence thus strongly suggests that implied CDS spreads contain valuable information for
forecasting future credit spreads that is not already impounded into post-call CDS levels. To
understand this result, we show that credit scores also contain information about future CDS market
risk and future corporate fundamentals. Higher credit scores forecast lower CDS risk -- across a
variety of measures -- over the subsequent year. Furthermore, higher credit scores forecast positive
changes in future corporate profitability and declines in firm leverage.  Importantly, these
findings are consistent with the negative forecasting coefficient of credit score for future CDS
changes: lower risk, higher profitability, and lower leverage are associated with lower future CDS
spreads.

There are two potential channels that can explain the ability of credit scores to forecast CDS
spread changes. First, it is possible that market participants are fully aware of the information
content of credit scores for future risk and profitability, but that CDS still rationally responds
with a lag. Unlike stock prices, which immediately respond to all future anticipated changes in cash
flows and discount rates, CDS contracts have a fixed maturity, and so the five-year fixed maturity
CDS spread may evolve predictably as a firm's fundamentals slowly change. For example, a firm that
plans to delever over the next several years will see a lower CDS spread immediately, but may be
expected to see an even lower five-year CDS in one year, and a lower one yet in two years, as credit
risk for the company continues to fall and each successive five-year CDS contract thus reflects a
lower average credit risk over its life. That credit scores forecast future credit risk and firm
fundamentals is consistent with this explanation. The alternative explanation is that market
participants do not fully respond to the credit-relevant information content of earnings calls
because they are capacity constrained (as in Sims 2011) and do not fully internalize all relevant
information. We refer to the two channels as the delayed rational response and the capacity
constrained investors hypotheses, respectively.

Under the rational delayed response hypothesis, characteristics that may proxy for call or firm
complexity should not impact the degree of predictability from credit scores to future CDS
changes. Under the capacity constrained investor hypothesis, more complex calls and calls about more
complex firms should result in greater predictability from credit scores to future CDS spread
changes. Furthermore, under the rational delayed response hypothesis, knowing the credit score of a
firm should not lead to profitable trading strategies because the forecasted CDS spread change is a
publicly known, rational reflection of anticipated future changes in firms' risks or fundamentals.

To identify whether complexity amplifies the forecasting power of credit scores, we follow the
accounting literature and modify our empirical specification to interact credit score with measures
that proxy for the informational environment: the dispersion of analyst earnings forecasts, the
number of analysts covering a given firm, a measure of the language complexity of the earnings call
itself, and the length of the earnings call transcript.%
\footnote{Bhushan (1989) finds that more analysts are associated with larger firms, that have higher
  stock return volatility and more institutional ownership. Lehavy, Li, and Merkley (2011) show that
  firms with less readable communications are associated with more analyst coverage. You and Zhang
  (2009) and Loughran and McDonald (2020) interpret the length of 10-K's as a proxy for either
  complexity or lack of readability.}
We find that more analyst coverage, which proxies for firm complexity, and greater call transcript
length, which proxies for call complexity, increase the predictive power of credit score, by making
the credit score coefficient for future spread changes even more negative.  The other two
interactions also lead to an increase in the magnitude of the credit score coefficient, but the
results are not significant. We interpret these results as evidence supporting the capacity
constrained investor interpretation.

To assess the practical economic impact of credit scores, we turn to an out-of-sample analysis using
the rolling text model, with both the token selection and lasso stages done without using future
information. The output of the rolling text model would have been available to market participants
in real time.  We form long-short credit portfolios that contain the maximally mispriced firms in
every month, while maintaining a zero credit exposure (in a sense to be explained in Section
\ref{s:oos}). Maximally mispriced means that the long side contains firms with the highest credit
scores (whose credit spreads are forecasted to tighten) and the short side contains the credits with
the lowest credit scores (that are forecasted to widen).  Our construction generates a long-short
portfolio that isolates differences in credit scores while maintaining minimal overall credit
exposure. 
{This portfolio construction, as well as a modified version of our forecasting
  regression which includes firm fixed effects, point to the importance of cross-sectional variation
  in the relationship between credit scores and future CDS market outcomes.} More general trading
strategies may take into account other characteristics shown in the literature to forecast returns,
but our approach zeros in specifically on the information content of credit scores.

We evaluate the performance of different parameterizations of this strategy against portfolio
simulations conducted under the null of no predictability, which provides a natural benchmark
against which to compare our backtested returns. We find that different parameterizations of our
trading strategies generate returns that systematically outperform the simulated distribution of
these statistics under the no-predictability null for both IG and HY firms. The outperformance is
statistically significant and economically large, as the strategies can add 3 to 4\% of annualized
return to long-only corporate bond portfolios without increasing their credit betas, which in the
context of the credit asset class is very large economic effect. This provides further evidence in
support of the capacity constrained investor hypothesis, and suggests that the information content
of earnings calls would have been a valuable, real-time tool for credit investors.

\subsection{Relationship to the Literature} \label{s:lit}

Collin-Dufresne, Goldstein, and Martin (CGM, 2001) focus on explaining changes in credit
spreads. They show that several factors suggested by the Merton (1974) model -- changing interest
rates, stock returns, and changes in implied volatility -- can only explain 25\% of monthly spread
changes of corporate bonds. They show that an aggregate credit factor accounts for much of the
variation in the model residuals, and conjecture these movements are caused by common supply-demand
conditions in corporate bond markets. Ericsson, Jacobs, and Oviedo (2009) perform an analysis
similar to Collin-Dufresne et al. (2001) but use CDS data. They argue that CDS data are a cleaner
measure of credit risk than corporate bond spreads, and find that the CGM factors explain a similar
mid-20\% of the variation of CDS spread changes. While the residuals from the CDS version of the
analysis did not have the pronounced common factor found by CGM in corporate bond data, much of the
variation in CDS spread changes still went unexplained.%
\footnote{Ericsson et al. (2009) and Campbell and Taksler (2003) show there is considerably more
  explanatory power for the {\it levels} of CDS spreads or corporate bond yields, as opposed to
  spread changes.}
Bao, Pan, and Wang (2011) show that illiquidity explains a good deal of time series and cross
sectional variation in corporate bond spreads from 2003 to 2009, and its explanatory power rises
during the Global Financial Crisis, lending support to the conjectured CGM mechanism. Our results
suggest that a portion of credit spread change residuals can be explained by our forward-looking
credit score measure, and other control variables introduced since CGM. In our most complete
specification in Table \ref{t:contemp-full sample-12-PVLGD}, the $R^2$ rises to 38.7\%.

There is a large, and growing, literature on the pricing of corporate credit risk. Guo, Lin, Wu, and
Zhou (2021, GLWZ) show that corporate bond sentiment is an important forecaster of future returns on
corporate bonds. They measure bond sentiment as minus the difference between a bond's current credit
spread and the credit spread implied by a fair-value model estimated in rolling windows. High
sentiment negatively predicts future returns, and the opposite holds for low sentiment. Bond
portfolios that short high- and long low-sentiment bonds earn high risk-adjusted returns. Our credit
score-based trading strategy has a similar flavor but uses earnings call implied spreads as the
fair-value benchmark. While we intentionally focus only on variation in credit score to isolate the
value of this information, our trading strategy can be modified to take into account information
from other forecasting variables by using a rolling fair-value approach as in GLWZ.

Bali et al. (2022) apply the dimensionality reduction methods of Gu, Kelly, and Xiu (2020) to a
large number of bond characteristics, and show that the most important predictors of month-ahead
corporate bond returns are liquidity and downside risk, while bond duration and past returns also
matter.%
\footnote{van Binsbergen and Schwert emphasize the importance of duration-matching for return
  measurement.}
They find that imposing dependence between bond and stock characteristics, as suggested by the
Merton (1974) model, improves the forecasting performance of pure machine learning approaches.
Bali, Subrahmanyam, and Wen (2021) show that corporate bond losers over the past 36 months tend to
outperform past corporate bond winners. Bartram, Grinblatt, and Nozawa (2020) document another
mean-reversal pattern in corporate bonds, proxied for by the bond book-to-market ratio, defined as
the bond price over the bond's face value (and closely related to PVLGD). Chung, Wang, and Wu (2019)
study the impact of volatility on the cross-section of corporate bond returns, and show bonds that
hedge volatility increases have low expected returns.  Cao et al. (2022) find that corporate bonds
with large increases in implied volatility over past month have lower future returns relative to
bonds with decreases in implied volatility. Bai, Bali, and Wen (BBW, 2019) show that downside risk,
measured as the 5th percentile monthly return on a corporate bond over the prior 36 months, is a
priced risk factor, with higher historical downside risk forecasting higher future returns. They
also find a liquidity premium in corporate bonds, and a short-term reversal effect. Bai, Bali, and
Wen (2021) show that the systematic risk implied by the BBW (2019) factor model is priced in the
cross section of bond returns, whereas idiosyncratic risk relative to their factor model is not. In
addition, corporate bond market returns are predicted by lagged corporate bond return variance.
Kelly, Palhares, and Pruitt (2021) use instrumented principal component analysis (Kelly et al. 2020)
to jointly estimate factors capable of explaining the cross-section of corporate bond returns and
the time-varying loadings of bonds on these factors. They find that the IPCA model is closely
approximated by a static five-factor model consisting of the bond spread, bond volatility, duration,
and value long-short factors, as well as an equal-weighted corporate bond portfolio.

Relative to the rest of the corporate credit forecasting literature, our key contribution is to
systematically capture credit-relevant information from corporate earnings calls and show that this
information is a useful forecaster of future credit spread changes, CDS market risk, and firm
fundamentals, even when controlling for an extensive set of other corporate bond forecasting
variables, both markets- and fundamentals-based, suggested in the prior work cited above. Our
results show that earnings calls are an important source of forward-looking credit information,
which is not spanned by other predictors of corporate bond returns and risk, and which is not
immediately reflected in market prices.

Our NLP methodology is similar to a recent literature that, rather than specifying word tone a
priori, seeks to extract the tonality of words by using market data. A related approach to ours is
Manela and Moreira (2017) who use a support vector regression (a close cousin of the lasso) to
estimate a mapping from the text of Wall Street Journal articles to the level of the VIX
index. Other related papers that use market responses to extract word tone are Jegadeesh and Wu
(2013), Ke, Kelly, and Xiu (2021), Garcia, Hu, and Rohrer (2022), and Calomiris, Harris, Mamaysky,
and Tessari (2022). Our application of these techniques to the analysis of credit markets is novel
in the literature.

We contrast our study with the recent, independent work of Donovan et al. (2021), who construct a
measure of credit risk by mapping information from firms' conference calls and the Management
Discussion \& Analysis sections of 10-Ks to CDS spreads. They show this information predicts future
credit events including downgrades, bankruptcy, and the level of credit spreads on private deals.
Our work differs from theirs in several important ways. First, we focus on the forecasting power of
our credit score measure for \textit{changes} in firms' CDS spreads, which are harder to predict
than future spread levels, as we have already pointed out. Second, in assessing the forecasting
power of credit score for future CDS spread changes, we control for the information content of
current CDS spreads, credit ratings, and many other firm characteristics, whereas Donovan et
al. (2021) explicitly focus only on firms where CDS spreads and credit ratings are not available.
Third, we investigate potential channels for the forecasting power of our credit measure, and find
evidence for both the delayed rational response and constrained investor hypotheses. Such an
investigation is not possible in the Donovan et al. (2021) setting because they analyze firms with
no traded CDS contracts, and thus have no source of market information to serve as a benchmark
forecast. We also develop a portfolio strategy to test the out-of-sample economic significance of
our measure. Methodologically, we propose a forward selection algorithm to identify tokens with high
explanatory power for firms' credit conditions. This, and our focus on only the parts of earnings
calls in the vicinity of credit terms, reduces the dimensionality of our model and facilitates its
implementation in real time. Due to the differing focus and set of analyzed companies, our paper and
the work of Donovan et al. (2021) are complementary.

The mechanism we document -- that credit-relevant information from earnings calls is not fully
absorbed by market participants -- is similar to Cohen, Malloy, and Nguyen (2020), who show that
quarter-over-quarter changes in the text of 10-Qs and 10-Ks predict negative future corporate
fundamentals (lower earnings and profitability, higher credit risk) and negative stock returns, but
with no announcement day effect. They conclude that investors are inattentive to changes in the
language that firms release in their 10-Ks and -Qs. A similar finding was reported by You and
Zhang (2009), who found a 12-month stock price drift following 10-K filings, suggesting that
investor reaction to the information content of 10-Ks ``seems sluggish.'' Furthermore, You and Zhang
(2009) document that more complex 10-K reports are associated with greater underreaction.%
\footnote{Jiang et al. (2019) show that an index of aggregate manager sentiment based on the text of
  firms' 10-Ks, 10-Qs, and conference calls negatively forecasts stock returns, suggesting stock
  investors overreact to the information content of management communication. Reconciling the
  findings of underreaction to specific parts of earnings calls and 10-Ks with this overreaction
  result to the overall tone of corporate disclosures is an interesting area for future work.}
We provide evidence that investors are similarly inattentive to the credit-relevant portions of
earnings calls, and that this information forecasts future risk and fundamentals, and provides
information that leads to economically meaningful out-of-sample trading performance.

The rest of the paper proceeds as follows.  Section \ref{s:data} describes our data set and control
variable construction. Section \ref{s:text} describes our text data and our credit score
methodology. Section \ref{s:emp} presents the contemporaneous and pure forecasting regression
results. Section \ref{s:channels} discusses mechanisms which can explain our findings.  Section
\ref{s:oos} evaluates our signal's out-of-sample performance. Section \ref{s:conc} concludes.

\section{Data} \label{s:data}

We follow Ericsson, Jacobs, and Oviedo (2009) in using five-year CDS, rather than corporate bond,
data in our analysis. Corporate bonds issued by the same company will often have idiosyncratic
features -- like change of control puts,  different call schedules, or different levels of
seniority -- making comparisons across bonds difficult. Many corporate bonds are very illiquid, and
the implicit assumption made in the literature that trading is possible at or near reported trade
prices is frequently unwarranted.%
\footnote{The Trade Reporting and Compliance Engine (TRACE), run by FINRA, reflects all transactions
  in U.S. corporate bonds. However, many of these transactions, especially those involving retail
  investors, often happen far away from the prevailing institutional prices. For example, a retail
  investor buying from a dealer will pay the dealer's offer price, which tends to embed a large
  transaction cost especially for small trades (Edwards, Harris, and Piwowar 2007). The implicit
  assumption made in much of the literature that a trading strategy can sell bonds at or near this
  transacted price overstates the actual profitability from trading corporate bonds. First,
  non-dealers cannot sell at dealer's offer prices -- they can only buy there. Second, the available
  liquidity at many observed prices, even for a trading strategy willing to buy at this price, is
  likely minimal as a large fraction of all corporate bonds experience little secondary market
  trading. Finally, short selling corporate bonds is often not possible because of lack of
  borrow. \label{foot:TRACE}}
CDS contracts are less sensitive to the idiosyncrasies of particular bonds because of their
cheapest-to-deliver feature: the CDS (or protection) buyer has the right to deliver any of a class
of bonds -- presumably the cheapest -- to the CDS seller in the case of a default event and be paid
the bond's full face value. CDS prices, which we obtain from IHS Markit, are composites of
end-of-day bid-offer quotes submitted by dealers, and are more reflective of market conditions than
bond trade prices, especially when the latter come from small trades. Oehmke and Zawadowski (2017)
argue that speculative credit trading concentrates in CDS markets exactly because their
standardization makes CDS contracts more liquid than the underlying corporate bonds.%
\footnote{While CDS contracts have differing liquidity levels, the variation is less pronounced than
  for bonds, and dealers are willing to trade at least several million dollars notional at the
  stated bid-offer prices.}
Panel B of Figure~\ref{f:summ-stats} shows the number of CDS contracts in our
sample over time.%
\footnote{Rarely CDS contracts leave our sample due to either default or M\&A activity. Section
  \ref{s:cds-exit} of the Online Appendix argues that our CDS sample is nevertheless free of
  survivorship bias as CDS contracts anticipate future credit-relevant events {\it prior} to the
  underlying firms exiting the sample.}
We use five-year CDS levels because, as Ericsson, Jacobs, and Oviedo (2009) show, these represent
the bulk of outstanding CDS contracts.

We transform CDS spreads to a measure we call {\it PVLGD}, which calculates the risk-neutral
expected present value of the future losses ensured by the CDS contract. The quoted CDS spread $S$
is relates to the PVLGD of a CDS contract via
\begin{equation} \label{eq:cds}
  S \times \text{PV01} = \text{PVLGD}.
\end{equation}
The derivation of \eqref{eq:cds} is shown in Section \ref{s:pvlgd} of the Online Appendix. PV01
determines the risk-neutral expected present value of receiving a single basis point annuity over
the life of the CDS contract, where the annuity stops paying in the case of default. With complete
markets, \eqref{eq:cds} shows there are two equivalent ways of buying insurance against default via
CDS. One is to pay the spread of $S$ basis points over the life of the contract, or until default
occurs. The other is to pay PVLGD upfront to the seller of protection. From the seller's point of
view, these two income streams are equally valuable.

The PVLGD can most intuitively be interpreted as follows. Consider a five-year Treasury bond with a
3\% coupon that trades at par. Now consider a risky corporate bond $B$, with the same five year
maturity, the same 3\% coupon, and a price of $P_B$. Since $B$ is risky, $P_B$ is less than
\$100. But how much less? Consider a five-year CDS contract which references bond $B$ and which
trades a PVLGD of $PV_B$. Abstracting away from some modeling and institutional details, buying $B$
and paying $PV_B$ upfront to buy CDS protection provides the equivalent payout to that of the
Treasury bond, and therefore
\begin{equation} \label{eq:noarb}
  P_B + PV_B = \$100.
\end{equation}
The PVLGD of a CDS can thus be interpreted as the discount from par of a corporate bond with the
same maturity and coupon as a par Treasury, and with the same default risk as the CDS contract.  In
this sense, the PVLGD is very similar to the bond book-to-market measure of Bartram, Grinblatt, and
Nozawa (2020).%
\footnote{{The PVLGD of a CDS contract is analogous to the implied volatility of an option. It
    is a model-based transformation of a market price which renders the latter more
    interpretable. As we show in Section \ref{s:robust}, replacing PVLGD with log CDS spreads -- a
    model-free measure -- leaves our results largely unchanged.}}

Using PVLGD and not $S$ in empirical analysis is important because of the large convexity in CDS
spreads.  The PV01 of a CDS contact falls quickly with $S$. Table \ref{t:pvlgds} shows some
representative CDS spreads (in basis points), and the associated PV01s and PVLGDs.
\begin{table}
  \centering
  \def\arraystretch{1.15}
  \begin{tabular}{lll} \hline
    S & PV01 & PVLGD        \\ \hline
    100   & 0.045 & 4.517   \\
    200   & 0.043 & 8.665   \\
    500   & 0.038 & 19.192  \\
    600   & 0.037 & 22.150  \\
    2500  & 0.020 & 49.619  \\
    2600  & 0.019 & 50.208  \\
    \hline
  \end{tabular}
  \caption{This shows the mapping from CDS spread $S$ (in basis points) to PV01 and PVLGD from
    \eqref{eq:cds} using assumptions detailed in Section \ref{s:pvlgd} of the Online
    Appendix.} \label{t:pvlgds}
\end{table}
From \eqref{eq:cds}, to a first-order approximation, the change in the PVLGD of a CDS contract is
given by $\Delta \text{PVLGD} \approx \text{PV01} \times \Delta S$. Since a spread increase of 100
basis points has a much larger PVLGD impact starting at a low spread than starting at a high spread
(e.g., consider the PVLGD impact of a $100 \to 200$ spread move versus $2500 \to 2600$), changes in
CDS spreads are a poor measure of the underlying default risk of a corporate bond in
\eqref{eq:noarb}. Changes in PVLGD provide a much better measure.
As we show in Table \ref{t:pvlgd-bbg-comp} and Figure \ref{f:bbg-cdsw} of the Online
Appendix, our PVLGD methodology exactly matches the industry-standard dollar settlement
calculation for CDS trades.

We use several sources of company-level data: balance sheet, income, and cash flow data from
Compustat; equity price data from CRSP; implied volatility data from OptionMetrics; analyst data
from I/B/E/S; and earnings call data from SP Global (the earnings call data are described in Section
\ref{s:text}). For Compustat, the data {\it report dates} are unavailable for roughly half of the
dataset. To avoid losing these observations, we timestamp Compustat data using the {\it data date}
plus three months: we assume data date $t$ observations are available only as of
$(t + \text{3 months})$ or after. This ensures that we do not use forward-looking information while
allowing us to retain the majority of our data. Creating a mapping between all these datasets is an
involved process. The mapping methodology and other data issues are discussed in detail in Section
\ref{s:datamap} of the Online Appendix.

Markit CDS data are classified into ten sectors: basic materials, utility, financials, consumer
services, technology, energy, consumer goods, industrial, telecommunications, and others. We drop
all financials because many of the controls (discussed below) do not apply to them. Each
firm-quarter observation is also classified into IG or HY using the firm's most recent {\it average
  rating} field from Markit.%
\footnote{Markit credit ratings and S\&P ratings available from Compustat match very closely.}
IG includes AAA, AA, A, and BBB ratings, and HY includes the others (BB, B, CCC, D,
unclassified). In total, there are \numobsig{} monthly observations in the IG group and \numobshy{}
in the HY group.

To control for known determinants of corporate bond returns and credit spread changes, we construct
an extensive set of predictor variables that have been proposed in the literature.  These are
summarized in Table \ref{t:controls}, and more detailed information about their construction is
available in Section \ref{s:controls} of the Online Appendix. Table \ref{t:summ} contains summary
statistics for these control variables. Figure \ref{f:freqfactorCorr} shows the cross-sectional
average of firm-level correlation matrices of our control variables, as well as the PVLGD and credit
score (defined in Section \ref{s:cred-score}). With several exceptions, most correlations between
explanatory variables are quite low, suggesting these capture distinct aspects of a firm's credit
environment. While PVLGD has several moderate correlations with controls (e.g., it is lower for
larger firms and higher for value firms and lower-rated firms), credit score is positively
correlated with PVLGD but largely uncorrelated with all other controls.

\section{Earnings Calls and Text Model Estimation} \label{s:text}

Our sample consists of 202,788 earnings call transcripts obtained from {\it S\&P Global} that took
place from January 2009 and December 2020. Each transcript undergoes several rounds of revisions,
and we use the most recently available version of each transcript, which typically includes
corrections to transcription errors that may have occurred in earlier versions.%
\footnote{SP Global delivers earnings calls transcript in four versions. Ranking by how soon a version
  is available after the call, there are: Spell Checked (minutes after the call), Edited (3 hours
  after the call), Proofed (2 hours after the Edited copy), Audited (audited after the Proofed copy,
  no real timing). In our analysis, for each earnings call, we use the version that is the latest
  available.}
We date an earnings call as having occurred on day $t$ if its announcement date-time took place
between 4 PM on day $t-1$ and 4 PM on day $t$.%
\footnote{Throughout, we refer to business days, not calendar days. For example, if day $t$ is a
  Friday, day $t+1$ is the subsequent Monday.}
Earnings calls that take place after 4 PM on day $t$ are therefore dated as of day $t+1$. For our
analysis, we need to match a firm's quarterly earnings call with its CDS in that quarter (see
Section \ref{s:datamap} of the Online Appendix). Panel C of Figure~\ref{f:summ-stats} shows the
number of earnings calls that can be matched to CDS data in each quarter of our sample.  There are a
total of \numobs{} firm-quarter observations with matched earnings call, CDS, and control variable
data.

To extract credit-relevant information from earnings calls we first combine their presentation and
Q\&A sections into one.%
\footnote{We tried a version of our analysis using only the Q\&A portion of earnings calls, but
  found this approach worked less well than using both the presentation and Q\&A sections.}
Because earnings calls are very long (transcripts are often between 20-30 pages), we need to
identify the portions of earnings calls that contain credit relevant information. To do this, we
manually collected a list of credit words and phrases that are indicative of discussions about a
firm's creditworthiness.  We started with a short list of seed words, like ``credit,'' ``credit
line,'' ``debt,'' and so on, and then identified frequently co-occurring words and phrases by
reading hundreds of call segments containing the initial list of seed words. Sometimes a credit word
is used in a non-credit context. For example, ``rate'' may refer to a company's financing rate, but
when used in the phrase ``exchange rate'', it no longer conveys the correct meaning. To address
this, we identified a list of excluded phrases so that any credit word that appears in an excluded
phrase does not indicate a credit-relevant part of the earnings call. The credit words and excluded
phrases lists are shown in Table \ref{t:word-lists} of the Online Appendix.

For a given earnings call, we then identify all credit sentences, which are those containing one or
more of the credit words from our list. However, if all the credit words in a sentence come from
excluded phrases, that sentence is not identified as a credit sentence. We then take the union of
all sentences that occur five sentences before or after credit sentences. All other parts of the
earnings call are dropped. We then clean the text in the retained sentences by stemming all words
using the Snowball stemmer from Python's \texttt{NLTK} package and replacing numbers with tokens
that indicate magnitude: \texttt{\_bln\_} for numbers in the billions, \texttt{\_mln\_} for numbers
in the millions, and \texttt{\_num\_} for numbers smaller than a million.%
\footnote{Section \ref{s:text-proc} in the Online Appendix gives details.}
We generate a document term matrix (DTM) for each earnings call using the cleaned, retained
sentences. Each row of the document term matrix corresponds to the earnings calls of firm $i$ on day
$t$, and includes the counts of the unigrams (words), bigrams (two-word phrases), and trigrams
(three-word phrases) that appeared in that call (we refer to each as token $j$). In our analysis, we
retain the top $N \in \{2000,5000\}$ highest-frequency terms once the DTM has been constructed
(without this pruning, the DTM would have 947,243 terms).  The resultant DTM has \numobs{} rows and
either 2000 or 5000 columns, is relatively sparse,%
\footnote{For $N = 5000\, (2000)$, 17\% (31\%) of the DTM entries are non-zero.}
and reflects {\it credit-related} language across our earnings call corpus.

\subsection{Ranking Tokens} \label{s:fwd-sel}

The core of our text methodology is to estimate a mapping from the credit-related language of a day
$t$ earnings call to the closest subsequent closing CDS level -- measured via the PVLGD
transformation from (\ref{eq:cds}).  If a call occurs prior to 4 PM on day $t$, the associated CDS
will be the day $t$ close; if the call takes place after 4 PM on day $t$, the associated CDS will be
the closing level on day $t+1$.  Estimating a mapping with either 2000 or 5000 tokens is
challenging, especially in the out-of-sample version of our analysis where we perform our text
analysis in rolling windows (discussed below). One way to reduce the dimensionality of the problem
is to focus on a smaller subset of the most frequent tokens in the DTM. The disadvantage of this
approach is that some tokens occur frequently, for example the word ``said,'' and yet have very
little explanatory power for PVLGDs. Rather than choosing the most frequent terms in the DTM, we
focus on selecting the most informative ones using a {\it forward selection} procedure as follows.

Let $f_{i,t,j}$ denote the number of times token $j$ appears in the earnings call of firm $i$ at time
$t$. The associated PVLGD is $PV_{i,t}$, derived using firm $i$'s closing CDS on day $t$.  We first
find the token $j^{(1)}$ with the highest absolute correlation,
$|\corr\left(PV_{i,t},f_{i,t,j^{(1)}}\right)|$, with the set of pooled PVLGDs. Once the first token is
selected, we regress its frequency out of the PVLGDs using the pooled regression
\begin{equation*}
PV_{i,t} = \alpha^{(1)} + \beta^{(1)} f_{i,t,j^{(1)}} + \xi_{i,t}^{(1)}.
\end{equation*}
Here $\xi_{i,t}^{(1)}$ is the residual once the count of token $j^{(1)}$ has been removed from all
PVLGDs. We select the next token $j^{(2)}$ as the one which has the highest correlation with
$\xi_{i,t}^{(1)}$, and then regress the count of that token out of $\xi_{i,t}^{(2)}$. The $n$th token
in this process is selected as the one with the highest absolute correlation between its count and
the residual from the $n-1$st regression, i.e., the largest
$|\corr(\xi_{i,t}^{(n-1)},f_{i,t,j^{(n)}})|$. We then calculate the residuals $\xi_{i,t}^{(n)}$ from the
regression
\begin{equation} \label{eq:xi-res}
  \xi_{i,t}^{(n-1)}= \alpha^{(n)} + \beta^{(n)} f_{i,t,j^{(n)}} + \xi_{i,t}^{(n)}.
\end{equation}
We repeat the above step until the desired number of tokens is selected.

This recursive procedure is similar to a traditional forward selection regression model, where a new
token is selected in the $n$th step as the one which maximizes the incremental $R^2$ in a
specification that includes the prior $n-1$ selected terms. However, this method is orders of
magnitude slower than ours because of the need to run a high-dimensional regression for each
candidate token in the $n$th step of the process. Our procedure essentially only involves
calculating a correlation between each candidate variable and the residual from the $n-1$st
step. The downside of our method is that the residual $\xi^{(n)}$ from the $n$th step is not
guaranteed to be orthogonal to the first $n-1$ selected tokens (though it is orthogonal to token
$j^{(n)}$ by construction), whereas in the traditional forward selection approach such orthogonality
is guaranteed. However, in unreported results, we verified that correlations between $\xi^{(n)}$ and
the already chosen tokens are close to zero. The sizable computational gain of our method far
outweighs this slight suboptimality. This computational speedup is less important for the
full-sample analysis, but is crucial for out out-of-sample tests in which the forward selection
procedure is run in rolling windows.

We implement the above procedure for investment grade and high yield names separately because these
groups of companies typically use different language in the credit-related portion of their earnings
calls.  Table \ref{t:lasso-words} shows the first 20 positive and negative correlation tokens -- in
terms of the signs of their $\beta^{(n)}$ coefficient from (\ref{eq:xi-res}) -- identified in the
forward selection process for the IG and HY samples. For IG some of the top-selected positive
correlation tokens include ``invest\_grade,'' ``coven'' (the stemmed form of covenant), and
``loan.'' The presence of these tokens is generally associated with higher PVLGDs (weaker
credit). Some of the top-selected negatively correlated tokens, i.e., those associated with lower
PVLGDs (stronger credit), are ``growth,'' ``sale.\_num\_,'' and ``quarter.increas.\_num\_.''  For
HY, the top-selected positively correlated tokens are ``matur,'' ``amend,'' and ``net.loss,'' all of
which are associated with wider credit spreads (worse credit). The top-selected HY tokens associated
with lower credit spreads (better credit) are ``growth,'' ``share\_repurchase,'' and
``earn.\_num\_.''%
\footnote{Bigrams with a (\_) separator are selected credit phrases from Table \ref{t:word-lists} of
  the Online Appendix, while those with a (.) separator were identified in the set of
  frequently-occurring tokens.}
We further discuss these selected words and phrases, and explain the Coefs column in Table
\ref{t:lasso-words}, in Section \ref{s:text-imp}.

Some tokens identified in the forward selection process are typically used by firms from a single
sector (e.g., ``bakken.shale''). Such tokens are functionally equivalent to a sector fixed effect,
which is included explicitly in our analysis, as discussed below. We therefore drop from the DTM
those tokens that are very prevalent in a single sector. For each token $j$, we calculate its
maximum sector concentration via
\begin{equation} \label{eq:c_j}
  c_j = \max_k \frac{\sum_{i,t} f_{i,t,j} \one[i \in k]}{\sum_{i,t} f_{i,t,j}},
\end{equation}
where $f_{i,t,j}$ is the number of time token $j$ appears in firm $i$'s time $t$ earnings call and
$\one[i \in k]$ is an indicator for whether firm $i$ belongs to (the set of firms in) sector
$k$. The $c_j$ measure shows the fraction of all occurrences of token $j$ that took place in the
most prevalent sector for that token.%
\footnote{Some examples of full-sample $c_j$ values: bakken 0.958, oil.equival 0.975,
  medicar.advantag 0.973, eagl.ford 0.972, kroger (a supermarket chain) 0.996, nucor (a steel
  manufacturer) 0.978, airplan 0.952, nike 0.914, therapi 0.908, restaur 0.900, capital.alloc 0.155,
  investor 0.163, cash\_flow 0.195. Since we have nine sectors, $c_j$ ranges from $1/9$ to $1$ by
  construction.}
The subsequent analysis drops tokens with $c_j > t_c$ for some fixed threshold $t_c$. Since
$c_j \leq 1$, when $t_c = 1$ the filter does not drop any tokens.

\subsection{Credit Score} \label{s:cred-score}

The prior section showed how we can rank tokens by their informativeness for explaining the panel of
PVLGDs and how we can measure sector concentrations of particular tokens. We now turn to our core
task of estimating a mapping from the credit-related language of earnings calls to PVLGDs. 

We refer to the DTM whose columns correspond to the retained tokens as $L$. We then estimate
a mapping between PVLGDs and this DTM as follows
\begin{equation} \label{eq:lasso}
  PV_{i,t} = S_{k(i)} + L_{i,t} \beta + \varepsilon_{i,t},
\end{equation}
where $PV_{i,t}$ is the PVLGD of the $i$th firm on day $t$ (following the pre-/post-4 PM timing
convention described at the start of Section \ref{s:fwd-sel}), $S_{k(i)}$ represents the sector
fixed effect for sector $k(i)$ of firm $i$, $L_{i,t}$ is the row of the DTM corresponding to this
earnings call, and $\beta$ is an $N_{FS}$-dimensional column vector.%
\footnote{We tried a version of (\ref{eq:lasso}) with PVLGD changes, i.e,
  $PV_{i,t} - PV_{i,t-\delta}$ where $\delta \in \{1,5,21,92,183,365\}$ is the number of days from
  the day $t$ earnings call. We also ran the model with quarter-over-quarter changes
  in $L_{i,t}$ as the explanatory variable. Neither variant worked as well as the one in
  \eqref{eq:lasso}. }
Even with the dimensionality reduction entailed in the choice of $N_{FS}$, the regression in
(\ref{eq:lasso}) is still too high-dimensional to yield stable estimates using traditional
methods. Because of this, we estimate (\ref{eq:lasso}) using a lasso regression with a penalty
parameter selected using 10-fold cross-validation.%
\footnote{This procedure is implemented using Python's \texttt{scikit-learn} toolkit. All regressors
  in (\ref{eq:lasso}) can potentially be excluded by the lasso.  A future direction is to explore a
  neural network approach for estimating \eqref{eq:lasso}.  One advantage of our lasso approach is
  greater interpretability of the text model.}

The model-implied PVLGD is defined as the fitted value from the regression in (\ref{eq:lasso}), i.e.,
$\hat{PV}_{i,t} = \hat S_{k(i)} + L_{i,t} \hat \beta$.  In the full-sample analysis, we refer to the
residual from (\ref{eq:lasso}) as the {\it credit score} associated with the earnings call:
\begin{equation} \label{eq:cred-score}
CS_{i,t} = PV_{i,t} - \hat{PV}_{i,t} = \hat \varepsilon_{i,t}.
\end{equation}
The credit score is thus the difference between the actual and model-implied PVLGD. Note that the
actual PVLGD, $PV_{i,t}$, is observed {\it after} the earnings call takes place, and should thus
reflect the information content of the call. In the rolling analysis (described below), the credit
score is still defined as in (\ref{eq:cred-score}), but the implied PVLGD $\hat{PV}_{i,t}$ is
calculated using a model estimated with only historical data applied to the frequency count,
$L_{i,t}$, of the present earnings call.

A positive credit score means the actual PVLGD is higher than the one implied by the earnings call,
and thus the earnings call contained language that is typically associated with tighter credit
spreads than those prevailing in the market immediately after the call. A negative credit score
means that the earnings call language called for higher credit spreads than what prevailed in the
market immediately after the call. The credit score thus captures information that is present in the
firm's earnings call, but not in market CDS prices. The credit score is our main variable of
interest in subsequent analysis.

\subsection{Text Model Implementation} \label{s:text-imp}

We perform the forward selection and model estimation from Sections \ref{s:fwd-sel} and
\ref{s:cred-score} both for the full data sample, as well as in rolling windows for out-of-sample
analysis. The rolling analysis uses no forward looking information, and produces a text model that
would have been available to investors in real time. We perform our main analysis based on
full-sample credit scores because these best capture the information in earnings calls that is
relevant for future credit outcomes. For robustness, we also run versions of our regressions using
the rolling text model. Section \ref{s:oos} develops a trading strategy based on rolling credit
scores to show the out-of-sample validity of our approach.

Our estimation has several parameters. One of these is $N \in \{2000,5000\}$ which determines
whether our starting DTM contains the most frequent 2000 or 5000 terms. The second parameter is
$N_{FS} \in \{250, 500, 1000\}$ which refers to the number of the most informative words from the
recursive forward selection procedure of the prior section that are retained for the mapping.
Finally, we allow the concentration threshold to vary with $t_c \in \{1/3, 1/2, 1\}$: any token
whose $c_j$ (from eq. \ref{eq:c_j} estimated using the entire corpus) is greater than $t_c$ is
dropped from the analysis and is replaced by the next highest-ranked explanatory token from the
recursive selection method with $c_j < t_c$ until $N_{FS}$ tokens are reached. These parameter
choices generate 18 possible versions of the analysis, each of which contains its own set of tokens,
though there is substantial overlap among the variants; Panel A in Table \ref{t:nlp-params} summarizes the
possible parameter combinations.

\begin{table}
  \centering
    \begin{tabular}{c|c}
    \toprule
      \multicolumn{2}{c}{\bf Panel A: Full-sample and rolling parameters} \\ \midrule
      DTM {most frequent} words ($N$) & 2000, 5000 \\
      Sector concentration threshold ($t_c$) & 1/3, 1/2, 1 \\
      Selected words ($N_{FS}$) & 250, 500, 1000 \\ \midrule
      \multicolumn{2}{c}{\bf Panel B: Rolling parameters} \\ \midrule
      Annual updating & 2-, 3-, 5-year, and expanding windows \\
      Monthly updating & 3-year window \\
      \bottomrule
    \end{tabular}
    \caption{Specification parameters for the full-sample and rolling text models.}
    \label{t:nlp-params}
\end{table}

The 18 model variants described above are used for our full-sample analysis.  For the rolling
analysis, we add two additional parameters. The first determines the lookback period over which we
perform forward selection and over which we estimate the text-PVLGD mapping from
(\ref{eq:lasso}). The second parameter, the updating frequency for the model, determines whether our
rolling window estimation moves forward by one month or one year at every step. For annual updating,
we use four different training horizons in the rolling analysis: 2-, 3- and 5-year rolling windows,
and an expanding window from the start of the sample to the present. For monthly updating, we only
use 3-year rolling windows because of the computational costs involved. This means that our of
out-of-sample analysis has $18 \times 5 = 90$ text model variants, which are summarized in Panel B
of Table \ref{t:nlp-params}.

For the full-sample and out-of-sample analyses, we estimate the text model separately for the IG and
HY samples. We do so because the credit-related language used by investment grade and high yield
firms tends to be quite different, as Table~\ref{t:lasso-words} demonstrates. For each ratings class
(IG or HY), we select the full-sample model (out of the 18 variants) with the highest $R^2$ from the
lasso regression in (\ref{eq:lasso}).%
\footnote{The lasso $R^2$ is calculated as $1 - \var(\hat
  \varepsilon_{i,t})/\var(PV_{i,t})$.  The highest $R^2$ full-sample text models for the IG and
    HY subsamples coincide at $N=5000$, $N_{FS}=1000$, and $t_c=1$.}
We then use the implied PVLGDs and credit scores from the selected models in subsequent analysis.

For the rolling analysis, we select in each period the model (out of 90 candidates) with the most
predictive credit score in the training window.  We run a forecasting regression for 12-month ahead
PVLGD changes using data in the traininig window, {and ensuring that the dependent variables
  do not extend outside the training window}. The forecasting regression, described in Section
\ref{s:forecast}, includes as regressors the credit score from a given text model estimated in the
training window, the current PVLGD, and an extensive set of other controls. We select the text model
whose credit score coefficient has the largest magnitude t-statistic. {We conjecture that text
  models whose credit scores are good predictors of PVLGD changes in training windows are also the
ones whose credit scores are good out-of-sample predictors.}

When the training window moves forward and the forecasting model is re-estimated, we choose a new
text model following the same procedure. Each period in our out-of-sample analysis may thus use
$\hat S_{k(i)}$ dummies and a $\hat \beta$ vector obtained from a different text model from the set
of 90 variants.  We provide further details about the rolling analysis in Section \ref{s:rolling} of
the Online Appendix, but emphasize that both the model estimation and selection are done without
using any future information. The signals resulting from this procedure would be available to
investors in real time. Section \ref{s:valid_ap} of the Online Appendix shows that PVLGDs and
implied PVLGDs (both full-sample and rolling) from \eqref{eq:lasso} are related, but imperfectly so,
which suggests that deviations between these two measure might be informative, which we begin to
investigate in Section \ref{s:emp}.

\subsection{Word Tone}

Table \ref{t:lasso-words} shows the full-sample lasso coefficients associated with the first 20
positive and negative tokens chosen via the recursive forward selection method of Section
\ref{s:fwd-sel} applied to the IG and HY subsamples. In all cases, the full-sample lasso coefficient
from (\ref{eq:lasso}) has the same signs as the $\beta^{(n)}$ coefficient in (\ref{eq:xi-res}),
which argues that the non-orthogonality of residuals issue in our recursive forward selection method
is not problematic in practice. For example, when an IG firm mentions ``invest\_grade'' in its
earnings calls, this tends to increase its credit spread since emphasizing its credit rating may
indicate a negative signal about its credit condition.  On the other hand, mentioning ``growth'' in
the earnings call decreases the credit spread for both IG and HY firms, as it is likely suggesting
good prospect for the firm. For IG firms, discussions of sales (``sale.\_num\_'') and investor
relations (``investor.relat'') are interpreted positively and lower credit spreads, {while the
  word ``vendor'' raises credit spreads, perhaps because it is associated with firms' working
  capital issues}. But the word ``amend'' (stemmed form of ``amendment'' or ``amending'') is very
negative for HY credit spreads (i.e., pushes them higher), as the market does not like to hear
management teams discuss amending things like bond covenants on conference calls. The same is true
of ``matur'' which may indicate investor nervousness when HY management teams discuss maturity
extensions or are overly focused on the maturity structure of their debt. On the other hand, and
perhaps surprisingly, ``share\_repurchas'' is associated with lower HY spreads, as the market
believes HY companies would only buy back their shares if their management teams were confident in
the company's creditworthiness. Our method's ability to endogenously determine the meaning of credit
words, and to allow this meaning to change over time in the rolling version of our model, is a key
feature of our approach.%
\footnote{The coefficients in Table~\ref{t:lasso-words} show the marginal impact of each token, but
  the global impact also depends on the correlation structure of token incidence. For example, if
  the tokens ``faster'' and ``growth'' frequently co-occur, the coefficient of either one in
  isolation in \eqref{eq:lasso} may be misleading.}

\subsection{Event Studies} \label{s:es}

We perform an event study to demonstrate the effectiveness of our credit score measure.  We bucket
firm-quarter PVLGD and credit score observations into deciles using independent sorts. We analyze
four types of non-overlapping {\it extreme quarters}: where the PVLGD is in the bottom decile (i.e.,
low) but the credit score is not in the bottom decile; where the credit score is in the bottom
decile (i.e., very negative) but the PVLGD is not in the bottom decile; where the PVLGD, but not the
credit score, is in the top decile; and where the credit score, but not the PVLGD, is in the top
decile.  We look at the behavior of firm-level PVLGD in the eight quarters before and after the
occurrence of {\it extreme} quarters. We use the full-sample text model for the analysis. All PVLGDs
are normalized by dividing by the extreme quarter PVLGD, so the time zero level is one by
construction.

Figure \ref{f:es} shows the event studies for the four types of extreme events. Each
event study is an average across all extreme firm-quarter observations for which all 17 quarters of
data exist. The top left chart in the figure shows that extreme low PVLGD quarters are preceded by
dropping PVLGDs and then followed by increasing PVLGDs that largely undo the prior eight quarters of
tightening. The double sorts ensure that these events are not contaminated by extremely low credit
scores, and so emphasize the effect of only having a low PVLGD. The top right chart shows the same
analysis for the lowest decile of credit score firm-quarter observations which are not also in the
bottom PVLGD decile. Low credit score quarters are also preceded by tightening PVLGDs and then
followed by widening PVLGDs. The size of the effect is comparable to the lowest PVLGD decile events
(the y-axis range in the two charts is not the same, but the post-event reaction is very similar).

The bottom left chart of Figure \ref{f:es} shows that the top decile of PVLGD firm-quarter
observations, which are not in the top decile of credit score quarters, are not preceded by credit
spread widening.  The state of having high PVLGDs that are consistent with management language
reflected in implied PVLGDs -- thus the credit score levels are not extreme -- is persistent. But
conditional on having extreme high PVLGDs, there is evidence of tightening in the subsequent eight
quarters.%
\footnote{In bottom left chart of Figure \ref{f:es}, many events have PVLGDs in the preceding eight
  quarters that are also in the top PVLGD decile. But these preceding quarters happen in the early
  part of our sample (which starts during the global financial crisis of 2007-2009) and thus do not
  show up as events because they lack the required 17 quarters of data to be included.}
The bottom right chart of the figure shows that extreme high credit score firm-quarter observations,
which are not extreme high PVLGD quarters, are preceded by credit spread widening, and then followed
by strong tightening with a drop in PVLGD of roughly 25\% of the event quarter PVLGD in the
subsequent two years. This is a much larger effect than the roughly 15\% tightening that follows
quarters with extreme high PVLGDs but not extreme credit scores.

The evidence from the event studies in Figure \ref{f:es} suggests that both extreme PVLGD and credit
score observations are followed by PVLGD reversals. There is a distinct effect for both extreme
credit scores and extreme PVLGDs, and the evidence of mean reversion appears strongest following
extreme firm-quarter credit score observations. That future PVLGD changes go in the direction of
implied PVLGDs suggests that our text-based implied PVLGD measure contains valuable information that
is not already reflected in post-call CDS levels. Section \ref{s:emp} establishes the forecasting
ability of credit scores rigorously by controlling for many other predictors of credit returns.

In Table \ref{t:event_example}, we show one example each of an outlier event in deciles 1 (low credit
scores) and 10 (high credit scores) of the extreme credit score, but not-extreme PVLGD, quarters. An
outlier credit score often results from a large quarter-over-quarter divergence between implied and
actual PVLGDs. The panels in the table correspond to firm-quarter observations, and show the top
ten words or phrases sorted by their contribution to the change in that firm's implied PVLGD from
the prior to the current (outlier) quarter. For example, if a firm uses the word ``liquid'' four
more times in the outlier quarter relative to the prior one, and if the language model coefficient
associated with ``liquid'' is 0.32, this would contribute to a quarter-over-quarter implied PVLGD
increase of 1.28.

The decile 1 example is the New York Times (NYT) credit score from July to November 2017. The
implied PVLGD of the company rose by over eight points, while its PVLGD remained almost
unchanged. Contributing to the increase in implied PVLGD was the less frequent use of the phrase
``revenue expectations'' which is associated with lower PVLGD scores (hence the negative coefficient
in the table), more frequent use of the word ``obligation'' (root ``oblig'') which is associated
with higher PVLGD scores, and less frequent use of words like ``user,'' ``growth,'' ``workforce,''
and ``acquire'' all of which are associated with lower PVLGDs. Based on our full-sample text model
this difference in conference call credit-related language should have been associated with a large
rise in NYT's PVLGD. This likely did not happen because, at the time of this conference call, NYT
has no bonds that were deliverable into the CDS contract (the latter can still trade at positive
spreads in anticipation of bond issuance) with the last NYT bond having matured in December
2016. So even though management language was credit-negative, we conjecture there was no CDS
reaction because market participants did not anticipate imminent bond issuance from the company.

The decile 10 example is KB Home (KBH), a publicly traded home builder. From January 2016 to March
2016, KBH's PVLGD fell by three points, but its implied PVLGD fell by over nine points. The largest
contributors to the fall in implied PVLGD were more frequent use of the phrases ``share repurchase''
which is associated with lower credit spreads (perhaps because companies repurchase shares when
their balance sheets are strong), ``share,'' ``revenue expectation,'' ``operating income margin,''
and less frequent use of the word ``liquid'' (or ``liquidity'') which is associated with higher
credit spreads. Managerial tone during KBH's conference call suggested that credit spreads should
have fallen more quarter-over-quarter than they actually did.

\section{Empirical Results} \label{s:emp}

Our empirical analysis consists of two parts. First, we analyze a specification that regresses
changes in PVLGD on contemporaneous changes in several key explanatory variables. We show that the
presence of lagged credit score improves the explanatory power in this specification, which is
surprising given that credit score is a lagged covariate. We then drop the contemporaneous terms and
analyze a pure forecasting version of the regression. We again conclude that credit score is an
important forecasting variable even after controlling for an extensive collection of other credit
forecasting variables. We start off using the full-sample text model, which best reflects the actual
information content of earnings calls, and then test the robustness of our results using the rolling
model in Section \ref{s:robust}.

In the subsequent analysis, we drop 267 observations with PVLGD larger than 30, corresponding to a
CDS spread higher than 2350 basis points on average.  These outliers account for 1.9\% of our final
sample.  They are usually companies in very distressed conditions, for which our credit score may
not be an informative measure.

\subsection{Contemporaneous Analysis} \label{s:contemp}

We first analyze the degree to which changes in PVLGD can be explained by contemporaneous variables
{that are suggested by Merton (1974); this connects our paper to the work on explaining
  contemporaneous changes in credit spreads done by Collin-Dufresne, Goldstein and Martin (2001) and
  Ericsson, Jacobs and Oviedo (2009).}  Because we use credit score as a regressor, we restrict the
sample to firm-months pairs $(i,t)$ where an earnings call exists for firm $i$ in month
$t$. $PV_{i,t}$ denotes firm $i$'s PVLGD level in month $t$. Here and in subsequent analysis, we use
$t$ to refer to months, and not the days of earnings calls as before. We consider the 12-month
changes in PVLGD: $\Delta PV_{i,t+\ell} = PV_{i,t+\ell} - PV_{i,t}$ with $\ell=12$.  To explain
$\Delta PV_{i,t+\ell}$, we first use the contemporaneous changes in three key variables: firm $i$'s
leverage $LEV_{i,t}$, defined as its long-term debt to long-term debt plus market capitalization
ratio, the risk-free rate $R_t^{(f)}$, and the implied volatility of firm $i$'s options $IV_{i,t}$
({see Table \ref{t:controls} for variable definitions}). Denote the 12-month changes in these
variables by $\Delta LEV_{i,t+\ell}$, $\Delta R_{t+\ell}^{(f)}$, and $\Delta IV_{i,t+\ell}$,
respectively.  These three variables follow from the structural model in Merton (1974), which shows
a firm's credit spread can be fully explained by the firm's leverage, asset volatility, and the
risk-free rate.  Cremers, Driessen, Maenhout, and David (2008) further find the option implied
volatility is more effective in explaining the CDS spread than realized stock volatility.

To capture potential momentum or mean-reversion in CDS spreads, we include the current PVLGD level
$PV_{i,t}$ in the regression.  We use the credit score $CS_{i,t}$ defined in (\ref{eq:cred-score})
to test if it can explain future changes in CDS spreads. Finally, we control for a set of
fundamental, equity, and credit factors ($X_{i,t}$) that have been shown to explain bond yields or
credit spreads in different settings; these are listed in Table \ref{t:controls}. The full
specification of the contemporaneous regression is given by:
\begin{align}
  \Delta PV_{i,t+\ell} &= \alpha + \beta_{lev} \Delta LEV_{i,t+\ell} + \beta_{rf}\Delta
                         R_{t+\ell}^{(f)} + \beta_{vol} \Delta
                         IV_{i,t+\ell} \label{eq:contemp-reg} \\
                       & \quad + \beta_{pv}PV_{i,t}+\beta_{cs} CS_{i,t}+\beta_{fac}^{\top} X_{i,t} +
                         \varepsilon_{i,t}. \nonumber
\end{align}
Our interest is in $\beta_{cs}$, which reflects the impact of the credit score, after controlling
for many other forecasting variables.  We cluster standard errors in (\ref{eq:contemp-reg}) at the
firm-month level. Table \ref{t:summ} shows summary statistics for the variables involved in this
analysis.

The regression results for 12-month ahead PVLGD changes are reported in Table \ref{t:contemp-full
  sample-12-PVLGD}.  The t-statistics of the coefficients are shown in parentheses, and only control
variables that are significant in at least one specification are retained. This regression uses the
full-sample credit score (results using the rolling text model are discussed in Section
\ref{s:robust}). In the first column, we only include the contemporaneous changes in the three
variables: risk-free rate, implied volatility, and leverage. We see all the three variables are
statistically significant with the expected signs in the full sample. Specifically, PVLGD decreases
in contemporaneous changes in the risk-free rate, but increases with implied volatility and leverage
ratio. The $R^2$ from the regression is 26.1\%, suggesting a large proportion of the variation in
credit spread remains unexplained by the structural model. This is consistent with the findings in
Collin-Dufresne, Goldstein, and Martin (2001) and Ericsson, Jacobs, and Oviedo (2009). The results
are similar when we add other known predictors of credit spread changes, as shown in the second
column.

In column (3), we add the current PVLGD level to the regression. We find the coefficient of
current PVLGD is negative and statistically significant, suggesting PVLGD has a mean-reverting
tendency: a one point increase in current PVLGD is associated with a 0.223 point decrease in PVLGDs
over a 12-month horizon.  Finally, the specification in column (4) further adds the credit score.
We find it has a negative coefficient of $-0.177$ and is statistically significant with a
t-statistics of $-9.75$.  Recall the credit score is defined as the actual PVLGD minus the implied
PVLGD from the earnings call. Thus, the negative coefficient of credit score suggests that when the
implied PVLGD is lower than the actual PVLGD (credit score is positive), the PVLGD tends to decrease
in the next 12 months. That is, the PVLGD moves {\it towards} the level implied by the earnings call
over a time-horizon of 12 months. A negative credit score (implied higher than actual) will tend to
increase PVLGDs over the next 12 months by 18\% of the deviation of implied from actual PVLGD at the
time of the call, which is a very large impact. This suggests that the credit score contains
information that is not spanned by other market variables -- including {\it future} changes in firm
$i$'s leverage, implied volatility, and interest rates -- and which is useful for forecasting future
changes in PVLGD in the direction of the current earnings-call implied PVLGD.

Our results contribute to the resolution of -- though do not fully resolve -- a long-standing
question in the credit literature: What drives time-series {\it changes} in credit spreads? Part of
what drives changes in credit spreads is a mean-reverting tendency of PVLGDs and another part has to
do with communication from management teams about the creditworthiness of their firms. The
incremental contribution to $R^2$ of this information, as well as of the other control variables
(going from column 1 to 4 in Table \ref{t:contemp-full sample-12-PVLGD}) is 12.6\%, bringing the
full model $R^2$ to 38.7\%.

\subsection{Forecasting Analysis}\label{s:forecast}

The regression in (\ref{eq:contemp-reg}) -- which simply mirrors the specification in
Collin-Dufresne, Goldstein, and Martin (2001) and Ericsson, Jacobs, and Oviedo (2009) -- provides an
interesting correlation analysis between changes in credit spreads and contemporaneous changes in
other firm-level and market variables.  However, it cannot be used to forecast PVLGD as the
contemporaneous changes $\Delta LEV_{i,t+\ell}$, $\Delta R_{i,t+\ell}^{(f)}$, and
$\Delta IV_{i,t+\ell}$ are obviously not known ex-ante. Furthermore, the regression in
(\ref{eq:contemp-reg}) suffers from potential endogeneity issues. For example, the increase in a
firm's implied volatility may be caused by the deterioration in its credit condition, instead of the
other way around.

It is cleaner to drop the contemporaneous terms from (\ref{eq:contemp-reg}), which then leads to a
pure forecasting specification, with no econometric issues:
\begin{equation} \label{eq:forecast-reg}
  \Delta PV_{i,t+\ell} = \alpha + \beta_{pv}PV_{i,t}+\beta_{cs} CS_{i,t}+\beta_{fac}^{\top} X_{i,t}
  + \varepsilon_{i,t}.
\end{equation}
We retain the control vector $X_{i,t}$ in the forecasting model since these values are known at time
$t$. The coefficients $\beta_{pv}$ and $\beta_{cs}$ capture the impact on future PVLGD changes from
the current PVLGD and full-sample credit score, respectively.

The estimation result for (\ref{eq:forecast-reg}) is reported in column (5) of Table
\ref{t:contemp-full sample-12-PVLGD}.%
\footnote{To save space, Table \ref{t:contemp-full sample-12-PVLGD}, as well as Tables
  \ref{t:forecast full sample} and \ref{t:interaction-full sample-12-PVLGD-PV} (to be explained
  shortly), only report results for control variables in $X_{i,t}$ that are significant in at least
  one specification, although all controls in $X_{i,t}$ are included in the regressions.}
The $R^2$ of the regression is 18\%, which is not surprising given that we dropped three
contemporaneous market-based covariates. It shows our model is indeed effective in forecasting
changes in firm's credit spread.  The coefficients on current PVLGD and credit score are still
negative and statistically significant; in fact, the magnitude of the credit score coefficient slightly
increases.  For PVLGD, $\beta_{pv}=-0.206$, meaning that a one point increase in current PVLGD is
associated with 0.206 points decrease in PVLGD over the next 12 months.  The coefficient for the
credit score is $\beta_{cs} = -0.197$.%
\footnote{In standard deviation terms, a one standard deviation increase in credit score is
  associated with a 0.18 ($-$0.197$\times$2.41/2.60) standard deviation decrease in future
  PVLGDs. \label{foot:cs-norm}}
Thus, the PVLGD is expected to retrace just under one fifth of its deviation away from the earnings
call implied PVLGD level over the next 12 months. Both the PVLGD and credit score effects are larger
than those in the contemporaneous specification in (\ref{eq:contemp-reg}). The information content
of earnings calls is thus very useful for forecasting future changes in corporate credit
spreads.

\subsection{Robustness Checks} \label{s:robust}

We next perform several robustness checks.  First, we replace 12-month ahead PVLGD changes with
six-month ahead changes as the dependent variable. Table \ref{t:contemp-full sample-6-PVLGD} in the
Online Appendix shows that credit score is a robust forecaster of six-month ahead changes in credit
spreads as well.  The coefficient of credit score is $-0.121$ in the forecasting regression,
statistically significant at the 1\% level.

Next, we obtain consistent findings when using 12-month ahead changes in log CDS spreads as the
dependent variable in equations (\ref{eq:contemp-reg}) and (\ref{eq:forecast-reg}), instead of
changes in PVLGD.  Unlike PVLGD, the CDS spreads are directly observed in the data without any model
assumptions.  The results are reported in Table \ref{t:contemp-full sample-12-logCDS} of the Online
Appendix.  In the forecasting regression, the coefficient of credit score is $-0.027$ and
statistically significant at the 1\% level, suggesting a positive credit score forecasts a decrease
in CDS spreads.%
\footnote{Note that the units of the PVLGD coefficients in Table \ref{t:contemp-full
    sample-12-PVLGD} (PVLGD changes) and those of the log CDS coefficients in Table
  \ref{t:contemp-full sample-12-logCDS} (log CDS changes) are not directly comparable.}

We next add both firm and time (year-month) fixed effects into regressions (\ref{eq:contemp-reg})
and (\ref{eq:forecast-reg}). The results are reported in Table \ref{t:contemp-full
  sample-12-PVLGD-FE} of the Online Appendix. We find they are largely similar to our main
specification in Table \ref{t:contemp-full sample-12-PVLGD}. Specifically, in the forecasting
regression (\ref{eq:forecast-reg}), the credit score coefficient is $-0.095$ with a t-statistics of
$-5.77$. {The coefficient, which reflects only time-series mean-reversion due to the presence
  of a firm fixed effect, is roughly half the size of the credit score coefficient in our base
  regressions in Table \ref{t:contemp-full sample-12-PVLGD}. This suggests that a good portion of the
  predictability from credit score to PVLGD changes comes from cross-sectional effects.} Another
difference is in the PVLGD coefficient, which goes from $-0.206$ to $-0.526$ (both highly
significant) when fixed effects are included. This increase occurs because of pronounced
cross-sectional persistence in PVLGDs (i.e., some firms trade at higher credit spreads than others),
but after controlling for this with a firm fixed effect, there is considerable mean reversion for
within-firm spreads.

We then examine the rolling credit score versions of regressions (\ref{eq:contemp-reg}) and
(\ref{eq:forecast-reg}) for 12-month ahead PVLGD changes. These use a credit score model that would
have been known to investors in real time, in the sense that $\hat S_{k(i)}$ and $\hat \beta$ from
(\ref{eq:lasso}) are estimated using only information available prior to the current conference
call.  We run OLS regressions to estimate (\ref{eq:contemp-reg}) and (\ref{eq:forecast-reg}) using
the same set of regressors as in the full sample analysis, {except that credit score now comes
  from our rolling text models.} Tables \ref{t:contemp-rolling sample-12-PVLGD} and
\ref{t:contemp-rolling sample-6-PVLGD} of the Online Appendix report the results of these
regressions.  The rolling credit score remains a significant forecaster of future PVLGD changes,
though the magnitude of the effect is weaker than in Tables \ref{t:contemp-full sample-12-PVLGD} and
\ref{t:contemp-full sample-6-PVLGD}.  For example, the coefficient of credit score in the rolling
text model regression for 12-month PVLGD change $-0.041$, which is statistically significant at the
1\% level.

Our results, using the full-sample and rolling credit score measures, suggest that both PVLGD and
credit score are useful predictors of future changes in PVLGD. This confirms past findings of a
price-to-book (proxied for by the PVLGD) value effect in corporate credit (Bartram, Grinblatt, and
Nozawa, 2020; Bali, Subrahmanyam, and Wen, 2021). Further, we find the surprising result that
textual information from earnings calls is not fully incorporated into CDS prices, and is useful for
forecasting CDS spread changes over the ensuing 6 to 12 months.  {The forecasting success of the
  rolling version of the text model, which is free of forward-looking information, suggests that
  credit scores may contain economically useful information for investors, a topic we return to in
  Section \ref{s:oos}. But first, we explore potential mechanisms underlying our results.}

\section{Channels} \label{s:channels}

Why does the post-call PVLGD level not fully reflect the credit-relevant information content of
earnings calls? Two channels may be at play. First, it is possible that credit score forecasts of
future PVLGD changes are consistent with their forecasts of future credit market risk or
fundamentals. Then the ability of credit score to forecast future PVLGD changes may be due to a
rational, but slow, market response to new information. We refer to this as the delayed rational
response hypothesis and discuss the underlying mechanism below. An additional possibility is that
earnings calls contain credit-relevant information, but investors only internalize this information
with a lag. We refer to this as the constrained information processing hypothesis. We find evidence
supportive of both hypotheses, though, surprisingly, the constrained information hypothesis appears
empirically more important.

Why would CDS spreads react to publicly available credit-relevant information with a lag? Since our
CDS contracts reflect anticipated default risk and risk premia over the next five years, on-the-run
CDS spreads (i.e., those of the most recent five-year contract) may react to public fundamental
information slowly, if that information changes the firm's future, but not present, credit risk. For
example, consider a company that announces it will reduce its credit risk by issuing equity and
buying back debt in year six. The current five-year CDS may not change at all because it still covers
five credit-risky years. But the new five-year CDS in one year, which will cover the firm's high
credit risk for four years and one year of low credit risk, will predictably trade at a lower spread
than the current five-year CDS. In two years, the new five-year CDS contract will predictably trade
at a still lower spread as its ratio of high- to low-risk years is even lower, and so on. CDS
contracts may thus rationally react more slowly to public information than stock prices, because the
latter reflect {\it all} future cash flows, and not only those over the next five years.

We use the full-sample text model in this section. Investors who fully paid attention to the credit
relevant portions of earnings calls would have been able to extract much more information than our
full-sample text model can capture; but this model is as close as we can get to the full information
set that would have been available to attentive investors.

\subsection{Delayed Rational Response Hypothesis} \label{s:delayed}

We first check whether the credit score forecasts changes in future credit risk in a way consistent
with the delayed rational response hypothesis. We consider three market-based risk measures. The
first one is the realized volatility of monthly PVLGD changes in the next 12 months, defined the
same as RVCredit (see Table \ref{t:controls}), except that it is calculated based on observations
12-month ahead. In addition, we consider the maximum monthly change and maximum cumulative change in
a firm's PVLGD over the next 12 months. Recall an increase in the PVLGD means an increase in
perceived credit risk or default risk premium, and a decrease in the value of the associated
corporate bonds.  We forecast credit risk using the following regression:
\begin{equation} \label{eq:risk-fund-reg}
  R_{i,t+12} = \alpha + \beta_{pv}PV_{i,t}+\beta_{cs} CS_{i,t}+\beta_{fac}^{\top} X_{i,t} +\varepsilon_{i,t},
\end{equation}
where $R_{i,t+12}$ is one of the three risk measures: realized volatility (RiskVol, i.e., RVCredit
for the next 12 months), maximum monthly change (RiskMaxIncr), or maximum cumulative change
(RiskCumSum). $X_{i,t}$ contains the same control variables as in equation
(\ref{eq:contemp-reg}). In particular, {to adjust for serial correlation in risk measures}, lagged
realized volatility is included in the controls via the RVCredit variable.%
\footnote{Replacing RVCredit with maximum monthly or cumulative PVLGD changes in the prior 12 months
  as controls leaves our results unchanged.}
If the ability of credit score to negatively forecast PVLGD changes stems from its ability to
forecast future risk, we would expect the coefficient $\beta_{cs}$ in (\ref{eq:risk-fund-reg}) to be
negative.

The regression results for (\ref{eq:risk-fund-reg}) are reported in the first three columns of Table
\ref{t:forecast full sample}.  The credit score, PVLGD and control variables are standardized so
each coefficient estimate shown in the table denotes the change in the dependent variable in units
of its interquartile range (IQR) due to a unit IQR change in the independent variable. The IQR,
defined as the difference between the 75th and 25th percentile of a variable, is a measure similar
to standard deviation (in a normal distribution, the IQR equals 1.35 standard deviations), but the
IQR is much less sensitive to outliers. Standard errors are clustered by month and firm. All data
have been winsorized at the 1\% and 99\% levels. Only those control variables that are significant
in at least one specification are shown in the table.

Our first finding is that the current PVLGD level forecasts higher future risk. This is to be
expected since higher CDS spreads reflect higher credit, and market, risk. Consistent with the
delayed rational response hypothesis, we find that $\beta_{cs}$ is significantly negative for all
three risk measures, even after controlling for the effects captured by $X_{i,t}$. The effect, in
units of IQR, ranges from $-$0.081 to $-$0.280, which is a large impact, and in line with the
predictive power of credit score for future PVLGD changes (see footnote \ref{foot:cs-norm}).  When
the call-implied PVLGD is lower than the market PVLGD, i.e., when credit score is positive, we
expect lower risk in the future. A delayed, rational reaction to this decreased risk level -- which
in unreported results persists longer than one-year -- may partially explain why credit score
negatively forecasts future changes in PVLGD.

Another possible channel for the forecasting power of credit score is through firm fundamentals.  If
a higher credit score forecasts improved firm fundamentals, it can lead to predictably lower future
credit spreads according to the delayed rational response hypothesis, as long as this change in
fundamentals takes place over a long time period. We now rerun the specification in
(\ref{eq:risk-fund-reg}) but where the dependent variable is the change in one of the following
fundamental measures: profitability as defined in Fama and French (2008), the firm's market leverage
(LEV), defined as long-term debt divided by the sum of long-term debt and the market value of
equity, and the logarithm of firm's total assets. We calculate the changes in these variables over
the next eight quarters to allow for the rational delayed response to play out, starting with the
first post-call quarterly observation of each dependent variable and ending with the eighth.%
\footnote{The results are qualitatively similar when using one- to five-year ahead windows.}
In the last three columns of Table \ref{t:forecast full sample}, we report the standardized
coefficients -- the change in units of IQR of each dependent variable for a one unit IQR change in
forecasting variable -- for the impact of PVLGD and credit score on these three fundamental
measures. The winsorization and standard errors are calculated in the same way as with the risk
measures.

Profitability, defined as the ratio of net income to sales as in Gompers, Ishii, and Metrick (2003),
is positively forecasted by credit score (at the 5\% level), suggesting that when implied PVLGD is
lower than PVLGD (i.e., management conversation is more sanguine than market CDS spreads) this is
good news for future profitability. This holds even after controlling for a large set of other
forecasting variables, and is consistent with the delayed rational response hypothesis. The economic
magnitude of this effect is 0.075 in standardized terms, which is in line with, though smaller than,
the impact of credit score on future PVLGD changes.  Credit score also negatively forecasts firms'
market leverage -- a positive for creditworthiness -- with a smaller magnitude ($-$0.033 in unit of
IQR) and lower statistical significance (at the 10\% level).  Profitability and market leverage are
not forecastable by PVLGD, again pointing to the unique information content of credit scores.
Credit score does not forecast log asset growth, while PVLGD forecasts asset growth negatively,
though the effect is not significant.

We therefore find evidence that credit scores negatively forecast future market risk and firm
leverage but positively forecast future firm profitability.  These results are consistent with the
delayed rational response hypothesis, with lower risk, higher profitability, and lower leverage
being associated with gradual declines in future PVLGD.

\subsection{Constrained Information Hypothesis}\label{s:info}

We now test the hypothesis that credit scores forecast future PVLGD changes because CDS spreads do
not fully react to the information content of earnings calls. To do so, we investigate how the
forecastability of PVLGD by credit scores depends on a firm's information environment. We consider
four information measures, each of which proxies for how difficult it may be for capacity
constrained investors, as in Sims (2011), to fully absorb all credit-relevant information from
earnings calls.

First, we measure the length of each earnings call (TransLen) using the number of words in the call
transcript. The longer the call, the harder it is for capacity-constrained investors to extract the
credit-relevant portions of the call. Thus, such credit-relevant information may be incorporated
into prices more slowly for longer calls. Next, we calculate the Flesch-Kincaid grade level
(FKGrade) for each call. This is a heuristic measure of the reading difficulty of English articles
expressed as the number of years of education generally needed to understand the text. The
Flesch-Kincaid score is given by
\begin{equation*}
  \text{FKGrade} = 0.39\left({\frac {\mbox{total words}}{\mbox{total sentences}}}\right) +
    11.8\left({\frac {\mbox{total syllables}}{\mbox{total words}}}\right)-15.59.
\end{equation*}
We conjecture that the higher the Flesch-Kincaid score of a call transcript, the harder it will be
for capacity constrained investors to fully react to its information content, which will lead to
more predictability of PVLGD changes by credit scores. Loughran and McDonald (2020) argue that
Flesch-Kincaid-type scores are imperfect measures of readability, and suggest using text length
instead. In light of this, we expect call length to be a better proxy for call complexity.

Next, we look at two measures related to analyst coverage of firms. We obtain from the I/B/E/S
database the number of analysts (NumAnlst) following each firm's stock. We define NumAnlst$_{i,t}$
as the total number of individual analysts that have given at least one price forecast for firm $i$
within the 12 months preceding month $t$.  We set NumAnlst$_{i,t}$ to zero if we cannot find any
analyst with a price forecast for firm $i$ in the 12 months preceding $t$. Finally, we calculate the
I/B/E/S analyst dispersion (DispAnlst) in price forecasts for a firm's stock.  DispAnlst$_{i,t}$ is
defined as the standard deviation of stock price forecasts for firm $i$ divided by the mean of the
forecasts in the year preceding month $t$.%
\footnote{Section \ref{s:analystdisp} of the Online Appendix details the calculation of analyst
  dispersion.}
Lehavy, Li, and Merkley (2011) show that firms with more complex investor communications attract more
analyst coverage; we therefore interpret the number of analysts covering a firm as a proxy for the
firm's business complexity. After all, employing an analyst is costly, and if a firm is easy to
understand, it is inefficient for many analysts to cover this firm only to produce non-differentiated
research; large analyst coverage is only justified if each analyst produces unique insights, which
is possible if the firm being analyzed is complex. Analyst disagreement may also proxy for firm
complexity. Under the capacity constrained investor thesis, earnings calls of more complex firms are
harder to fully understand, and thus it should take the market longer to respond to the information
content of calls of firms with more analysts and with higher analysis disagreement.

Each of the four information measures is classified into a decile bin based on the cross-section of
observations over the past 12 months. For example, we compare the number of analysts covering firm
$i$'s stock in the year prior to month $t$ to all NumAnlst$_{j,s}$ observations for
$s \in \{t-11,t\}$ and then assign NumAnlst$_{i,t}$ to a decile bin based on these
observations. This binning controls for outliers and potential nonlinearities in the raw
measures. Section \ref{s:decile} of the Online Appendix gives more details on the procedure. We then
interact the decile form of each variable with the credit score in the forecasting regression
(\ref{eq:forecast-reg}) for future PVLGD changes.  The interaction coefficient indicates whether a
given measure of the firm's information environment increases or decreases the forecasting power of
credit score for future PVLGD changes. The interacted version of (\ref{eq:forecast-reg}) is
\begin{equation} \label{eq:reg-int}
  \Delta PV_{i,t+\ell} = \alpha+ \beta_{pv}PV_{i,t} + \beta_{cs} CS_{i,t} + \beta_{md} MD_{i,t}
  + \beta_{cs\times md} CS_{i,t}\times MD_{i,t} + \beta_{fac}^{\top} X_{i,t} + \varepsilon_{i,t},
\end{equation}
where $MD_{i,t}$ is the demeaned decile variable, using the full-sample mean of all decile
levels. Our focus is on $\beta_{cs\times md}$ which measures the impact of the information
environment on the forecasting power of credit score.%
\footnote{The results with change in log CDS spread as dependent variable in (\ref{eq:reg-int}) are
  qualitatively similar.}

The results for regression (\ref{eq:reg-int}) are in Table \ref{t:interaction-full
  sample-12-PVLGD-PV}.  The decile variable in the interaction term in (\ref{eq:reg-int}) is labeled
as TransLen, FKGrade, NumAnlst, and DispAnlst respectively. For comparison, we also report the
regression results without the interactions, i.e., dropping the terms $\beta_{md} MD_{i,t}$ and
$\beta_{cs\times md} CS_{i,t}\times MD_{i,t}$ in (\ref{eq:reg-int}). The regression without the
interaction term for each information measure is constrained to have the same observations as the
with-interaction regression, which means the sample size changes slightly for the four different
specifications.  The factors $X_{i,t}$ from (\ref{eq:forecast-reg}) are included in all the
specifications, but only ones that are significant in at least one specification are shown in the
table.

We see that the information interaction term $\beta_{cs\times md}$ is insignificant for the
Flesch-Kincaid grade and for analyst dispersion, though both terms are negative suggesting that
higher analyst disagreement and higher Flesch-Kincaid scores increase the impact of credit score for
future PVLGD changes. The insignificance of the Flesch-Kincaid interaction may be due to the fact
that Flesch-Kincaid-type measures are poor representations of readability, as Loughran and McDonald
(2020) argue.

However, the credit score-information interaction terms are significantly negative for the number of
analysts ($-0.026$) and for transcript length ($-0.033$), both are significant at the 1\% level.  
These
effects are economically large.  For example, if the transcript length of an earnings call is in the
top decile -- so $MD_{i,t}\approx 4.5$ since it is demeaned -- the effect of credit score on future
PVLGD changes equals $\beta_{cs}+\beta_{cs\times md} MD_{i,t} = -0.204-0.033\times 4.5=-0.352$, which
is 72\% larger than the average effect $\beta_{cs} = -0.204$. The number of analysts has a similarly
large impact on the forecasting power of credit score.

These results support the constrained information hypothesis for the forecastability of PVLGD by
credit scores.  They show that the market may not fully incorporate credit-relevant information from
the earnings calls of firms where the call is long (and thus difficult to fully digest) or where the
firm's business is very complex (as proxied by analyst coverage). Thus, automated measures of the
credit impact of earnings calls help address the capacity constraint faced by investors and may be
of great practical value. Our information constraint result is most closely related to the
underreaction mechanism in You and Zhang (2009) and Cohen, Malloy, and Nguyen (2020), who provide
evidence that stock investors are inattentive to the language in firms' 10-K filings. Other related
papers document stock price underreaction to the information content of news (Tetlock,
Saar-Tsechansky, and Macskassy, 2008; Heston and Sinha, 2017; Ke, Kelly, and Xiu, 2021; Glasserman,
Li, and Mamaysky, 2022).

Overall, our results suggest that the rational delayed response of fixed-maturity CDS spreads to
public news plays a role in explaining why credit scores forecast future PLVGD changes. However, the
evidence also points to informationally constrained investors who do not respond to the totality of
credit-relevant information in earnings calls. Under the rational delayed response hypothesis, the
information in credit scores should not lead to profitable trading strategies because, controlling
for other firm characteristics, the risk-reward embedded in CDSs with far-from-zero and
close-to-zero credit scores should be the same. However, under the informationally constrained
investor hypothesis, knowing credit scores should lead to profitable trading strategies because this
information has not yet been fully reflected in market prices. To better understand the economic
magnitude of -- and the underlying mechanism behind -- our findings, we next turn our attention to
an out-of-sample analysis of PVLGD predictability by credit scores.

\section{Out-of-Sample Tests} \label{s:oos}

We use trading simulations to show the out-of-sample economic impact of our credit score measures.
Recall the credit score represents the discrepancy between the market PVLGD and the model-implied
PVLGD (see equation \ref{eq:cred-score}). It therefore is a measure of disagreement between the
market and the model.  We found in Section \ref{s:emp} that higher credit scores forecast future
declines in a firm's PVLGD. Thus, on average, when the model and the market disagree, the market
tends to gravitate in the direction of the model over time. We now use the rolling text model (see
Section \ref{s:text-imp}), which employs a credit score that would have been known to investors in
real time because it only uses historical information. The rolling text model allows us to evaluate
the out-of-sample benefit of using credit score information. Our trading strategies seek to isolate
the credit score dimension of information. Richer trading strategies may use other information shown
in the literature to forecast credit returns, but our focus is narrower than this and we analyze
strategies keyed only off credit score.

Before detailing our portfolio construction, a brief aside on nomenclature: In our trading strategy,
we go long firms whose CDSs have positive credit scores, in anticipation of dropping future PVLGDs,
and we go short firms with negative credit scores. Going long credit through CDS means selling
protection, and thus betting against default. If protection is sold at a CDS spread corresponding to
a 10 PVLGD, and the credit spread of the firm improves (i.e., falls), the PVLGD will drop meaning
that the CDS can be bought back at a lower price than 10 and a gain is realized. Going short credit
via CDS means buying protection, and betting on increased default risk. If one buys protection via
CDS at a PVLGD of 10 and the credit spread of the firm increases, the PVLGD will increase and the
CDS contract can be unwound by selling at a price higher than 10, thus realizing a profit. This is
the opposite of what happens with corporate bonds, where going long means buying bonds and going
short means borrowing and selling bonds.

Given the tendency of future PVLGDs to trend in the direction of the earnings call-implied PVLGDs,
we interpret the credit score as a measure of mispricing. We then seek to construct maximally
mispriced portfolios -- i.e., those with the highest possible weighted credit score -- subject to
certain constraints. In each month $t$, our portfolio consists of three equally weighted
sub-portfolios constructed at the end of months $t-1$, $t-2$, and $t-3$. We use three sub-portfolios
to allow time for markets to react to earnings call news. At the end of month $t$, the positions in
the sub-portfolio from month $t-3$ are liquidated and a new, month $t$ sub-portfolio is added.

At the end of month $t$, we select companies that had at least one earnings call in the past three
months.%
\footnote{In order to calculate strategy returns, we only consider firm-month observations for which
  we have CDS data in months $t$ through $t+3$. We address any potential survivorship bias due to
  firms leaving our sample in Section \ref{s:cds-exit} of the Online Appendix.}
For purposes of the trading strategy, each company's credit score is calculated by
(\ref{eq:cred-score}) using its PVLGD at the end of month $t$ and the implied PVLGD from its most
recent earnings call (which could be up to three months ago). Index the companies by
$i=1,2,\ldots,n_t$ and denote their weights in the portfolio by $w_{i,t}$. We allow both long and
short positions: a positive $w_{i,t}$ means we go long the credit, and thus sell protection via CDS;
and a negative $w_{i,t}$ means we go short the credit by buying CDS. Given credit scores for month
$t$ -- defined as above -- we would like to maximize the total mispricing of the portfolio, given by
$\sum_{i=1}^{n_t} w_{i,t}CS_{i,t}$, subject to following constraints: First, we require the
portfolio to be self-financing by having a zero total PVLGD (more on this below); second we impose
concentration limits by setting a lower and upper bound for each individual weight $w_{i,t}$;
finally, we add a constraint on the total long and short position of the portfolio, which limits the
leverage allowed for the portfolio.

Formally, the portfolio optimization problem is given by
\begin{eqnarray}
  \max &&\ \sum_{i=1}^{n_{t}} w_{i,t} CS_{i,t} \qquad \text{(Total mispricing)} \label{port_obj} \\
  \textrm{s.t.}& & \sum_{i=1}^{n_{t}} w_{i,t}PV_{i,t}=0, \label{constr_sumPV} \\
       & & w_{i,t}\leq u, \,\, w_{i,t}\geq l, \,\,\forall i,\label{constr_ul} \\
       & &\sum_{i=1}^{n_{t}} w_{i,t}\one[w_{i,t}>0]\leq U,  \,\, \sum_{i=1}^{n_{t}}
           w_{i,t}\one[w_{i,t}<0]\geq L. \label{constr_UL} 
\end{eqnarray}
The constraint in (\ref{constr_sumPV}) can be thought of as controlling for other important CDS
characteristics, to the extent that these are captured by a firm's PVLGD. The constants $l$ and $u$
in (\ref{constr_ul}) denote the lower and upper bound for the individual positions; $U$ and $L$ in
(\ref{constr_UL}) denote the limit for the total long and short position of the portfolio. The
constraint (\ref{constr_UL}) is non-linear as the indicator function $\one[\cdot]$ is involved. In
Section \ref{s:opt-dets} of the Online Appendix we show how this constraint can be linearized. This
optimization is solved using the \texttt{gurobi} package in Python. The resultant portfolio isolates
firms' credit score exposure while ensuring minimal exposure to overall credit markets. We also
investigated a version of the above portfolio problem which constrains portfolio turnover, and found
the results to be qualitatively similar to the unconstrained results that we report below.

We assume that \$100 of capital supports the portfolio resulting from the above optimization. Since
PVLGD in (\ref{eq:cds}) is calculated for CDS contracts with \$100 principal, the weights in
(\ref{port_obj}-\ref{constr_UL}) give the fraction of the \$100 that is invested in a single CDS
contract; $l,u,L,U$ should be interpreted in this context. It will simplify matters to assume that
all CDS contracts trade with a coupon of zero, which means that the buyer of protection needs to
pay, and the seller thus receives, an upfront amount equal to the PVLGD from (\ref{eq:cds}). With
this convention, the portfolio is self-financing via the constraint in (\ref{constr_sumPV}) because
any short CDS position is purchased via PVLGD received from the sale of protection in a long
position. The \$100 capital is assumed to earn the risk-free rate, which can be ignored when
calculating excess returns.

For example, consider a sub-portfolio with $\{w_1=0.5,w_2=0.5,w_3=-2\}$ where $u=0.5$, $U=1$,
$l=L=-2$, and the PVLGDs are $\{10,10,5\}$ respectively. This sub-portfolio satisfies the
constraints of the optimization problem in (\ref{port_obj}--\ref{constr_UL}). Consider the excess
return of the portfolio if the PVLGDs in the next month become $\{8,8,4.5\}$ respectively. The two
long positions in securities 1 and 2 (long positions are indicated via positive weights) will earn
\$1 each because the CDS was sold at a PVLGD of 10 and now trades at a PVLGD of 8, for a total gain
from the long side of the portfolio of \$2. The short side of the portfolio will experience a loss
of \$0.5 for each of two contracts (CDS was bought for 5 and now trades at 4.5), leading to a total
loss from the short side of \$1. Therefore the portfolio will gain \$1 on a capital base of
\$100. This portfolio will thus experience a 1\% excess return.

\subsection{Evaluating Strategy Performance} \label{s:perf-eval}

At the end of every month $t$, we solve the optimization in (\ref{port_obj}--\ref{constr_UL}) to
generate a new sub-portfolio and use it to replace the month $t-3$ sub-portfolio that is being
dropped. The change in the PVLGD for the overall portfolio in month $t$ is the sum over the three
equal-weighted sub-portfolios:
\begin{equation} \label{port_change}
  \Delta {PV}^{(tot)}_{t}  =  \frac{1}{3} \times \sum_{j=1}^3\Delta PV^{({sub})}_{t-j,t},
\end{equation}
where $ \Delta PV^{({sub})}_{t-j,t}$ denotes the month--$t$ change in the total PVLGD of the
sub-portfolio constructed at the end of month $t-j$, given by
\[
  \Delta PV^{({sub})}_{t-j,t} = \sum_{i=1}^{n_{t-j}}w_{i,t-j} (PV_{i,t} -PV_{i,t-1}).
\]
Here firm $i$'s weight $w_{i,t-j}$ was calculated at the end of month $t-j$. For the first month in
the sample, all three portfolios are set to the sub-portfolio from the initial optimization.

The portfolio's excess return in month $t$ is $R_t = -\Delta PV^{(tot)}_t/100$, where the
denominator reflects the \$100 capital supporting the CDS positions. The minus sign reflects the
convention that a long position (indicated with a positive weight $w_{i,t-j}$) makes money when
PVLGD declines, and short positions (indicated with negative weights) make money when the PVLGD
increases.  The annualized excess return of the strategy is
\begin{equation} \label{eq:port_return}
\bar{R} = \Big[\prod_{t=1}^T (1+R_{t})\Big]^{12/T}-1,
\end{equation}
where $T$ is the total number of months over which we run the strategy.

We implement our strategy separately for investment grade (IG) and high yield (HY) names, based on
the firm's ratings immediately preceding the earnings call. We set the limits on total long and
short positions as $(L,U) = (-4,4)$ for both IG and HY. This allows four-fold leverage for the
portfolio, which is not extreme since our portfolio has a zero total PVLGD and thus little exposure
to the aggregate credit market.  We consider 25 representative specifications of individual weight
limits $(l,u)$, reflecting different degrees of single-name concentration allowed in the portfolio.
Both symmetric and asymmetric weight limits are considered.  We show the portfolio test results in
Table \ref{t:porttest}, with Panels A and B for the IG and HY groups, respectively.  Each cell
reports the annualized return from the portfolio associated with a weight limit $(l,u)$, with the
values of $l$ and $u$ given in the table's rows and columns respectively.

We test the statistical significance of the annualized return using 100 simulated portfolios under
the null hypothesis of no-predictability, for the IG and HY portfolios for each $(l,u)$ combination
listed in the table.  Details of the simulation strategy are in Section \ref{s:null-dist} of the
Online Appendix. The 1\%, 5\%, and 10\% significance levels of the strategy return are represented
by $^{\ast\ast\ast}$, $^{\ast\ast}$, and $^{\ast}$, respectively.  For example, the strategy return
is significant at 1\% level if at most one of the 100 simulated strategies has a higher average
annualized return. 

Before discussing the results in Table \ref{t:porttest}, we make several observations. First, our
portfolio construction methodology, {and in particular the constraint in \eqref{constr_sumPV}},
isolates credit score exposure without having an exposure to aggregate credit markets
indexes.%
\footnote{For the 30 portfolio outcomes shown in Table \ref{t:porttest}, the range of monthly
  portfolio return betas to either the IG or HY indexes is $(0.01, 0.15)$. The IG and HY aggregate
  credit indexes are obtained from FRED website.}
{Thus our portfolio strategy hinges on the cross-sectional relationship between credit scores
  and future CDS market outcomes.} In practice, combining credit scores with other factors for
forecasting credit returns may further improve portfolio performance. Second, compared to much of
the corporate bond anomaly literature, our trading strategy is more readily implementable since CDS
contracts for a single firm are fungible and can be easily bought (going short) or sold (going
long). Footnote \ref{foot:TRACE} discusses some limitations of simulated strategies using corporate
bond data. Finally, because the CDS market is generally much more liquid than any one corporate
bond, there are fewer off-market CDS prices (Markit reports CDS spreads that are the consensus marks
across all dealers trading that contract) than there are untradable bond prices on TRACE. Our
portfolio simulations establish a natural benchmark under the null of no predictability, while
controlling for the underlying data quality, against which our CDS trading strategies can be
compared.%
\footnote{In future research, it would be interesting to apply our simulation approach to TRACE bond
  data and check whether some of the reported returns from long-short corporate bond portfolios
  remain significant.}

We find our trading strategy delivers a positive return in all cases.  For example, with a -0.3
limit on short and a 0.1 limit on long positions, the trading strategy has an annual return of
2.934\% for IG and 4.135\% for HY.  The p-values of both results relative to portfolio returns
simulated under the null of no predictability are better than 1\%, which means that at most one of
the 100 simulations had higher average annualized returns. Looking at the significance levels for
the returns of the IG and HY strategy variants (across the cells of Table \ref{t:porttest}), we see
that 24 of the 25 IG strategies are significant at the 5\% level or better, while 12 (14) of 25 HY
strategies are significant at the 5\% (10\%) level or better.  Specifically, the trading strategy
for HY generally delivers large and significant returns when the limit on the short positions is
large (e.g., the bottom two rows with $l=-0.3$ or $-0.4$).  This suggests that when trading HY, the
investor should hold large short positions on CDS with negative credit scores, essentially betting
their credit spreads will further widen.

We also test the joint statistical significance of the 25 trading strategies for IG/HY.  When
evaluating the joint significance, we need to account for the fact that the returns of the 25
trading strategies are not independent.  The details of the statistical test are included in Section
\ref{s:joint_test} of the Online Appendix.  The collective performance of our trading strategies is
statistically highly anomalous: Under the null hypothesis of no-predictability, the probability to
achieve the joint performance of our 25 trading strategies is smaller than 0.5\% for both IG and HY.

To gauge the economic importance of our trading strategies, consider a long-only IG (HY) credit
portfolio. According to data obtained from FRED, since 1972 (1986) the average annualized return of
the IG (HY) long-only corporate bond portfolio has been 7.14\% (7.55\%). Consider supplementing
these returns with those of the $(l,u) = (-0.3,0.1)$ strategies. The IG and HY returns would rise to
10.07\% and 11.69\% respectively. Such gains would place an IG or HY credit fund into the upper
echelon of its competitors. Using the \$100 of corporate bonds as collateral, the CDS long-short
strategies could be readily implemented by institutional investors. Of course, institutions could
simply tilt existing long-only portfolios by overweighting bonds with positive credit scores and
underweighting those with negative credit scores, which would produce more modest performance
improvements, but without requiring the use of any leverage.

We conclude that credit scores contain useful information for forecasting credit spread changes,
which translates to statistically and economically significant gains in an out-of-sample strategy
that can be readily implemented by institutional investors. This evidence lends further weight to
the capacity constrained investor hypothesis. If the predictability of credit score for future PVLGD
changes only reflected a delayed but rational response to the information content of credit scores,
such a rational reaction would not lead to profitable trading strategies, as the risk-reward
embedded in CDSs with large positive or negative credit scores would be identical to that of CDSs
with close-to-zero credit scores, once other firm characteristics are controlled for. However, we
find this is not the case.

\section{Conclusion} \label{s:conc}

In this paper we introduce credit score, a novel measure of corporate creditworthiness that we
extract from the text of quarterly earnings calls. Our measure is straightforward to implement, can
be calculated in real time, and endogenously determines the credit impact of words used in earnings
calls to discuss corporate credit. Our measure forecasts future changes in corporate credit spreads,
even after controlling for an extensive set of predictors for corporate bond and CDS returns
identified in the literature. We have presented evidence that credit score forecasts future
firm-level outcomes, such as market risk, profitability, and firm leverage. Surprisingly, our
complexity-based tests suggest that market participants do not fully incorporate the credit-relevant
information in earnings calls into their credit assessments. In out-of-sample trading tests, we show
that credit scores lead to economically and statistically significant gains that can be achieved by
institutional investors trading in long-short CDS portfolios. This finding supports the
informationally constrained investor hypothesis and suggests that algorithmically-generated credit
scores should be of interest to both academic researchers and to credit investors.

In order to offer an economic assessment of the usefulness of our earnings call measure, our trading
strategies intentionally {focus only on cross-sectional information embedded in credit
  scores}. We do not claim that this trading strategy optimally uses all available information that
may be useful for forecasting credit returns. {For example, future work can consider potential
  time-series predictability from credit scores, which was not our focus.} Another interesting area
for future work is to extend our portfolio construction methodology to take into account other firm
characteristics that have been shown to forecast credit returns, and to assess the performance of
this expanded trading strategy on CDS and corporate bond data.

Our paper contains several methodological innovations. First, we argue PVLGD changes are a cleaner
measure of changes in market-perceived creditworthiness than are changes in credit spreads. Second,
the implied PVLGD measure that we obtain from the text of earnings calls is novel and provides a
template that can be applied in other contexts (for example, to assess the impact of earnings calls
on firm-level implied volatility). As part of obtaining an implied PVLGD, we introduce a highly
computationally efficient technique to identify important explanatory words for PVLGDs. Third, our
methodology for simulating portfolio returns under the null of no predictability is novel, and
provides a natural benchmark that controls for underlying data quality. Finally, as in Donovan et
al. (2021), our implied PVLGD measure can be used to extract CDS-like information for firms with
earnings-calls, but without active CDS markets. We hope that future researchers will find these
tools useful.

\appendix

\section*{References}

\begin{list}{}{\topsep 0pt \leftmargin .2in \listparindent -0.2in \itemindent
    -0.2in \parsep \parskip}

\item Altman, E., 1968, ``Financial ratios, discriminant analysis and the prediction of corporate
  bankruptcy,'' {\it Journal of Finance}, 23 (4), 589--609.
  
\item Bai, J., T. Bali, and Q. Wen, 2019, ``Common risk factors in the cross-section of corporate
  bond returns,'' {\it Journal of Financial Economics}, 131, 619--642.

\item Bai, J., T. Bali, and Q. Wen, 2021, ``Is there a risk-return tradeoff in the corporate bond
  market? Time-series and cross-sectional evidence'' {\it Journal of Financial Economics}, 142,
  1017--1037.

\item Bali, T., A. Goyal, D. Huang, F.  Jiang, and Q. Wen, 2022, ``Predicting corporate bond
  returns: Merton meets machine learning,'' working paper.
  
\item Bali, T., A. Subrahmanyam, and Q. Wen, 2021, ``Long-term reversals in corporate bond
  markets,'' {\it Journal of Financial Ecomomics}, 139, 656--677.

\item Bao, J., J. Pan, and J. Wang, 2011, ``The illiquidity of corporate bonds,'' {\it Journal of
    Finance}, 66 (3), 911--946.

\item Banz, R. W, 1981, ``The Relationship between Return and Market Value of Common Stocks,'' {\it
    Journal of Financial Economics}, 9 (1), 3--18.

\item Bartram, S., M. Grinblatt, and Y. Nozawa, 2020, ``Book-to-market, mispricing, and the
  cross-section of corporate bond returns,'' working paper.
  
\item Bernard, V. and J. Thomas, 1989, ``Post-earnings-announcement drift: Delayed price response or
  risk premium?'' {\it Journal of Accounting Research}, 27, 1--36.

\item Bharath, ST. and T. Shumway, 2008, ``Forecasting default with the Merton distance to default
  model,'' {\it Review of Financial Studies}, 21(3), 1339--69.

\item Bhushan, R., 1989, ``Firm characteristics and analyst following,'' {\it Journal of Accounting
    and Economics}, 11, 255--274.
  
\item Calomiris, C., J. Harris, H. Mamaysky, and C. Tessari, 2022, ``Fed implied market prices and
  risk premia,'' working paper.
  
\item Campbell, J. and G. Taksler, 2003, ``Equity volatility and corporate bond yields,'' {\it
    Journal of Finance}, 58 (6), 2321--2350.

\item Cao, J., A. Goyal, X. Xiao, and X. Zhan, 2022 ``Implied volatility changes and corporate bond
  returns,'' {\it Management Science}, forthcoming.

\item Chordia, T., A. Goyal, Y. Nozawa, A. Subrahmanyam, and Q. Tong, 2017, ``Are capital market
  anomalies common to equity and corporate bond markets? An empirical investigation,'' {\it Journal
    of Financial and Quantitative Analysis}, 52 (4), 1301--1342.

\item Chung, K., J. Wang, and C. Wu, 2019, ``Volatility and the cross-section of corporate bond
  returns,'' {\it Journal of Financial Economics}, 133 (2), 397--417.
  
\item Cohen, L., C. Malloy, and Q. Nguyen, 2020, ``Lazy prices,'' {\it Journal of Finance}, 75 (3),
  1371--1414.

\item Collin-Dufresne, P., R. Goldstein, and J.S. Martin, 2001, ``The determinants of credit spread
  changes,'' {\it Journal of Finance}, 56 (6), 2177--2207.

\item Cremers, M., J. Driessen,  P. Maenhout, and D. Weinbaum, 2008, ``Individual stock-option
  prices and credit spreads,'' {\it Journal of Banking \& Finance}, 32(12), 2706--2715.
  
\item Donovan, J., J. Jennings, K. Koharki, and J. Lee., 2021, ``Measuring credit risk using
  qualitative disclosure,'' {\it Review of Accounting Studies}, 26 (2), 815--863.
 
\item Edwards, A., L. Harris, and M. Piwowar, 2007, ``Corporate bond market transaction costs and
  transparency,'' {\it Journal of Finance}, 62 (3), 1421--1451.
  
\item Ericsson, J., K. Jacobs, and R. Oviedo, 2009, ``The determinants of credit default swap
  premia,'' {\it Journal of Financial and Quantitative Analysis}, 44 (1), 109--132.

\item Fama, E. F., and K. R. French, 1992, ``The Cross-Section of Expected Stock Returns,'' {\it
    Journal of Finance}, 47 (2), 427–-465.

\item Fama, E. F., and K. R. French, 2008, ``Dissecting Anomalies,'' {\it Journal of Finance}, 63
  (4), 1653–-1678.

\item Garcia, D., X. Hu, and M. Rohrer, 2022, ``The color of finance words,'' working paper.

\item Gompers, P., J. Ishii, and A. Metrick, 2003, ``Corporate governance and equity prices,'' {\it
    Quarterly Journal of Economics}, 118 (1), 107--155.
  
\item Gu, S., B. Kelly, and D. Xiu, 2020, ``Empirical asset pricing via machine learning,'' {\it
    Review of Financial Studies}, 33, 2223--2273.
  
\item Guo, X., H. Lin, C. Wu, and G. Zhou, 2021, ``Investor sentiment and the cross-section of
  corporate bond returns,'' working paper.

\item Hastie, T., R. Tibshirani, and J. Friedman, 2009, {\it The elements of statistical learning:
    Data mining, inference and prediction}, Springer.

\item Heston, S. and N. Sinha, 2017, ``News vs. sentiment: Predicting stock returns from news
  stories,'' {\it Financial Analysts Journal}, 73 (3), 67--83.

\item Jegadeesh, N., 1990, ``Evidence of predictable behavior of security returns,'' {\it 
    Journal of Finance}, 45 (3), 881--898.

\item Jegadeesh, N. and D. Wu, 2013, ``Word power: A new approach for content analysis,'' {\it
    Journal of Financial Economics}, 110, 712--729.

\item Jegadeesh, N., and S. Titman, 1993, ``Returns to buying winners and selling losers:
  implications for stock market efficiency,'' {\it Journal of Finance}, 48 (1), 65--91.
  
\item Jiang, F., J. Lee, X. Martin, and G. Zhou, 2019, ``Manager sentiment and stock returns,'' {\it
    Journal of Financial Economics}, 132 (1), 126--149.
  
\item Jostova, G., S. Nikolova, A. Philipov, and C. W. Stahel, 2013, ``Momentum in corporate bond
  returns,'' {\it Review of Financial Studies}, 26 (7), 1649--1693

\item Kelly, B. T., Pruitt, S., and Su, Y, 2020, ``Instrumented principal component analysis,''
  working paper.

\item Ke, Z., B. Kelly, and D. Xiu, 2021, ``Predicting returns with text data,'' working paper.

\item Kelly, B., D. Palhares, and S. Pruitt, 2023, ``Modeling corporate bond returns,'' {\it Journal
    of Finance}, 78 (4), 1967--2008.

\item Lehavy, R., F. Li, and K. Merkley, 2011, ``The effect of annual report readability on analyst
  following and the properties of their earnings forecasts,'' {\it  Accounting Review}, 86 (3),
  1087--1115.
  
\item Lehmann, B., 1990, ``Fads, martingales, and market efficiency,'' {\it  Quarterly Journal of
    Economics}, 105 (1), 1--28.
  
\item Lee, M., P. Meyer-Brauns, S. Rizova, and S. Wang, 2020, ``The cross-section of corporate bond
  returns,'' {\it Dimensional Fund Advisors working paper}.

\item Glasserman, P., F. Li, F., and H. Mamaysky, 2022, ``Time variation in the news-returns
  relationship,'' working paper.

\item Loughran, T. and B. McDonald, 2020, ``Textual analysis in finance,'' {\it Annual Review of
    Financial Economics}, 12, 357--375.
  
\item Manela, A. and A. Moreira, 2017, ``News implied volatility and disaster concerns,'' {\it
    Journal of Financial Economics}, 123 (1), 137--162.

\item Merton, R., 1974, ``On the pricing of corporate debt: The risk structure of interest rates,''
  {\it Journal of Finance}, 29, 449--470.

\item Nozawa, Y., 2017, ``What drives the cross-section of credit spreads?: A variance decomposition
  approach,'' {\it Journal of Finance}, 72 (5), 2045--2071.

\item Oehmke, M. and A. Zawadowski, 2017, ``The anatomy of the CDS Market,'' {\it Review of
    Financial Studies}, 30 (1), 80--119.
  
\item Penman, S., F. Reggiani, S. A. Richardson, I. Tuna, 2014, ``An accounting-based characteristic
  model for asset pricing'', working paper.

\item Polbennikov, S., M. Dubois, and A. Descl\'{e}e, 2020, ``Have systematic credit factors lived
  up to expectations? Performance through Covid-19,'' Barclays Credit Research, April 2020.

\item Sims, C., 2011, ``Rational inattention and monetary economics,'' Chapter 4 in {\it Handbook of
    Monetary Economics}, Vol. 3A, Elsevier.

\item Tetlock, P., M. Saar-Tsechansky, and S. Macskassy, 2008, ``More than words: Quantifying
  language to measure firms' fundamentals,'' {\it Journal of Finance}, 63 (3), 1437--1467.

\item van Binsbergen, J. and M. Schwert, 2022, ``Duration-based valuation of corporate bonds,''
  working paper.
  
\item You, H. and Z.-j. Zhang, 2009, ``Financial reporting complexity and investor underreaction to
  10-K information,'' {\it Review of Accounting Studies}, 14, 559--586.
  
\end{list}

\clearpage


\begin{figure}
  \centering
  {\bf Panel A: Quarterly number of earnings calls} \\[7pt]
  \includegraphics[width=0.76\textwidth,trim={0 0 0 1.8cm},clip]{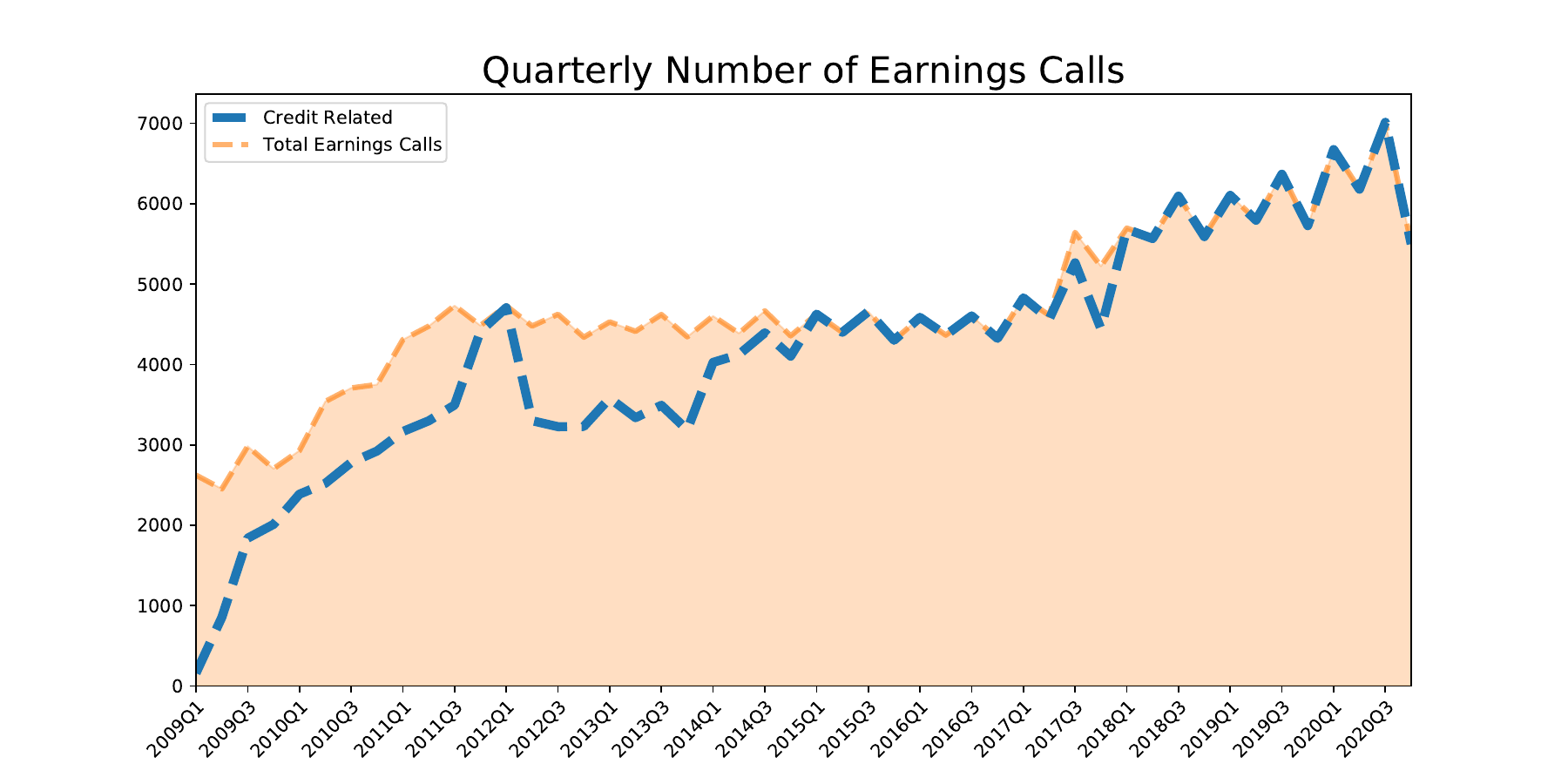} \\
  {\bf Panel B: Quarterly number of CDS observations} \\[7pt]
  \includegraphics[width=0.76\textwidth,trim={0 0 0 1.8cm},clip]{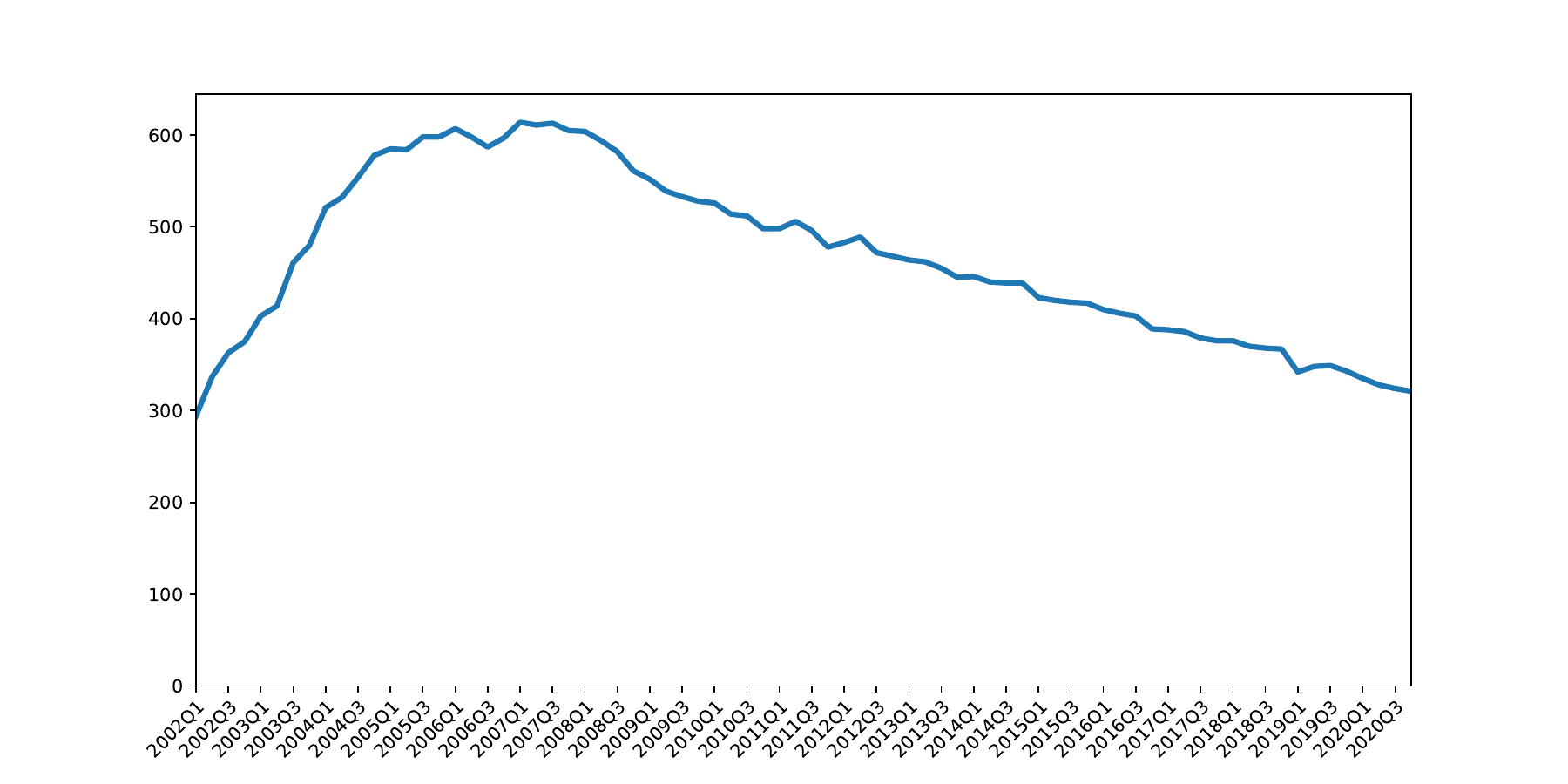} \\
  {\bf Panel C: Quarterly number of earnings calls with a matched CDS} \\[7pt]
  \includegraphics[width=0.76\textwidth,trim={0 0 0 1.8cm},clip]{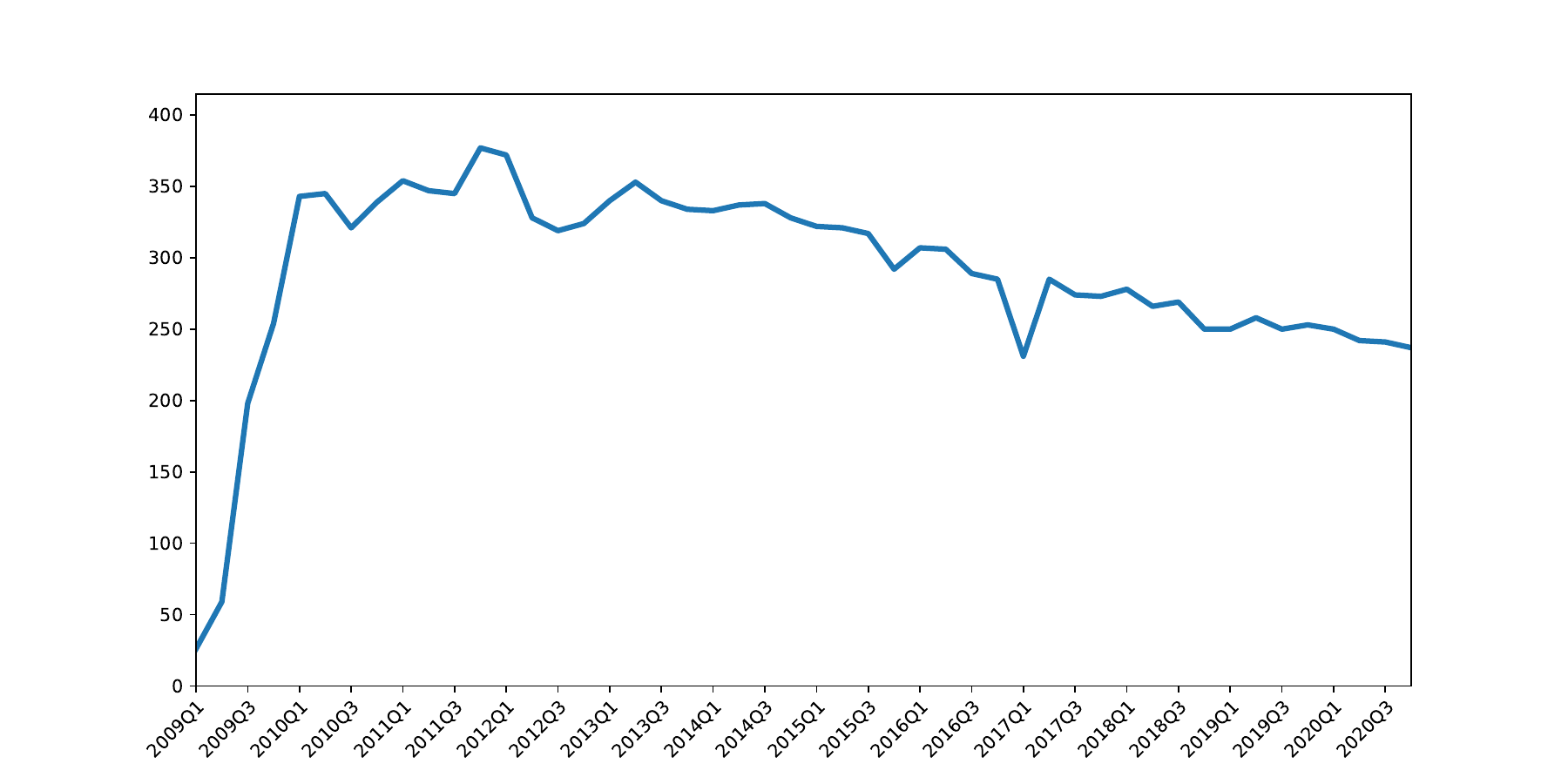} \\
  \caption{The top panel shows the number of quarterly earnings calls in the data set, as well as
    the number of calls that have a credit-related discussion.  The middle panel shows the number of
    quarterly credit default swap observations (we count each firm at most once within each
    quarter).  The bottom panel shows the number of quarterly earnings calls that can be matched to
    a credit default swap in that quarter.}
  \label{f:summ-stats}
\end{figure}

\begin{figure}
  \centering
  {\bf Correlation Matrix of Regression Controls} \\[5pt]
  \includegraphics[width=1.1\textwidth,trim={0.25cm 0 0 0},clip]{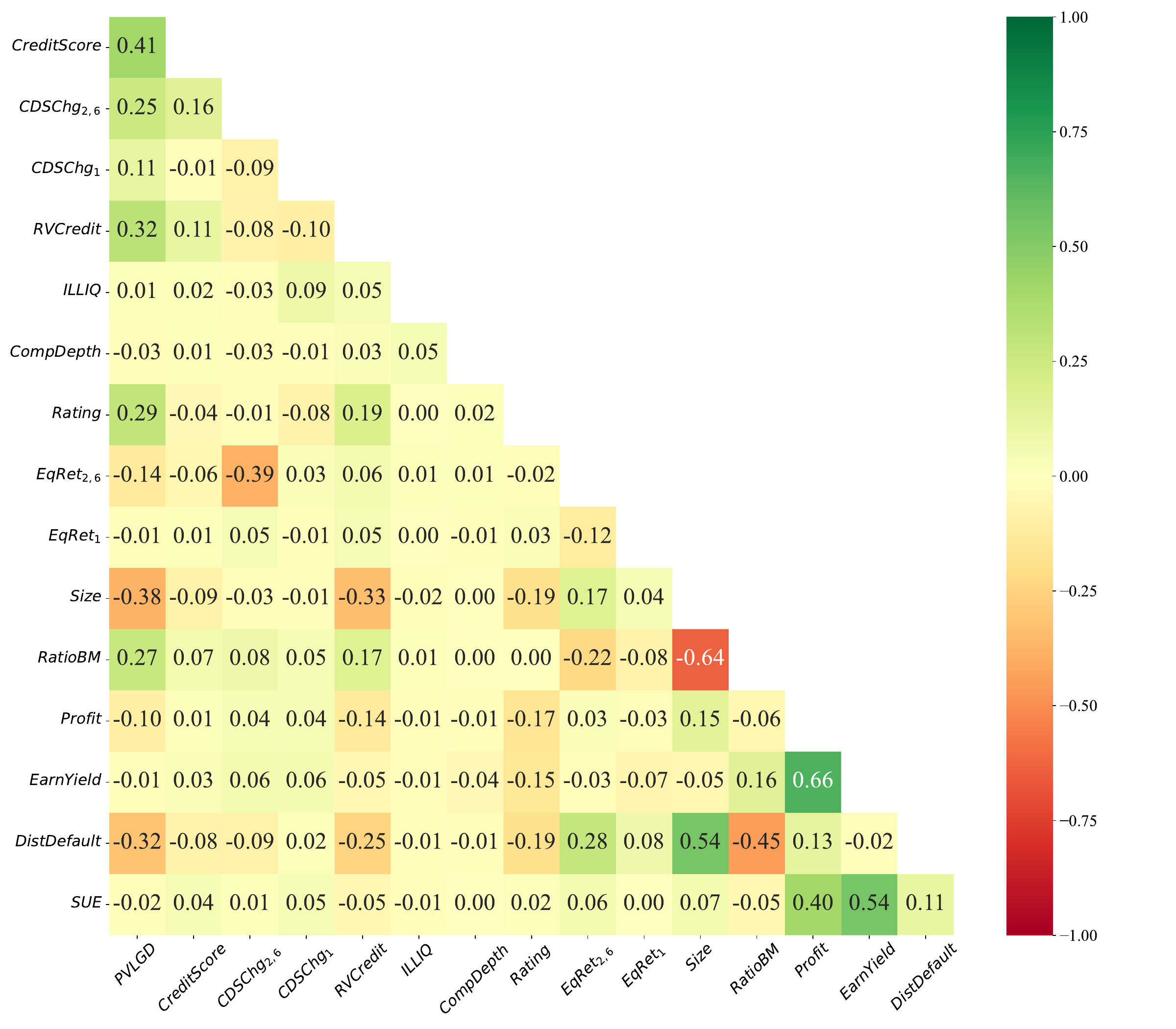} 
  \caption{This figure shows the correlation of the regression control variables calculated as the
    mean across firms of the time series correlations of variables for each firm using the full
    sample of data. The variables are shown in Table \ref{t:controls} and described fully in Section
    \ref{s:controls} of the Online Appendix. All statistics
      are based on full sample series winsorized at 1\% and 99\% percentiles.}
  \label{f:freqfactorCorr}
\end{figure}

\begin{figure}
  \centering
  {\bf Event studies around PVLGD and credit score outliers} \\[5pt]
  {\bf \small Panel A: Low PVLGD and credit score studies} \\[3pt]
  \mbox{\includegraphics[width=0.54\textwidth]{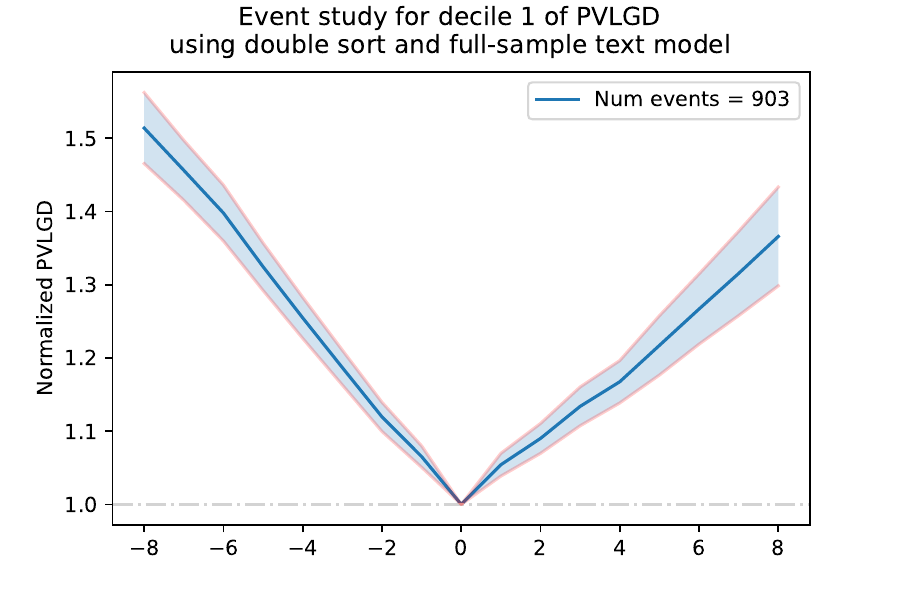}
    \hspace{-30pt}
    \includegraphics[width=0.54\textwidth]{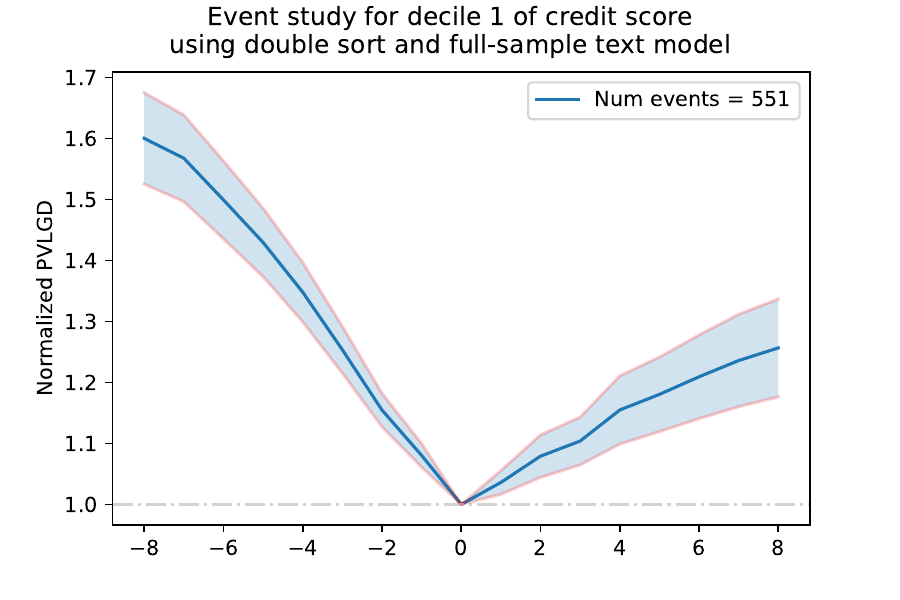}}
  {\bf \small Panel B: High PVLGD and credit score studies} \\[3pt]
  \mbox{\includegraphics[width=0.54\textwidth]{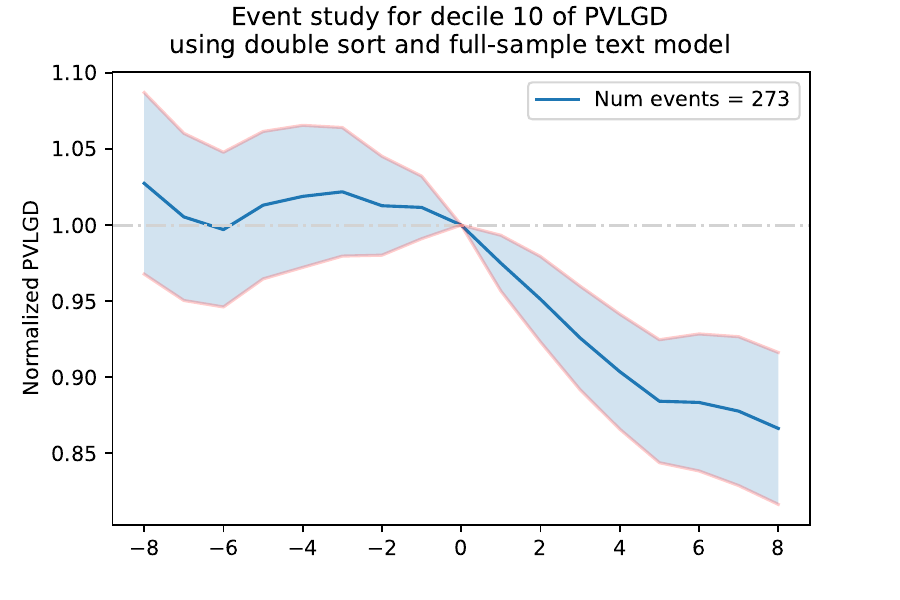}
    \hspace{-30pt}
    \includegraphics[width=0.54\textwidth]{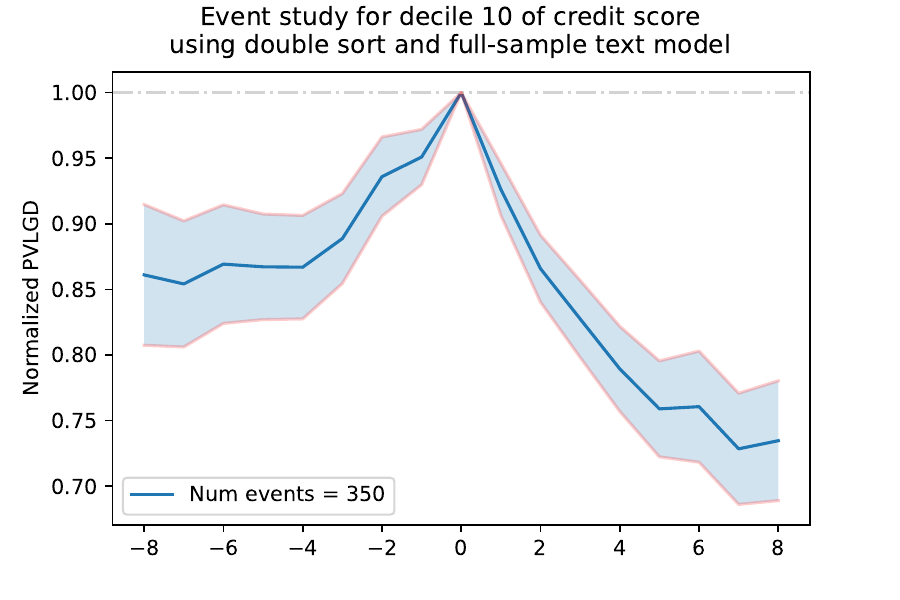}}
  \caption{The left plot of Panel A shows the normalized PVLGD response in eight quarter windows
    around name-date pairs whose PVLGD falls into the bottom decile of all name-date PVLGD
    observations, but whose credit score does not fall into the bottom decile of all name-date
    credit score observations. The right plot of Panel A shows the analogous normalized PVLGD event
    study for bottom decile name-date credit score pairs whose PVLGD is not in the bottom decile of
    name-date PVLGD observations. The left plot of Panel B shows the analogous PVLGD response for
    name-date observations in the top decile of PVLGD but not in the top decile of credit score. The
    right plot of Panel B shows the analogous PVLGD response for name-date observations in the top
    decile of credit score that are not also top decile observations of PVLGD. The number of events
    in each event study is shown in the legend; an event enters the study sample only if the full 17
    quarters of data are available. The 95\% confidence bands assume independence.}
  \label{f:es}
\end{figure}

\clearpage

\begin{landscape}
  \begin{table}[!htb]
    \centering
    \def\arraystretch{1.1}
    {\bf Control Variable Definitions} \\[5pt]
    {\footnotesize
      \begin{tabular}{lll} \hline
        Variable & Description & Relevant Literature        \\ \hline
        CDSChg$_{2,6}$  &    CDS momentum in month $t-6$ to $t-1$  & Jostova et al. (2013), Lee et al. (2021)   \\
        CDSChg$_1$   & CDS reversal in the most recent one month & Bartram, Grinblatt, and Nozawa (2020)   \\
        RVCredit   & CDS realized volatility in prior 12 months & Chordia et al. (2017) \\
        ILLIQ   & {CDS illiquidity, measured by the negative of covariance of } & Bao, Pan, and Wang (2011)  \\
                 &  daily change in PVLGD and its one-day lag & \\
        CompDepth & Composite depth as a proxy for CDS market depth, another &  Bao, Pan, and Wang (2011) \\
                 &  measure for market liquidity & \\
        Rating  & Average credit rating from Markit & {Bali et al. (2022), Guo, Lin, Wu, and Zhou (2021),} \\
                 & & Bao, Pan, and Wang (2021)\\
        EqRet$_{2,6}$ & Equity momentum in month $t-6$ to $t-1$ & Jegadeesh and Titman (1993)\\ 
        EqRet$_{1}$ & Equity reversal in the most recent one month & Jegadeesh (1990), Lehmann (1990),  Chordia et al. (2017)\\
        Size & Firm size, logarithm of market value of firm equity & Fama and French (1992), Chordia et al. (2017)\\
        RatioBM & Firm value, book to market ratio & Fama and French (2008) \\
        Profit & Firm profitability, ratio of equity income to sales (quarterly) & {Gompers et al. (2003)} \\
        EarnYield &{Earnings yield, ratio between diluted earnings per share and} & {Penman, Richardson, Reggiani, and Tuna (2014),}  \\
                 &  stock price & Bartram, Grinblatt, and Nozawa (2020)\\
        DistDefault & Distance-to-default measure in Bharath and Shumway (2008) &  Bharath and Shumway (2008)\\
        SUE & Standardized unexpected earnings in Bernard and Thomas (1989) &   Bartram, Grinblatt, and Nozawa (2020)\\
        $R^f$ & Risk-free rate as the 10-year US treasury bond interest rate & Collin-Dufresne, Goldstein, and Martin (2001)\\
        $LEV$ & Firm leverage, long-term debt divided by long-term debt plus & Ericsson, Jacobs, and Oviedo (2009) \\
                 &  firm's market value & \\
        $IV$ & Average implied volatility of firm's 30-day put and call options & Cremers et al. (2008)\\
        \hline
      \end{tabular}
    }
    \caption{This table summarizes the regression control variables used in our study. Equity
      measures refer to the stock of the firm whose bonds are being referenced by the CDS. All
      measures are calculated as of month-end. $R^f$, $LEV$, and $IV$ are only used as covariates in
      the contemporaneous regressions of Section \ref{s:emp}. Section \ref{s:controls} of the Online
      Appendix provides detailed explanations of all of these variables.} \label{t:controls}
  \end{table}
\end{landscape}
  
\begin{table}
    \centering
    {\bf Summary Statistics} \\[5pt]
    \small
    \begin{tabular}{@{}lcccccccc@{}l}
\toprule
        Variable &  Count &    Mean &     Std &    0.05 &    0.25 &     0.5 &    0.75 &     0.95 \\
\midrule
           PVLGD &  13517 &    6.46 &    5.64 &    1.31 &    2.57 &    4.35 &    8.17 &    19.63 \\
 CreditScore &  13817 &   $-$0.01 &    2.41 &   $-$3.67 &   $-$1.27 &   $-$0.19 &    1.03 &     4.40 \\
             $R^f$ &  13517 &    2.26 &    0.70 &    0.87 &    1.82 &    2.26 &    2.71 &     3.56 \\
         IV &  13512 &    0.29 &    0.13 &    0.15 &    0.20 &    0.26 &    0.35 &     0.55 \\
          LEV &  13517 &    0.44 &    0.18 &    0.17 &    0.29 &    0.41 &    0.57 &     0.77 \\
      CDSChg$_{2,6}$ &  13140 &   $-$0.12 &    1.94 &   $-$3.17 &   $-$0.60 &   $-$0.03 &    0.34 &     2.82 \\
       CDSChg$_{1}$ &  13478 &   $-$0.08 &    0.88 &   $-$1.55 &   $-$0.25 &   $-$0.02 &    0.12 &     1.17 \\
\bottomrule
\end{tabular} 
    \caption{\small This table includes all variables in our analysis and their summary statistics.
      We annotate actual scales of some variables if necessary. PVLGD is the measure of the price of
      a CDS contract as defined by equation (\ref{eq:cds}), CreditScore is the measure of
      creditworthiness as in equation (\ref{eq:cred-score}).  See Section \ref{s:controls} for
      detailed definitions of interest rate $R^{f}$, implied volatility IV, leverage LEV, CDS
      reversal CDSChg$_1$, CDS momentum CDSChg$_{2,6}$, realized volatility RVCredit, illiquidity
      ILLIQ, market depth CompDepth, average credit rating Rating, equity momentum EqRet$_{2,6}$,
      equity reversal EqRet$_1$, firm size Size, firm value RatioBM, firm profitability Profit,
      earnings yield EarnYield, distance-to-default DistDefault, and standardized unexpected
      earnings SUE.  PVLGDChg$_{1y}$ is the yearly change of PVLGD, RiskVol, RiskMaxIncr, RiskCumSum
      is the realized volatility, maximum monthly change, and maximum cumulative changes of PVLGD in
      next 12 months, respectively; LEVChg$_{2y}$, ProfChg$_{2y}$, and AssetChg$_{2y}$ (log) denote
      the two-year ahead changes in market leverage ratio, profit as in Gompers, Ishii, and Metrick
    (2003), and
      the logarithm of total asset, respectively. NumAnlst, DispAnlst, FKGrade, and TransLen are
      information proxies defined as in Sections \ref{s:delayed} and \ref{s:info}. All statistics
      are based on full sample series winsorized at 1\% and 99\% percentiles.}
\label{t:summ}
\end{table}


\begin{landscape}
  \begin{table}[!htb]
    \begin{center}
      \bf Forward Selected Words and Full-Sample Lasso Regression Coefficients\\[5pt]
    \end{center}  
    \small
    \begin{minipage}{.5\linewidth}
      \begin{center}
        \textbf{Panel A: Investment Grade}
      \end{center} 
      \centering 
      \begin{tabular}{@{}cccc@{}c}
\toprule
             Words &  Coefs &                 Words &  Coefs \\
\midrule
      invest\_grade &  0.316 &                growth & $-$0.020 \\
       \_num\_.\_mln\_ &  0.030 &           \_num\_.\_bln\_ & $-$0.051 \\
             coven &  0.510 &                 train & $-$0.118 \\
              loan &  0.103 &            sale.\_num\_ & $-$0.044 \\
          missouri &  0.173 & quarter.increas.\_num\_ & $-$0.137 \\
            suppli &  0.092 &               weather & $-$0.038 \\
        chief.oper &  0.361 &              worldwid & $-$0.094 \\
     price.environ &  0.130 &                impact & $-$0.022 \\
             pound &  0.195 &        investor.relat & $-$0.228 \\
    year.year.basi &  0.136 &               realize & $-$0.105 \\
       report.form &  0.429 &     ratio.\_num\_.\_num\_ & $-$0.151 \\
  loss.\_num\_.\_num\_ &  0.210 &                   gas & $-$0.029 \\
            vendor &  0.235 &   net.interest\_expens & $-$0.168 \\
   current.quarter &  0.299 &          suppli.chain & $-$0.114 \\
adjust.oper.margin &  0.168 &                 truck & $-$0.081 \\
           softwar &  0.051 &              file.sec & $-$0.315 \\
              list &  0.146 &           credit\_rate & $-$0.072 \\
              ford &  0.091 &          year.quarter & $-$0.087 \\
        \_mln\_.cash &  0.173 &                recycl & $-$0.096 \\
            liquid &  0.059 &                   jay & $-$0.108 \\
\bottomrule
\end{tabular}
    \end{minipage}
    \begin{minipage}{.5\linewidth}
      \begin{center}
        \textbf{Panel B: High Yield}
      \end{center}  
      \centering
      \begin{tabular}{@{}cccc@{}c}
\toprule
      Words &  Coefs &               Words &  Coefs \\
\midrule
      matur &  0.722 &              growth & $-$0.100 \\
      amend &  1.044 &     share\_repurchas & $-$0.384 \\
   net.loss &  1.413 &          earn.\_num\_ & $-$0.567 \\
       game &  0.329 & strong.balanc\_sheet & $-$0.960 \\
     liquid &  0.320 &           oper.rate & $-$0.455 \\
  transform &  0.406 &              averag & $-$0.149 \\
     reduct &  0.222 &              pandem & $-$0.583 \\
     atlant &  0.891 &              extend & $-$0.322 \\
     airlin &  0.500 &               share & $-$0.115 \\
factor.caus &  0.668 &                user & $-$0.564 \\
      trend &  0.142 &           residenti & $-$0.599 \\
       cash &  0.168 &        invest\_grade & $-$0.501 \\
    freight &  0.540 &         matur.\_num\_ & $-$0.445 \\
      fresh &  1.065 &       \_num\_.present & $-$0.835 \\
     repres &  0.276 &               great & $-$0.195 \\
       mail &  0.514 &            net.debt & $-$0.508 \\
       hill &  1.128 &               incom & $-$0.116 \\
      bundl &  0.734 & long.term.sharehold & $-$1.203 \\
    network &  0.195 &       revenu.expect & $-$1.628 \\
trail.\_num\_ &  0.798 &                load & $-$0.249 \\
\bottomrule
\end{tabular}
    \end{minipage}  
    \caption{This table shows the top twenty words identified by the correlation-based forward
      selection method of Section \ref{s:fwd-sel}, sorted by the order of selection. The sign of the
      variables is determined from their $\beta^{(n)}$ coefficient from \eqref{eq:xi-res}.  Also
      shows are the corresponding coefficients from the full-sample lasso regression from
      \ref{s:cred-score}. Bi- and trigrams are indicated by a period (.) separating the individual
      words. Bigrams with an underscore (\_) separator indicate that the bigram is from the credit
      words list in Table \ref{t:word-lists} in the Online Appendix.} \label{t:lasso-words}
  \end{table}
\end{landscape}

\begin{table}
  \centering
  {\bf Extreme Credit Score Examples: Deciles 1 and 10} \\[5pt]
  \footnotesize
  \scalebox{0.94}{    \begin{tabular}{cccm{1cm}ccc}
\\
    \multicolumn{3}{l}{\textbf{Decile 1 Example}} &       & \multicolumn{3}{l}{\textbf{Decile 10 Example}} \\
\cmidrule{1-3}\cmidrule{5-7}    \multicolumn{3}{c}{Company:  New York Times Co. (NYT)} &       & \multicolumn{3}{c}{Company: KB Home (KBH)} \\
    \multicolumn{3}{c}{Date: 2017.07.27 to 2017.11.01} &       & \multicolumn{3}{c}{Date: 2016.01.07 to 2016.03.23} \\
    \multicolumn{3}{c}{PVLGD: 2.303 to 1.873} &       & \multicolumn{3}{c}{PVLGD: 19.155 to 16.185} \\
    \multicolumn{3}{c}{Implied PV: 7.323 to 15.706} &       & \multicolumn{3}{c}{Implied PV: 17.529 to 8.304} \\
\cmidrule{1-3}\cmidrule{5-7}    Word  & Contribution & Coefficient &       & Word  & Contribution & Coefficient \\
\cmidrule{1-3}\cmidrule{5-7}    revenu.expect & 4.883 & $-$1.628 &       & share\_repurchas & $-$3.840 & $-$0.384 \\
    oblig & 2.483 & 0.497 &       & share & $-$1.952 & $-$0.115 \\
    user  & 1.692 & $-$0.564 &       & revenu.expect & $-$1.628 & $-$1.628 \\
    growth & 0.797 & $-$0.100 &       & oper.incom.margin & $-$1.377 & $-$0.689 \\
    workforc & 0.563 & $-$0.563 &       & statement.mean & $-$1.303 & $-$1.303 \\
    quarter.year & 0.532 & $-$0.133 &       & liquid & $-$1.280 & 0.320 \\
    foundat & 0.471 & $-$0.471 &       & presid.chief.execut & $-$0.934 & $-$0.934 \\
    acquir & 0.420  & $-$0.210 &       & today.discuss & $-$0.780 & $-$0.780 \\
    reduc & 0.405 & 0.135 &       & profit & $-$0.742 & 0.046 \\
    plan.\_num\_.\_num\_ & 0.402 & 0.402 &       & situat & $-$0.726 & 0.363 \\
\cmidrule{1-3}\cmidrule{5-7}    \end{tabular}%
}
  \caption{This table shows an example of credit scores in deciles 1 and 10 of the extreme credit
    score but not-extreme PVLGD double sorts described in Section \ref{s:es}. The implied PVLGDs
    come from the full-sample model. Each row corresponds to a token that is a top implied PVLGD
    change contributors in the quarter. The contribution column shows the magnitude of the
    contribution (change in token frequency times token coefficient), and the coefficient column
    shows the full-sample implied PVLGD model coefficient associated with the token. For example,
    tokens with negative coefficients will contribute positively to credit score if they occur less
    frequently.}
  \label{t:event_example}
\end{table}


  \begin{table}
    \centering
    {\bf Dependence of {12}-month {PVLGD} Changes on Forecasting Variables:\\ \capitalisewords{full sample}
      Text Model} \\[5pt]
    \footnotesize \begin{tabular}{@{}lccccccccc@{}c}
\toprule
        &      (1) &     (2) &      (3) &     (4) &     (5) \\
\midrule
         $R^f$ &   $-$0.423 &  $-$0.451 &   $-$0.321 &  $-$0.307 &         \\
             &  ($-$4.73) & ($-$4.59) &  ($-$3.73) & ($-$3.77) &         \\
     IV &    4.625 &   4.179 &    3.995 &   3.951 &         \\
             &   (5.98) &  (5.53) &   (5.36) &  (5.49) &         \\
      LEV &   12.349 &  11.538 &   11.671 &  11.616 &         \\
             &  (13.59) & (12.14) &  (12.55) & (12.76) &         \\
       PVLGD &          &         &   $-$0.223 &  $-$0.180 &  $-$0.206 \\
\bottomrule
\end{tabular}
    \caption{This table reports regressions results of {12}-month {PVLGD} changes on the
      contemporaneous changes in company market leverage, risk-free rate, and option implied
      volatility, adding {PVLGD} level, implied credit score and factors. The full specification is
      given in equation (\ref{eq:contemp-reg}). The table reports results based on the full sample implied
      credit score. Column (1) presents results with contemporaneous changes, column (2) adds
      control factors, column (3) adds {PVLGD} levels, column (4) adds credit scores, and column (5)
      drops contemporaneous changes. We cluster standard errors by entity and month, the
      t-statistics are reported in parenthesis. Only control variables that are statistically
      significant in at least one specifications are included. All statistics
      are based on full sample series winsorized at 1\% and 99\% percentiles.}
    \label{t:contemp-full sample-12-PVLGD}
  \end{table}

\begin{table}
  \centering
  {\bf Forecast Regressions on Risk and Fundamental Measure Changes:\\
    Full Sample Text Model} \\[5pt]
  \scalebox{0.8}{\begin{tabular}{@{}lcccccc@{}c}
\toprule
        &  RiskVol &  RiskMaxIncr &  RiskCumSum  & ProfChg$_{2y}$ & LEVChg$_{2y}$ &   AssetChg$_{2y}$ (log) \\
\midrule
 CreditScore &  $-$0.081 &  $-$0.141 &  $-$0.280 &     0.075 &   $-$0.033 &   0.006 \\
             & ($-$4.98) & ($-$6.13) & ($-$7.78) &    (2.01) &  ($-$1.79) &  (0.31) \\
       PVLGD &   0.324 &   0.198 &   0.010 &    $-$0.078 &   $-$0.046 &  $-$0.073 \\
             &  (5.91) &  (2.95) &  (0.13) &   ($-$0.86) &  ($-$0.99) & ($-$1.40) \\
   CDSChg$_{1}$ &   0.034 &   0.041 &   0.051 &    $-$0.002 &    0.016 &   0.001 \\
             &  (2.34) &  (1.75) &  (1.64) &   ($-$0.08) &   (1.60) &  (0.13) \\
    RVCredit &   0.379 &   0.419 &   0.365 &     0.081 &   $-$0.035 &   0.061 \\
\bottomrule
\end{tabular}}
  \caption{This table reports regression results of the 12-month risk and 2-year changes in
    fundamental measures on full sample implied credit score, PVLGD, and control
    variables. Coefficients reflect the change in the dependent variable in units of its
    interquartile range (IQR) due to a unit IQR change in the independent variable. T-statistics are
    reported in parentheses. We cluster standard errors by entity and month. The first three
    dependent variables reflect 12-month ahead outcomes and the last three reflect 24-month ahead
    ones. RiskVol is defined as the realized volatility of PVLGD, RiskMaxIncr is the maximum monthly
    PVLGD change, RiskCumSum is the maximum cumulative PVLGD change over the next year,
    ProfChg$_{2y}$ is the two-year change in profitability defined as in Gompers, Ishii, and Metrick
    (2003),
    LEVChg$_{2y}$ is the two-year change in the market leverage ratio, AssetChg$_{2y}$ (log) is the
    two-year change of the logarithm of total assets. Only control variables that are statistically
    significant in at least one specifications are shown. We winsorize both dependent and
    independent variables at 1\% and 99\% percentiles.}
  \label{t:forecast full sample}
\end{table}


\begin{table}
  \centering
  {\bf Forecast Regressions on {12}-Month PVLGD Changes with Interactions:\\ \capitalisewords{full sample} Text Model} \\[5pt]
  \small
  \begin{tabular}{@{}lcccccccc@{}c}
\toprule
          &     (1) &     (2) &     (3) &     (4) &     (5) &      (6) &     (7) &     (8) \\
\midrule
   CreditScore &  $-$0.193 &  $-$0.185 &  $-$0.190 &  $-$0.189 &  $-$0.193 &   $-$0.210 &  $-$0.190 &  $-$0.204 \\
               & ($-$8.93) & ($-$8.69) & ($-$8.90) & ($-$8.88) & ($-$8.89) & ($-$10.20) & ($-$8.92) & ($-$9.74) \\
         PVLGD &  $-$0.193 &  $-$0.204 &  $-$0.193 &  $-$0.193 &  $-$0.193 &   $-$0.198 &  $-$0.192 &  $-$0.197 \\
               & ($-$7.50) & ($-$7.83) & ($-$7.44) & ($-$7.45) & ($-$7.50) &  ($-$7.90) & ($-$7.43) & ($-$7.78) \\
    DispAnlst &         &   0.065 &         &         &         &          &         &         \\
               &         &  (3.71) &         &         &         &          &         &         \\
 CS\_DispAnlst &         &  $-$0.002 &         &         &         &          &         &         \\
\bottomrule
\end{tabular} 
  \caption{This table reports regression results of the {12}-month {PVLGD} changes on the {PVLGD} level,
    full sample implied credit score, control variables and their interactions with credit score. Even
    columns report results with factors and interections, and odd columns report corresponding
    results without interactions.  (Full specification is:
    $\Delta {PV}_{i,t+\ell} = \alpha+ \beta_{pv}{PV}_{i,t}+\beta_{cs} CS_{i,t}+ \beta_{cv} CV_{i,t} +
    \beta_{cs \times cv} CS_{i,t}\times CV_{i,t} + \beta_{fac}^{\top} X_{i,t}
    +\varepsilon_{i,t}$). (1)-(2) are for results when the control variable is analyst dispersion
    during the past 12 months; (3)-(4) for Flesch-Kincaid Grade; (5)-(6) for number of analysts
    that have made at least 1 estimate during the past 12 months; (7)-(8) for transcript length of
    earnings calls. We cluster standard errors by firm and month, the t-statistics are reported
    in parenthesis. Only control variables that are statistically significant in at least one
    specifications are included. All statistics
      are based on full sample series winsorized at 1\% and 99\% percentiles.}
  \label{t:interaction-full sample-12-PVLGD-PV}
\end{table}



\newpage
\begin{table}
  \centering
  {\bf Trading Strategy Simulations using Rolling Text Model} \\[5pt]
  {
    \begin{tabular}{c|ccccccc}
    \multicolumn{8}{c}{Panel A: Investment Grade}\\
    \toprule
          & 0.01  & 0.02  & 0.05  & 0.1   & 0.2   & 0.3   & 0.4 \\
    \midrule
     $-$0.01  &   --     &      --  &    --    & 0.865$^{\ast\ast\ast}$ & 0.615$^{\ast\ast\ast}$ &     --   & --  \\
     $-$0.02  &   --     &    --    &   --     & 1.669$^{\ast\ast}$ & 1.747$^{\ast\ast\ast}$ &      --  & --  \\
    $-$0.05  &   --     &   --     & 1.806$^{\ast\ast\ast}$ & 2.406$^{\ast\ast\ast}$ & 2.141$^{\ast\ast\ast}$ & 1.662$^{\ast\ast\ast}$ & 1.289 \\
     $-$0.1   & 1.228$^{\ast\ast\ast}$ & 1.626$^{\ast\ast\ast}$ & 2.241$^{\ast\ast\ast}$ & 2.779$^{\ast\ast\ast}$ & 2.716$^{\ast\ast\ast}$ & 2.180$^{\ast\ast\ast}$ & 1.897$^{\ast\ast\ast}$ \\
     $-$0.2   & 1.460$^{\ast\ast\ast}$ & 2.001$^{\ast\ast\ast}$ & 2.447$^{\ast\ast\ast}$ & 3.073$^{\ast\ast\ast}$ & 3.413$^{\ast\ast\ast}$ &    --    &  -- \\
     $-$0.3   &  --      &   --     & 2.546$^{\ast\ast\ast}$ & 2.934$^{\ast\ast\ast}$ &   --     &  --      &  -- \\
     $-$0.4   & --       &    --    & 2.626$^{\ast\ast\ast}$ & 2.867$^{\ast\ast\ast}$ &    --    &  --      &--   \\
     \bottomrule
     \multicolumn{8}{l}{}\\
     \multicolumn{8}{l}{}\\
\multicolumn{8}{c}{Panel B: High Yield}\\
    \toprule
          & 0.01  & 0.02  & 0.05  & 0.1   & 0.2   & 0.3   & 0.4 \\
    \midrule
    $-$0.01 & --     & --     & --     & 0.235$^{\ast\ast}$ & 0.249$^{\ast\ast}$ & --     & -- \\
     $-$0.02 & --     & --     & --     & 0.369 & 0.463$^{\ast\ast}$ & --     & -- \\
     $-$0.05 & --     & --     & 1.327 & 1.500 & 0.944 & 0.505 & 0.655 \\
     $-$0.1  & 1.203$^{\ast\ast\ast}$ & 1.714$^{\ast\ast\ast}$ & 2.608$^{\ast\ast}$ & 2.687 & 2.630$^{\ast}$ & 2.071 & 1.727 \\
     $-$0.2  & 1.105 & 2.400$^{\ast\ast\ast}$ & 3.437$^{\ast\ast}$ & 3.360 & 3.014$^{\ast}$ & --     & -- \\
     $-$0.3  & --     & --     & 3.691$^{\ast\ast}$ & 4.135$^{\ast\ast\ast}$ & --     & --     & -- \\
     $-$0.4  & --     & --     & 4.027$^{\ast\ast\ast}$ & 4.366$^{\ast\ast}$ & --     & --     & -- \\
    \bottomrule
    \end{tabular}%
    \caption{\small This table reports the annualized return from each portfolio test specification
      and its statistical significance under the null hypothesis of no predictability. Panel A and B
      report results for IG and HY samples respectively. The leverage constraints in
      (\ref{constr_UL}) for the two samples are set as $(-4,4)$.  Each cell corresponds to a weight
      limit specification $\{l,u\}$ in (\ref{constr_ul}), with the value of $l$ and $u$ given in the
      row and column respectively. P-values indicating 1\%, 5\%, and 10\% significance levels are
      represented by $^{\ast\ast\ast}$, $^{\ast\ast}$, and $^{\ast}$, respectively. More details of
      the simulation are in Section \ref{s:null-dist} of the Online Appendix.}
  \label{t:porttest}%
  }
\end{table}%
\makeatletter\@input{yy.tex}\makeatother
\end{document}


\title{Online Appendix: \paperName}

\author{Harry Mamaysky\thanks{Columbia Business School, hm2646@columbia.edu.}
  \and Yiwen Shen\thanks{HKUST Business School, yiwenshen@ust.hk.}
  \and Hongyu Wu\thanks{Yale School of Management, hw499@yale.edu.}}

\maketitle

Section \ref{s:pvlgd} explains how we calculate PVLGDs. Section \ref{s:datamap} explains our data
mapping procedure, and Section \ref{s:cds-exit} discusses cases when firms with CDS observations
left our sample. Section \ref{s:controls} details our control variable construction methodology.
Section \ref{s:textmod} gives details about our text model estimation and how we derive the earnings
call credit score.  Section \ref{s:valid_ap} analyzes the cross-sectional and time series
relationship between implied PVLGDs and PVLGDs. Section \ref{s:interactions} describes the
construction of decile levels for information related variables.  Section \ref{s:opt-dets} provides
the details for the trading simulation under the null hypothesis of no predictability.

\section{Calculation of PVLGD} \label{s:pvlgd}

This section explains the calculation of the PVLGD.
The relationship between the CDS spread, $S$, and the PVLGD is given by
\begin{equation} \label{eq:pvlgd}
  S \times \underbrace{\frac{ 1}{100} \sum_{i=1}^C \E^*\one[\tau >
    t_i] \times (t_i-t_{i-1}) \times e^{-r t_i}}_{\text{PV01}} = \underbrace{\sum_{i=1}^C
    \E^*\one[t_{i-1} < \tau \leq t_i] \times e^{-r t_i} \times L}_{\text{PVLGD}},
\end{equation}
where $C$ is the number of CDS coupon payments (typically paid quarterly), $t_i$ is the date of the
$i$th coupon payment with $t_0=0$, $r$ is the deterministic interest rate, $L$ is the deterministic
loss-given-default (i.e., the difference between the cheapest-to-deliver bond's price and par),
$\tau$ is the stopping time corresponding to a default event that triggers the CDS payout, $\E^*$ is
the risk-neutral expectation operator, and $\one$ is an indicator variable for the time of
default. We model $\E^*\one[\tau > t_i] = e^{-h t_i}$ for a constant default intensity $h$;
therefore $\E^*\one[t_{i-1} < \tau \leq t_i] = e^{-ht_{i-1}} - e^{-ht}$. In (\ref{eq:pvlgd}), $S$ is
expressed in basis points (e.g., 150 for 1.5\% spread), PV01 gives the change in the value of the
CDS for a one basis point increase in $S$ (e.g., PV01=$0.045$), and $L$ is the loss per \$100
principal value of CDS (e.g., \$60). Section \ref{s:numerical-pvlgd} gives the details of our
numerical implementation of (\ref{eq:pvlgd}).

Equation (\ref{eq:pvlgd}) states that the risk-neutral expectation of what the CDS buyer will pay
for default insurance, given by $S \times PV01$, equals the dollar value of the seller's
risk-neutral expected payout from selling protection, i.e., the PVLGD of the contract. 

\subsection{Numerical Implementation} \label{s:numerical-pvlgd}

We calculate the credits' PVLGDs from equation (\ref{eq:pvlgd}) as follows.  Given the constant
default intensity $h$, the default probabilities in (\ref{eq:pvlgd}) can be calculated as
\begin{equation*}
  \mathsf{E}^{\ast}\mathbf{1}[\tau>t_i] = e^{-ht_i}\, \text{ and } \,
  \mathsf{E}^{\ast}\mathbf{1}[t_{i-1}<\tau<t_i] = e^{-ht_{i-1}} - e^{-ht_i}.
\end{equation*}
Thus, we can drive the PV01 and PVLGD as
\begin{equation*}
    \text{PV01} = \frac{1}{100}\sum_{i=1}^C e^{-(r+h)t_i} \times (t_i - t_{i-1}) 
\end{equation*}
 and 
 \begin{equation}
    \text{PVLGD} = \sum_{i=1}^C \left(e^{-ht_{i-1}} - e^{-ht_i}\right) \times e^{-rt_i} \times L.\label{eq:PVLGD}
\end{equation}
Then for the default intensity $h$, the implied par spread is given by 
\begin{equation}
    \text{ParSpread} = \frac{\text{PVLGD} }{ \text{PV01}}=\frac{ \sum_{i=1}^C \left(e^{-ht_{i-1}} - e^{-ht_i}\right) \times e^{-rt_i} }{\sum_{i=1}^C e^{-(r+h)t_i} \times (t_i - t_{i-1})}\times 100 \times L.\label{eq:parspread}
\end{equation}
For standard CDS considered in our study, their maturity is five years and the coupons are paid
every quarter. So we have $t_i - t_{i-1} = 0.25$ and $C=5/0.25 = 20$.  The coupon payment dates are
given by $\{0.25,0.50,..., 5\}$.  In addition, we set the loss-given-default as 60, suggesting the
bond loses 60\% of the par value when the firm defaults.  The $r$ is the five-year risk-free rate on
the pricing date of CDS.

We use a grid of default intensity $h$ and calculate the par spread by (\ref{eq:parspread}) for each
$h$.  To balance precision and computational burden, we use a finer grid for small $h$ and a looser
grid for large $h$. The grid is given by 0 to 0.20 with a step of 0.0025, 0.24 to 4 with a step of
0.04, and 4.04 to 200 with a step of 0.5. We get the proper intensity $h$ associated with the
observed spread by interpolation, and then the credit's PVLGD by (\ref{eq:PVLGD}).%
\footnote{We also considered a second-order adjustment to account for the accrued coupon between
  coupon dates, We find adding this second-order adjustment makes almost no difference to the
  resulting PVLGD.}

In Table \ref{t:bbg_compare}, we report the points upfront ({\it Pts Upf}) amount from the Bloomberg
pricing calculator using the ISDA Standard Upfront Model for the spread levels in the $S$ column of
Table \ref{t:pvlgds}. This model is used by practitioners to cash settle CDS trades. The points
upfront represents the dollar amount paid by the seller of a CDS with a spread $S$ but having a zero
coupon. It is very close to our own PVLGD calculation.  The small discrepancy between our and
Bloomberg's values arises because we do not use the same interest rate curves (Bloomberg is using
the SOFR curve at the time of pricing, and we are using a flat rate equal to the sample average
shown in Table \ref{t:pvlgds}). We show sample Bloomberg pricing screens for CDS PVLGDs with $S=100$
or 500 (accessed via \texttt{CDSW} command) in Figure \ref{f:bbg-cdsw}.

\begin{table}[htbp]
  \centering
  \caption{Comparison of point upfront (\textit{Pts Upf}) from Bloomberg and our PVLGD measure,
    assuming $r=0.0226$ and $L=60$. The imputed default intensity $h$ from \eqref{eq:pvlgd} is also
    shown.} \label{t:pvlgd-bbg-comp}
    \begin{tabular}{cccc}
    \toprule
    S     & Pts Upf (BBG) & PVLGD & $h$ \\
    \midrule
    100   & 4.511 & 4.517   & 0.017\\
    200   & 8.666 & 8.665   & 0.033\\
    500   & 19.263 & 19.192 & 0.082\\
    600   & 22.254 & 22.150 & 0.099\\
    2500  & 50.033 & 49.619 & 0.396\\
    2600  & 50.631 & 50.208 & 0.411\\
    \bottomrule
    \end{tabular}
  \label{t:bbg_compare}
\end{table}

\section{Data Processing} \label{s:datamap}

This section details our data mapping procedures between CDS, text, stock price, fundamental, and implied
volatility data.

\subsection{Mapping CDS Data with Text Data}

Each earnings call from {\it SP Global} is associated with a {\it Company ID} and a release date,
while each CDS observation is associated with a \textit{cusip} and the corresponding date. We obtain
the matching files between \textit{cusip} and \textit{gvkey} from CRSP database and the matching
file between \textit{gvkey} and \textit{Company ID} from the Capital IQ database. Both databases are
on WRDS. The matching files are available upon requests.

When we match the data, we use \textit{gvkey} as the identifier to link \textit{cusip} with
\textit{Company ID} for each observation. These mappings have a time period during which they apply,
and we use {\it gvkey}-{\it cusip}-{\it Company ID} matches only during time periods when they are
valid. For each earnings call, if it happens before 4 PM, we associate it with the CDS from the
firm on the same trading day. Otherwise, we use the CDS from the next trading day.  This matching
results in the panel of observations with CDS and textual data.

\subsection{Mapping Text Data with Compustat Data}
 
Compustat data is associated with a \textit{gvkey} and a data date. Since we assume the fundamental
information is available no earlier than three months after the data date, we match the earnings
call data with the Compustat data from three months ago using the mapping file from \textit{gvkey}
to \textit{Company ID}.  Specifically, if an earnings call takes place in month $t$, we match it
with the Compustat data with data date in month $t-3$.

\subsection{Mapping Compustat Data with I/B/E/S Data}
I/B/E/S observations are identified by \textit{ncusip} and dates, whereas Compustat ones are
identified by \textit{gvkey} and data dates. We match these two data sets via the \textit{permno}
from the CRSP data. WRDS provides mapping files from \textit{ncusip} to \textit{permno} and from
\textit{permno} to \textit{gvkey}. The mapping files also include the validity period of each
mapping.  We use \textit{permno} as the median identifier to join \textit{ncusip} with
\textit{gvkey} for each observation. If there are multiple matches between different identifiers, we
drop those matches that are out of the validity periods.

\subsection{Firms Leaving the Sample} \label{s:cds-exit}

Here we briefly discuss our analysis about firms (CDS observations) exiting the sample, and whether
spread widening or tightening has already happened by the time of exit. The relative (to corporate
bonds) liquidity of CDS contracts allows them to quickly react in anticipation of future credit
shocks, and CDSs associated with names that exit our sample due to default or M\&A activity tend to
experience spread changes that fully reflect these credit events prior to exiting our
sample. Overall, the evidence suggests that our CDS sample is free of survivorship bias due to
credit events that were not anticipated by markets prior to their occurrence.

Throughout our sample, there are 59 firms whose data termination dates precede the end of the sample
(Dec 2020).%
\footnote{The full list of firms is available from the authors.}
We manually analyzed each of these cases around the exit date to identify the cause for exit, e.g.,
a credit event, name changes, etc. Among the 59 firms, the reasons for 51 exits are detailed in
Table \ref{t:real-loss}. We do not find obvious precipitating events for the remaining 8 cases. In
terms of the credit ratings, 31 firms were in IG (26 had BBB ratings) and 28 in HY. All 5
bankruptcy/default events take place among HY observations.

\begin{table}
  \centering
  \caption{CDS sample dropout cases classified by reason for exit. {\it Nothing} indicates that we
    could not identify a reason for the firm's exit from the sample. Each column reports the average
    of the indicated quantity over all members of the set shown in the Status column.  } \label{t:real-loss}
\begin{tabular}{@{}lccccc@{}l}
\toprule
             Status &  Count &  $\mu$(PV) &  $\mu$(PV \% Chg$_{3m}$) &  $\mu$(PV Chg$_{3m}$) \\
\midrule
                M\&A &     37 &   7.08 &       9.56 &      0.50 \\
            Nothing &      8 &  12.50 &       8.43 &     $-$0.14 \\
 Bankruptcy/default &      5 &  51.24 &     162.21 &     20.80 \\
        Name change &      5 &   7.23 &     $-$21.24 &     $-$3.08 \\
     Delayed report &      1 &  21.55 &      35.06 &      5.59 \\
Exit from a country &      1 &  22.89 &     103.77 &     11.66 \\
  Liquidity concern &      1 &  35.57 &     299.04 &     26.66 \\
         Separation &      1 &   6.97 &      31.94 &      1.69 \\
\bottomrule
\end{tabular}
\end{table}

For each sample-exit event, Table \ref{t:real-loss} summarizes the PVLGD of the firms at the time of
exit. It reports the occurrence of each type of event and the within-group mean of the following
variables:%
\footnote{For groups with one observation, we report the exact event.}
the PVLGD, defined in \eqref{eq:pvlgd}, on the exit day (PV); the percentage change in PV in the 3
months prior to the exit (PV \% Chg$_{3m}$); the change in PV in the 3 months prior to the exit
(PV Chg$_{3m}$). Table \ref{t:real-loss} shows that cases not involving distress (M\&A, Nothing, Name
change, Separation), the change in credit spread (or PVLGD) in the three months prior to exit was
minor. For the other exit categories associated with distress (Bankruptcy/default, Delayed report,
Exit from a country, Liquidity concern), the firms, on average, see a large increase in their PVLGDs
before exiting the sample. The fact that events associated with distress are associated with large
PVLGD increases prior to firms' leaving the sample and the fact that non-distress events do not see
a large change in PVLGD prior to sample exit suggest that survivorship bias is not an issue for our
analysis.

\section{Control variables} \label{s:controls}

This section specifies the details in our construction of the control variables listed in Section
\ref{s:data} in the main text. We calculate each measure at the end of the month.

\subsection{CDS related Factors}
\begin{enumerate}[label=\roman*)]
\item CDS Momentum ($\text{CDSChg}_{2,6}$)\\
  The cumulative 5-month changes in PVLGD starting from 6-months ago excluding the most recent month
  ($t-6,\dots,t-2$). Jostova et al. (2013) document significant momentum profitability in corporate
  bond markets that is not captured by the equity momentum. Lee et al. (2021) provide evidence on
  momentum profitability in CDS market for both IG and HY samples.

\item CDS Reverse ($\text{CDSChg}_1$)\\
  The most recent 1-month lagged PVLGD changes, computed from the end of month $t-2$ to the end of
  month $t-1$. Bartram, Grinblatt, and Nozawa (2020) find the bond return reversal to negatively
  forecast excess corporate bond returns in the cross section.

\item Realized Volatility (RVCredit)\\
  The realized volatility of monthly PVLGD changes calculated calculated as the standard deviation
  of the monthly PVLGD change over the past 12 months. This proxies for the volatility controls
  used by Chordia et al. (2017) for explaining corporate bond returns.

\item Illiquidity (ILLIQ)\\
  A proxy for market illiquidity, measured as the negative of the covariance of the daily change of
  log PVLGD and its one-day lag, calculated in rolling 1-month windows. Bao, Pan, and Wang (2011)
  find this explains a substantial part of time variation in corporate bond yields, at the aggregate
  and individual levels. This is computed on a monthly basis. Denote by $PV_s$ the PVLGD at the end
  of day $s$, we have
  \begin{center}
    $\text{ILLIQ}=- \text{Cov}(\log\left({PV_s/PV_{s-1}}\right),\, \log\left({PV_{s-1}/PV_{s-2}}\right))$.
  \end{center} 
  The covariance is taken for the daily changes within a given month. We treat the value as missing
  if there are less than 10 pairs of daily changes from day $t$ and day $t+1$ for a company within a
  month.

\item Composition Depth (CompDepth)\\
  A proxy for CDS market depth, measured as the number of distinct contributors of CDS levels,
  obtained from Markit. We use it as another proxy for market liquidity, in addition to ILLIQ.

\item Rating: average credit rating from Markit. The values are AAA, AA, A, BBB, BB, B, CCC, D, and
  unclassified. We convert these to integers with $\text{AAA}=1$, $\text{AA}=2$, \dots,
  $\text{D}=8$, and unclassified observations are set 9. Credit ratings are a commonly used control
  for corporate bond returns, see, for example, Bali et al. (2022), Guo, Lin, Wu, and Zhou (2021),
  Bao, Pan and Wang (2011).

\end{enumerate}

\subsection{Equity and Fundamental Factors}

We retrieve raw data from Compustat and construct different factors based on them. Later when
matched to CDS data and other measures, we lag the fundamental data by 3 months to account for the
fact that fundamental data are usually published with a lag of 1 quarter.  The variables numbered
\ref{item:rf} -- \ref{item:iv} are only included as covariates in the contemporaneous regressions of
Section \ref{s:emp} and are not shown in the correlation table.

\begin{enumerate}[label=\roman*)]
\item Equity Momentum ($\text{EqRet}_{2,6}$)\\
  Momentum of stock price, the return of stock from the end of the month $t-6$ to the end of the
  month $t-2$. Specifically, in terms of Compustat variable names, the stock price at the end of
  month $t$ is $prc_{i,t}$, and thus
  \[
    {\text{EqRet}_{2,6}}=\displaystyle\frac{prc_{i,t-2}-prc_{i,t-6}}{prc_{i,t-6}}
  \]
  This proxies for the momentum effect of Jegadeesh and Titman (1993).

\item Equity Reversal ($\text{EqRet}_{1}$)\\
  The most recent 1-month lagged equity returns, to control for the equity reversal effect
  documented in Jegadeesh (1990) and Lehmann (1990). EqRet$_1$ and EqRet$_{2,6}$ are used as
  controls in Chordia et al. (2017) and both positively predict monthly corporate bond returns.

  Specifically, in terms of Compustat variable names, 
  \[
    {\text{EqRet}_{1}} =\displaystyle\frac{prc_{i,t-1}-prc_{i,t-2}}{prc_{i,t-2}}.
  \]
  
\item Size (Size)\\
  Size of the company issuing a CDS. Measured as natural logarithm of a company's
  capitalization. Specifically, in terms of Compustat variable names:
  \begin{center}
    $\text{Size}=\text{log}(shrout_{i,t}\times prc_{i,t})$
  \end{center}
  This is the natural logarithm of the market value of the firm equity (in millions of dollars) and
  proxies for the size effect of Banz (1981) and Fama and French (1992). Size, value, and
  profitability were used by Chordia et al. (2017) in explaining the cross-section of corporate bond
  returns; they found all three variables to negatively predict monthly corporate bond returns.

\item Profitability (Profit)\\
  Profitability of the company issuing a CDS, as in {Gompers, Ishii, and Metrick
    (2003)}. It is calculated as the ratio of equity income (income before extraordinary items-dividend on
  preferred shares+deferred income taxes) to sales, all measured in the last quarter. Specifically,
  in terms of Compustat variable names:
  \begin{center}
    $\text{Profit}=\displaystyle\frac{ibadjq_{i,t}-dvpq_{i,t}+txdiq_{i,t}}{saleq_{i,t}}$
  \end{center}
  We treat the value as missing if the denominator is zero or negative.

\item BM (RatioBM)\\
  Natural logarithm of the ratio of the book value of equity to the market value of equity. See Fama
  and French (2008). Specifically, in terms of Compustat variable names:
  \begin{center}
    $\text{RatioBM}=\displaystyle\frac{atq_{i,t}-ltq_{i,t}}{shrout_{i,t} \times prc_{i,t}}$.
  \end{center}

\item Earnings Yield (EarnYield)\\
  Earnings yield of the company issuing a CDS. Calculated as the ratio between the diluted earnings
  per share and the stock price. See Penman, Richardson, Riggoni, and Tuna (2014). Bartram,
  Grinblatt, and Nozawa (2020) use this as controls for corporate bond returns. Specifically, in
  terms of Compustat variable names:
  \begin{center}
    $\text{EarnYield}=\displaystyle\frac{epsfiq_{i,t}}{prc_{i,t}}$.
  \end{center}

\item Distance-to-Default (DistDefault)\\
  Distance-to-default measure proposed by Bharath and Shumway (2008). It uses the functional form
  and variables in the Merton (1974) distance-to-default model. Specifically, in terms of Compustat
  variable names, it is calculated as:
  \begin{equation*}
    \text{DistDefault} = \frac{\log\left(\frac{E_{i,t}+D_{i,t}}{D_{i,t}}\right)+\big(r_{i,t}-(\sigma^V_{{i,t}})^2/2\big)T}{\sigma^V_{{i,t}}\sqrt{T}}.
  \end{equation*}
  where $E_{i,t}$ is the market cap of the firm; $D_{i,t} = dlcq_{i,t}+0.5\times dlttq_{i,t}$ is the
  firm's total face value of debt. We follow Vassalou and Xing (2004) to set it as the debt in
  current liabilities plus one-half of long-term debt. $r_{i,t}$ is the firm's stock return in the
  previous year. We set horizon $T=1$ year. The volatility $\sigma^V_{{i,t}}$ is given by:
  \begin{equation*}
    \sigma^V_{{i,t}}=\frac{E_{i,t}}{E_{i,t}+D_{i,t}}\sigma^E_{{i,t}}+\frac{D_{i,t}}{E_{i,t}+D_{i,t}}(0.05+0.25\sigma^E_{{i,t}}),
  \end{equation*}
  where $\sigma^E_{{i,t}}$ denotes the annualized percent standard deviation of stock returns in the
  prior year.  Bharath and Shumway (2008) show that this distance-to-default measure forecasts
  defaults better than Merton's distance-to-default model.
  
\item Standardized Unexpected Earnings (SUE)\\
  Standardized unexpected earnings of the firm issuing a CDS, defined as in Bernard and Thomas
  (1989). In terms of Compustat variable names, it is given by:
  \begin{equation*}
    \text{SUE}=\frac{epsfiq_{i,t}- epsfiq_{i,t-12}}{prc_{i,t}}.
  \end{equation*}
  Bartram, Grinblatt, and Nozawa (2020) find that SUE positively forecasts corporate bond returns.

\item Risk-free rate ($R^{f}$): the 10-year US treasury bond interest rate. Collin-Dufresne,
  Goldstein and Martin (2001) report a negative impact of $R^{(f)}$ on contemporaneous changes in
  credit spreads. \label{item:rf}

\item Leverage ($LEV$): the leverage of a firm defined as long-term debt divided by the sum of
  long-term debt and firm's market value. Ericsson, Jacobs and Oviedo (2009) document that the
  change in the leverage ratio is contemporaneously related to increasing CDS spreads.

\item Implied volatility ($IV$): the average implied volatility of the firm's 30-day put and call
  options, obtained from OptionMetrics. Cremers et al. (2008) show that implied volatility of stock
  options can explain variation in corporate bond yield spreads in both cross-sectional and time
  series contexts. \label{item:iv}
  
\end{enumerate}

\section{NLP Model for Credit Score} \label{s:textmod}

\subsection{Text Processing Procedures} \label{s:text-proc}

As described in Section \ref{s:text} of the main paper, we filter each earnings calls to get the
credit-related text as the five sentences before and after a credit sentence, which is a sentence
that includes a credit word. In addition, the sentences that include an excluded phrase are not
regarded as credit sentences, hence do not qualify as center sentences of credit-related
texts. However, these sentences may enter the corpus if they surround a credit sentence. We present
the credit words and excluded phrases in below.

\begin{table}[H]
  \centering
  \caption{Lists of credit words and excluded phrases.} \label{t:word-lists} \vskip 5pt
  \def\arraystretch{1.5}
  \begin{tabular}{p{1.75cm}p{13cm}}
    \hline
    List & Contents \\ \hline
    Credit words &
                   \textit{credit, credit line, bank line, bond, debt, leverage, rate, snp, moodys,
                   coverage ratio, leverage ratio, fitch, share repurchase, capital structure,
                   invest grade, high yield, liquidity, line credit, shareholder, leaseback,
                   creditworthiness, repayment, revolver, rating agency, covenants, cash
                   availability, cash balance, fully drawn, balance sheet, interest expense, credit
                   rating, financial profile, financial flexibility, returning cash, dividend,
                   delever, interest coverage, financing cost, refinance, refinancing, loan,
                   unsecured, secured, convertible bond, interest payment, basis point, cash flow}
    \\ \hline
    Excluded phrases &
                       \textit{exchange rate, bond rate, credit card, creditworthy customer, liquid
                       storage, shareholder meeting}  \\ \hline
  \end{tabular}
\end{table}

Now we explain how we extract numerical tokens \texttt{\_bln\_}, \texttt{\_mln\_},
\texttt{\_num\_}. We start with replacing all non-alphabet and non-numerical characters with a blank
placeholder. This will allow us to identifier words by splitting the text data by blanks. We apply
this process to all characters except for periods ($.$), as they separate sentences, and indicates
decimal points.%
\footnote{These two scenarios have distinct patterns in the text. When a period ($.$) separates
  sentences, it comes with a blank, and when it indicates a decimal point, it comes with a numerical
  character. We eliminate the latter case during the extraction of numerical tokens so that in the
  subsequent NLP analysis we can distinguish sentences by periods.}
Then we negate and stem all the words, both via the \texttt{NLTK} package of Python. Since negation
by \texttt{NLTK} attaches a \texttt{\_NEG} tag to a word, when stemming the negated words, we make
sure to stem the part of the word without the negation tag. Finally, we use regular expressions to
substitute sequence of numerical characters. For numbers with at least 10 consecutive numerical
characters before a period, we replace them by \texttt{\_bln\_}, for those with at least 7, we
replace them by \texttt{\_mln\_}, then we end up with substituting the rest of the numerical
sequences with \texttt{\_num\_}.

\subsection{Implementation of Rolling Sample Analysis} \label{s:rolling}

We now provide the implementation details for the rolling sample analysis described in Section
\ref{s:text}.  A rolling sample NLP model is defined by five hyperparameters $M$ = ($N$, $t_c$,
$N_{FS}$, $s_{lb}$, $f$), as given in Table \ref{t:nlp-params}.  The first three parameters ($N$,
$t_c$, $N_{FS}$) are used for constructing DTMs and selecting tokens for a given panel of CDS
spreads and calls.  The procedures are described in Sections \ref{s:fwd-sel} and \ref{s:cred-score}.
For the $s$th month in our sample, the \textit{lookback window}
$s_{lb} \in \{24, 36, 60, s\}$\footnote{The up-to-date lookback uses an expanding window from the
  start of the sample to the present. By its definition at month $s$, the lookback window is exactly
  $s_{lb}=s$.} determines the training window $[s-s_{lb},s)$, over which we implement forward
selection and train the PVLGD-text mapping model (\ref{eq:lasso}).  The \textit{updating frequency}
$f \in \{1, 12\}$ describes whether our training horizon moves forward by a month or a year. It also
determines the predicting window $[s,s+f)$ for which we use the trained model to construct credit
scores.

We use January 2012 (denoted by $s_0$) as the first training month regardless of the lookback
window.  Then, with the updating frequency $f$, we will need to train the lasso model and make
predictions on the following set of months: $S_f=\{s_0, s_0+f, s_0+2f, \dots\}$.  Since our text
data starts from Jan 2009, this initialization guarantees that there are enough data for 2-year,
3-year, and up-to-date lookback windows at January 2012.  For 5-year lookback window, we start with
a 3-year lookback window and march forward using an up-to-date window until there are five years of
data, For example, consider the case of 5-year lookback window ($s_{lb}=60$) with an annual updating
($f=12$).  We first estimate the model in January 2012 using the data in the three years of 2009,
2010, and 2011.  Then in January 2013, we estimate the model using four years' data from 2009 to
2012.  For 2014 and later years, we estimate the model using the prior five years' data as there are
now enough observations for the lookback window.

We summarize the steps for rolling sample analysis in the following. 

\begin{enumerate}[label=\arabic*)]
\item {Choose a specification}\\
  We first choose a specification  $M$ = ($N$, $t_c$, $N_{FS}$, $s_{lb}$, $f$) from Table \ref{t:nlp-params}. 
  The first three parameters determine how we construct DTM and select tokens from the training sample. 
  The last two parameters determine the lookback window and updating frequency. 
  We obtain the set of updating months as $S_f=\{s_0, s_0+f, s_0+2f, \dots\}$. 
\item {Construct training and predicting DTMs}\\ For each updating month $s \in S_f$, we construct
  the DTM and select tokens based on ($N$, $t_c$, $N_{FS}$) using the corpus in the lookback window
  $[s-s_{lb},s)$.  This generates the DTM for selected tokens in the training window, denoted by
  $L_{[s-s_{lb},s)}$.  Then we construct the DTM for the predicting window $L_{[s,s+f)}$.  It
  consists of the same set of selected tokens in $L_{[s-s_{lb},s)}$, but uses their counts in the
  earnings calls in the predicting window $[s,s+f)$.
\item {Calculate out-of-sample credit scores}\\
  We train a lasso model using the DTM $L_{[s-s_{lb},s)}$ and the associated PVLGDs in the lookback
  window $[s-s_{lb},s)$, i.e.,
  \begin{equation} \label{eq:rollinglasso}
  PV_{i,t\in [s-s_{lb},s)} = S_{k(i),[s-s_{lb},s)} + L_{i,t,[s-s_{lb},s)} \beta + \varepsilon_{i,t\in [s-s_{lb},s)}.
\end{equation}

This produces the estimated sector dummies $\hat{S}_{k(i),[s-s_{lb},s)}$ and the coefficient vector
$\hat{\beta}_{[s-s_{lb},s)}$.  We then compute the out-of-sample PVLGD in the predicting window
$[s,s+f)$ as
\[
  \hat{PV}_{i,t\in [s,s+f)} = \hat{S}_{k(i),[s-s_{lb},s)} + L_{i,t\in [s,s+f)} \hat{\beta}_{[s-s_{lb},s)}.
\] 
Finally, we obtain the credit scores for  $[s,s+f)$ as
\begin{equation} \label{eq:rollingcred-score}
  CS_{i,t\in [s,s+f)} = PV_{i,t\in [s,s+f)} -\hat{PV}_{i,t\in [s,s+f)} = \hat \varepsilon_{i,t\in [s,s+f)}.
\end{equation}


\item {Choose the best model specification}\\
  We then obtain the best predicted credit score series by comparing the forecasting strength from
  different specifications.  This is done for each month starting from January 2012 ($s_0$).  For
  each month $s$, we choose the best specification $M$ based on the t-statistics of the credit score
  in the forecasting regression (\ref{eq:forecast-reg}) using the data in the corresponding training
  window.  For example, consider the month of September 2018. For the specification with five-year
  lookback window and yearly updating, we use the t-statistics of credit score in
  (\ref{eq:forecast-reg}) from the training window January 2013 to December 2017 (we only include
  the 12-month ahead PVLGD change for the first four years of the window to ensure that the changes
  only involve the data in the training window).  This is because September 2018 is contained in the
  corresponding prediction window (January 2018 to December 2018).  On the other hand, for the
  specification with three-year lookback window and monthly updating, we use the t-statistics from
  the training window August 2015 to August 2018.  For each month, we select the model specification
  with the largest absolute t-statistics. Thus, the selected specification can change over time.

\item {Aggregate out-of-sample credit scores}\\
  Finally, we aggregate the credit scores for each month from its best specification to get the full
  series of out-of-sample credit scores.

\end{enumerate}

Note that out-of-sample credit scores $CS_{i,t}$ uses only information that is known to market
participants as of day $t$. In our out-of-sample forecasting tests, $CS_{i,t}$ will be asked to
forecast outcomes that happen strictly after day $t$.

\section{Validation of Credit Scores} \label{s:valid_ap}

Our text model is useful to the extent that the language used in firms' earnings calls
effectively reflects their credit quality.  In this section, we perform a more systematic
validation analysis for our credit score measures.

\begin{figure}
  \centering    
  {\bf Implied PVLGD by PVLGD sorts} \\[5pt]
  {\bf \small Panel A: Full-sample text model} \\[3pt]
  \mbox{\includegraphics[width=0.5\textwidth]
    {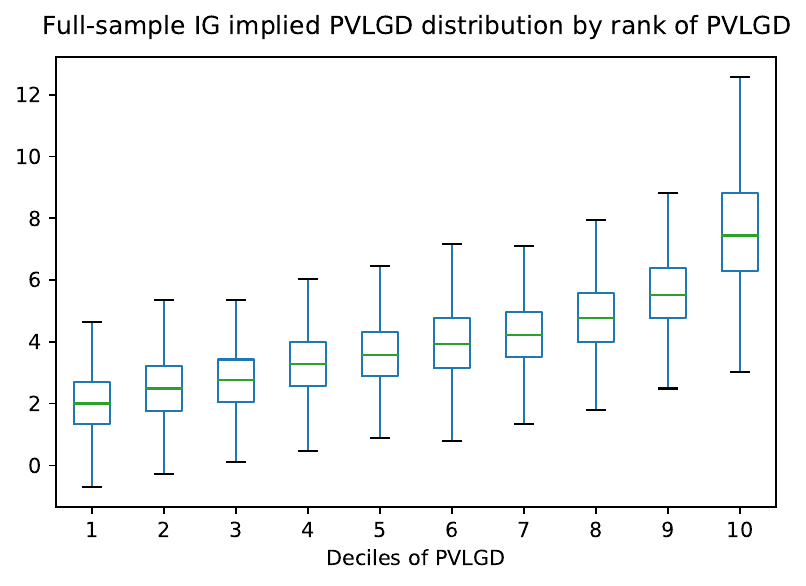}
    \includegraphics[width=0.5\textwidth]
    {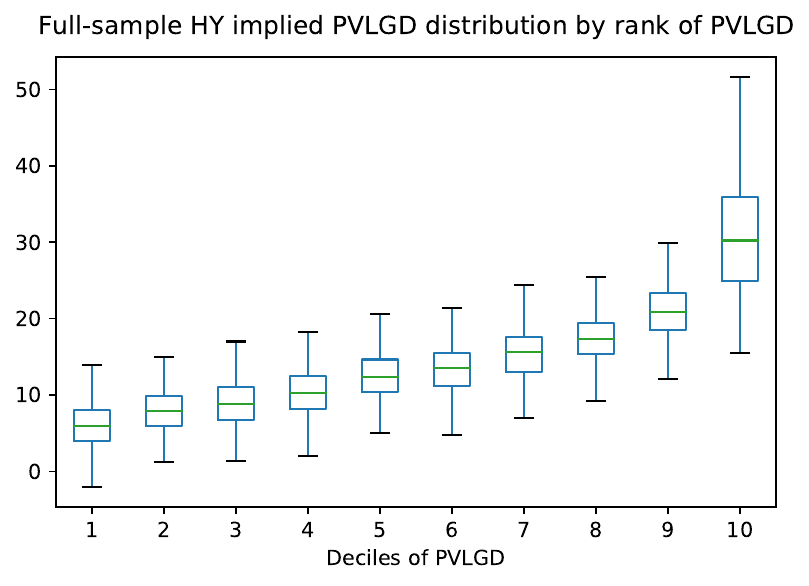}}
  {\bf \small Panel B: Rolling text model} \\[3pt]
  \mbox{\includegraphics[width=0.5\textwidth]
    {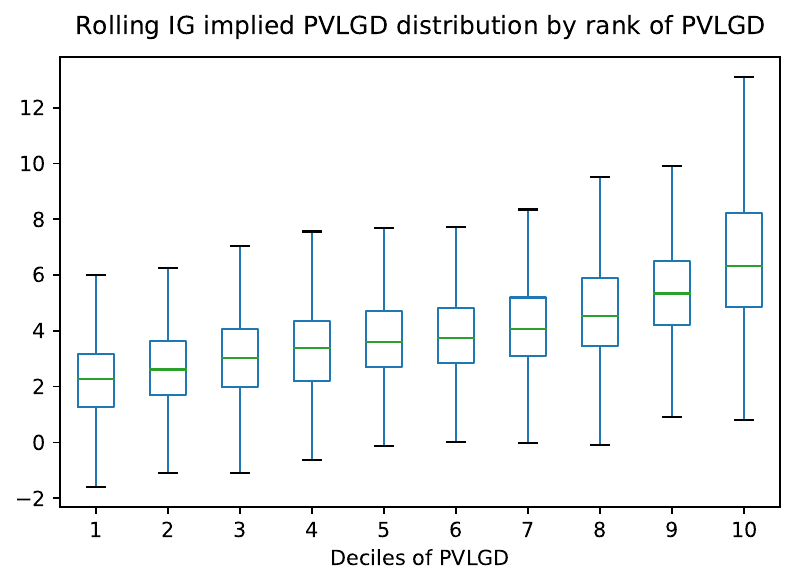}
    \includegraphics[width=0.5\textwidth]
    {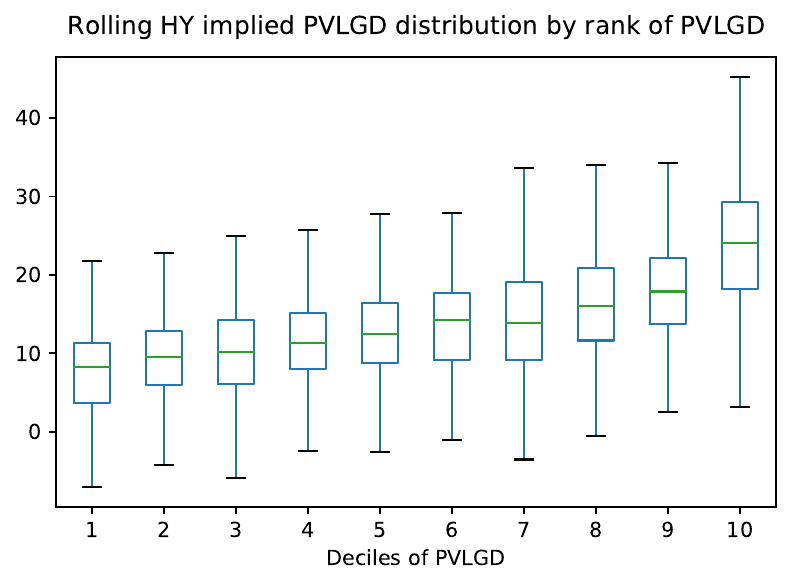}}
  \caption{The left chart in Panel A shows the distribution of full-sample implied PVLGD scores in
    buckets sorted by the post-call PVLGD for investment grade (IG) names, using the most recent
    rating prior to the call date. The right chart in Panel A shows the analogous full-sample
    results for high yield (HY) names. Panel B shows analogous results for implied PVLGD obtained
    from the out-of-sample text model. The implied PVLGDs come from the best-$R^2$ in- and
    out-of-sample models. The boxplots show the median, the interquartile range (IQR), and the first
    and third quartiles plus 1.5 times the IQR.} \label{f:boxplots}
\end{figure}

Note that if our method of capturing credit-related language in earnings calls is effective, then
ranking firms by their PVLGDs should produce a similar ranking in implied PVLGDs.  To investigate
this, all earnings calls are sorted into decile buckets based on their post-call PVLGDs. The top
panel of Figure \ref{f:boxplots} shows the distribution of implied PVLGDs from the full-sample text
model by PVLGD bucket (the left chart is for IG names, and the right chart is for HY). Both the IG
and HY charts show a strong monotonic relationship between earning calls' PVLGDs and their implied
counterparts. The language of earnings calls reflects the credit quality of firms, and our
full-sample text model can characterize the nature of this relationship.

The bottom panel of Figure \ref{f:boxplots} shows the same analysis, but using the rolling text
model instead. Since the rolling text model uses information available strictly prior to each call
to determine the implied PVLGD, this is a more stringent test of the ability of our text model to
capture credit-relevant information from earnings calls. The monotonic relationship between
post-call PVLGDs and the rolling text model implied PVLGDs suggests that the rolling text model,
available to investors in real time, can capture important aspects of how earnings call language
reflects credit quality.

Despite the monotonic relationship between single-name PVLGDs and implied PVLGDs, in each PVLGD
bucket there is a wide distribution of implied PVLGDs. Implied PVLGDs reflect credit quality on
average, but contain a large amount of other information. It is this other information that may
prove useful to investors in assessing credit quality of individual firms. Our credit score, which
is the difference between the post-call PVLGD and the call-implied PVLGD, captures this information.

\begin{figure}
  \centering
  \includegraphics[width=\textwidth]{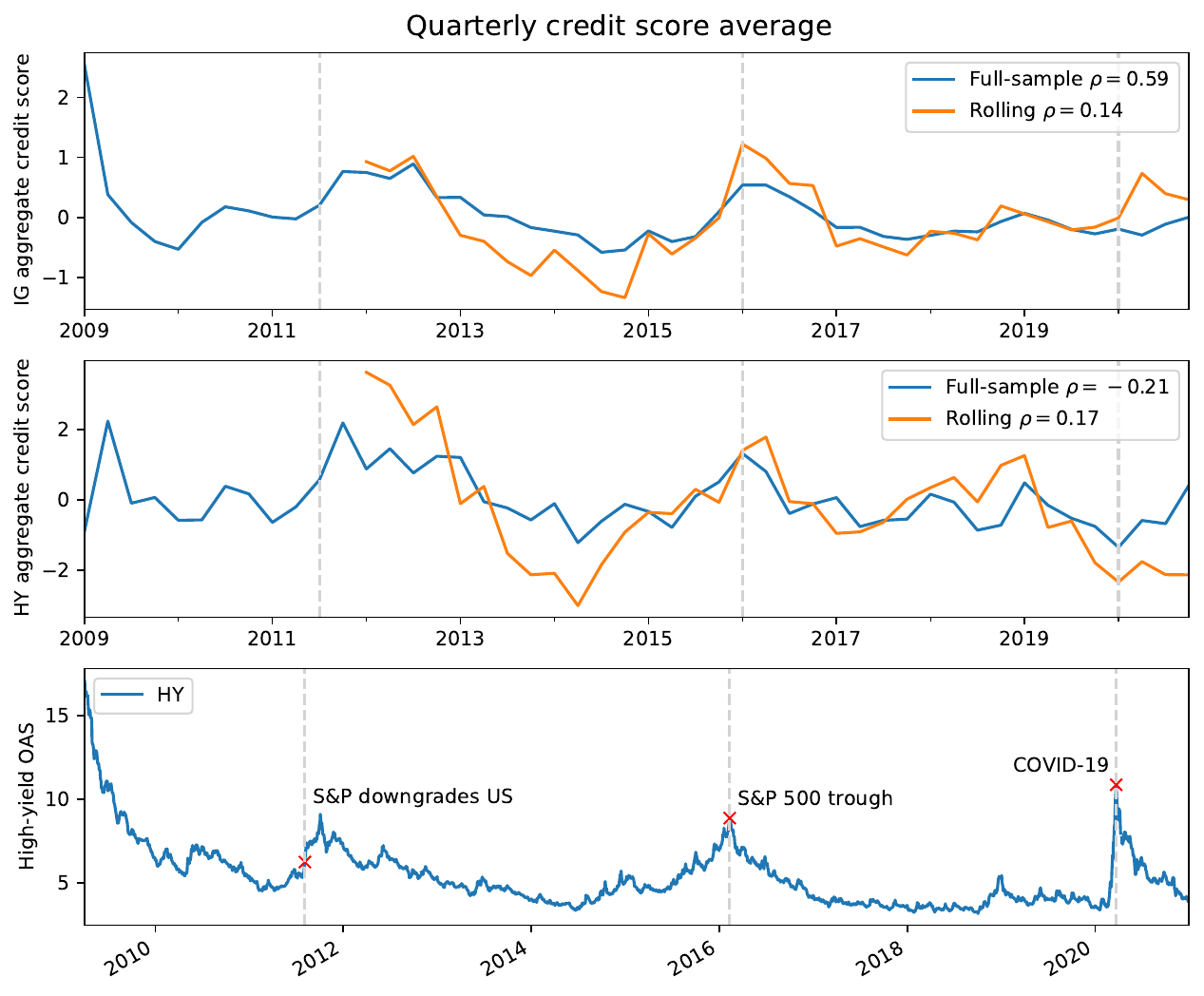}
  \caption{The top panel shows, in each quarter, the cross-sectional average earnings call credit
    score using the full-sample (blue line) and rolling (yellow, shorter line) text models for names
    whose most recent rating prior to each earnings call was investment grade.  The middle panel
    shows the analogous chart for high yield names.  The bottom panel shows a time series of option
    adjusted spreads for the ICE Bank of America high yield corporate bond index with the 2011 S\&P
    downgrade of U.S. government debt from AAA, the S\&P 500 index 2016 trough, and the trough of
    the COVID-19 market selloff highlighted. The vertical, dashed lines in the top two panels mark
    these same events. The correlation between quarterly changes in credit scores and HY option
    adjusted spreads is shown in each legend.} \label{f:ts-agg}
\end{figure}

To a large extent, the Figure \ref{f:boxplots} reflects cross-sectional variation in implied and
actual PVLGDs. But credit pricing also has pronounced time series variation, and we next check if
our model captures this aspect of the data. The top panel of Figure \ref{f:ts-agg} shows the time
series of credit scores from the full-sample and rolling text models for investment grade
names. Each quarterly observation is the average credit score of all IG names in that quarter. The
middle panel shows the analogous chart for high yield names. The bottom panel shows the option
adjusted spread (OAS) for the ICE Bank of America high yield corporate bond index, which captures
the majority of all U.S. dollar high-yield corporate bonds. Three events that were associated with
widening credit spreads -- the 2011 S\&P downgrade of U.S. government debt, the S\&P 500 index
trough in 2016, and the onset of COVID-19 in 2020 -- are marked with vertical gray lines.

At an aggregate level, the full-sample and rolling text models track each other closely. The IG and
HY credit score series reflect some of the fluctuations in the high yield OAS. For example, both the
OAS (bottom panel) and credit score series (top two panels) have local peaks around the S\&P
U.S. downgrade and the 2016 market trough. On the other hand, we find that a good portion of
fluctuations in credit spreads is captured by implied PVLGDs.  For example, at the start of the
COVID-19 pandemic, when credit spreads widened, the aggregate credit score was largely unchanged
because implied PVLGDs -- through the language of earnings calls -- reflected the factors that led
to a widening in credit spreads.  For this reason, the correlations between quarterly changes in
aggregate credit scores and HY option adjusted spreads are relatively low, suggesting that aggregate
credit scores proxy for different information than do aggregate credit spreads.%
\footnote{The highest correlation with HY OAS changes comes from quarterly changes in the
  full-sample credit score for IG names; even in this case, the amount of joint variation (in the
  $R^2$ sense) is only 0.35 ($=0.59^2$).}

\section{Construction of Information Variables} \label{s:interactions}
 
\subsection{Calculation of Analyst Dispersion}\label{s:analystdisp}

We describe how we calculate the analyst dispersion (DispAnlst) of price forecasts in the past year
for a firm. From the I/B/E/S data base, we can obtain the standard deviation, average, and number of
price forecasts in each month.  For firm $i$ and month $s$, denote them by $\sigma(P_{i,s})$,
$\bar{P}_{i,s}$, and $n_{i,s}$, respectively. Then, we can calculate the average price and price
squared for the past 12 months of $t$ as
\begin{equation*}
\text{Avg}(\{P_{i,s}\}_{t-12}^{t-1}) = \frac{\sum_{s=t-12}^{t-1}\bar{P}_{i,s}\times n_{i,s}}{\sum_{s=t-12}^{t-1}n_{i,s}},
\end{equation*}
and 
\begin{equation*}
\text{Avg}(\{P^2_{i,s}\}_{t-12}^{t-1}) = \frac{\sum_{s=t-12}^{t-1}\left( (\bar{P}_{i,s})^2+\sigma(P_{i,s})^2\right)\times n_{i,s}}{\sum_{s=t-12}^{t-1}n_{i,s}}.
\end{equation*}
This allows us to calculate the standard deviation of price forecasts in the past year as 
\begin{equation*}
\text{Std}(\{P_{i,s}\}_{t-12}^{t-1}) = \sqrt{\text{Avg}(\{P^2_{i,s}\}_{t-12}^{t-1}) - \text{Avg}(\{P_{i,s}\}_{t-12}^{t-1}) ^2}.
\end{equation*}
The analyst dispersion is given by 
\begin{equation}
\text{DispAnlst}_{i,t} = \frac{\text{Std}(\{P_{i,s}\}_{t-12}^{t-1})}{\text{Avg}(\{P_{i,s}\}_{t-12}^{t-1})}.
\end{equation}

\subsection{Converting to Decile Levels}\label{s:decile}

This section describes how we construct the demeaned decile levels for the information variables
used in regression (\ref{eq:reg-int}).  In short, the decile variables are constructed using the
corresponding data in the one-year window before the release of the earnings call.

First, the transcript length (TransLen) and Flesch-Kincaid grade (FKGrade) are calculated for each
earnings call in our sample.  For each month $t$, we get all earnings calls in our sample released
in the past year (excluding current month), i.e., in month $t-12$, $t-11$, ..., and $t-1$. Denote
the set of these earnings calls by $C_t$. Then, for each earnings call released in month $t$, we get
its decile level for transcript length or Flesch-Kincaid grade using the corresponding quantities of
the earnings calls in set $C_t$.

On the other hand, we note that the number of analysts (NumAnlst) and the analyst forecast
dispersion (DispAnlst) are defined on the firm-month level.  Thus, the construction of their decile
variables is a bit different with that for the TransLen and FKGrade.  For each month $t$, we
identify the set of firms with at least one earnings call in the past year, denoted by $F_t$.  For
each firm in $F_t$, we get its NumAnlst and DispAnlst at month $t$.  Recall NumAnlst
(resp. DispAnlst) denotes the total number of analysts (resp. dispersion in analysts forecasts) of
the firm in the past 12 months.  Thus, it suffices to calculate the NumAnlst and DispAnlst for firm
in $F_t$ at month $t$, even the firm does not have an earnings call in month $t$ (NumAnlst and
DispAnlst are not constructed on textual data).  Then, for each earnings call released in month $t$,
we get the decile level for its NumAnlst and DispAnlst using the corresponding quantities of firms
in $F_t$, which are all calculated at month $t$.

We perform the above steps for all months in our sample to construct the decile variables.  Then, we
demean the decile variables using the entire sample. The demeaned decile levels are then used in
regression (\ref{eq:reg-int}).

\section{Portfolio Optimization} \label{s:opt-dets}

We transform (\ref{constr_UL}) to linear form as follows.  Introduce additional decision variables
$w_{i,t}^{(+)}$ and $w_{i,t}^{(-)}$ for $i=1,2,\ldots,n_t$, which represent the positive and
negative parts of the weight $w_{i,t}$ respectively. Then, the constraint (\ref{constr_UL}) is
equivalent to
\begin{equation}
\sum_{i=1}^{n_{t}} w_{i,t}^{(+)}\leq U,\,\,\sum_{i=1}^{n_{t}} w_{i,t}^{(-)}\geq L \label{constr_UL_new}
\end{equation}
with 
\begin{equation}
 w^{(+)}_{i,t} \geq w_{i,t},\,\,\,\,w^{(+)}_{i,t} \geq 0,\,\,\,\, w^{(-)}_{i,t} \leq w_{i,t}, \text{ and }\,w^{(-)}_{i,t} \leq 0. \label{constr_w_pn}
\end{equation}
In summary, our portfolio optimization can be formulated as a linear programming problem with
objective (\ref{port_obj}) and constraints (\ref{constr_sumPV}), (\ref{constr_ul}),
(\ref{constr_UL_new}), and (\ref{constr_w_pn}).

\subsection{Null Distribution for Trading Strategy} \label{s:null-dist}

We use simulation to evaluate the performance of our trading strategy under the null hypothesis of
no predictability. In the simulation, we generate portfolio by assigning the optimal weights solved
from the model randomly to all candidate CDS.  This is implemented as follows. For each month $t$,
we first solve the optimization problem in (\ref{port_obj}--\ref{constr_UL}). We represent the
structure of the optimal weights by the vector $(n_u,n_l,n_o)$; $n_u$ and $n_l$ denote the number of
weights that are set to the upper and lower limit ($u$ and $l$), respectively, $n_o$ is the number
of weights that lie between $l$ and $u$. Thus, the sum $n_u+n_l+n_o$ equals to the non-zero weights
in the optimal portfolio.  We find that $n_o$ is generally very small in the optimal portfolio.
Specifically, we have $n_o=1$ (resp. $n_o=2$) in 72\% (resp. 26.8\%) of the months and specification
combinations, and the largest value of $n_o$ is 3 in all cases.  Thus, the structure of the optimal
portfolio can be interpreted as follows. We first select two groups of CDS according to their credit
scores and set their weights to the limits $u$ or $l$, respectively. We then add another few CDS to
the portfolio to set its total PVLGD to zero, as required by (\ref{constr_sumPV}).

For each month $t$, we generate a random portfolio that has a similar structure to the optimal one.
We first randomly select $n_u$ CDS (out of the set of all CDSs for which we have data in months $t$
through $t+3$ so that price changes can be calculated) and set their weights at the upper limit
$w_{i,t} = u$. Similarly, we randomly select $n_l$ CDS and set their weights at the lower limit
$w_{i,t}=l$. Denote the two sets of CDS by $S_u$ and $S_l$, respectively. Next, we pick another CDS
to satisfy the zero-PVLGD constraint (\ref{constr_sumPV}).  Note that the weight for the last CDS
must also lie between $l$ and $u$. The admissible set for CDS satisfying this condition is given by
\begin{equation} \label{eq:So}
S_o \equiv \{i\notin S_l \bigcup S_u | \,\,  l\cdot PV_{i,t} \leq -u \sum_{j\in S_u} PV_{j,t} -l
\sum_{j\in S_l}  PV_{j,t}\leq u\cdot PV_{i,t}\}.
\end{equation}

We randomly select one CDS from set $S_o$ and set its weight as
$w_{i,t} =-(u \sum_{j\in S_u} PV_{j,t} + l \sum_{j\in S_l}PV_{j,t} )/PV_{i,t} $.  Then it is easy to
verify that adding the currently selected CDS to the portfolio results in a zero total PVLGD. The
weights for other, non-selected CDS are set to zero. The newly generated portfolio has a similar
structure with the optimal one, as it has the same numbers of weights set at the upper and lower
limits.  However, here the weights are assigned randomly without using the credit scores.  In the
last step, we verify whether the leverage constraint on total long and short positions in
(\ref{constr_UL}) is satisfied. If not, we repeat the previous steps until the generated portfolio
is feasible, i.e., satisfying constraints (\ref{constr_sumPV}--\ref{constr_UL}).  In some unusual
cases, it is not possible to find the last CDS to satisfy the zero-PVLGD constraint, i.e., the
admissible set $S_o$ in (\ref{eq:So}) is empty.  We describe how we handle such cases in Section
\ref{ap:weight_gen} below.

We simulate 100 trials for the trading strategy. In each trial, we generate one portfolio using the
above algorithm for each month.  We then calculate the portfolio return for each trial by
(\ref{port_change}) and (\ref{eq:port_return}). Denote the returns by $\bar{R}^{(n)}$ for
$n=1,2,..,100$. The set $\{\bar{R}^{(n)}\}$ provides a distribution of the return from simulated
weights, and thus allows us to conduct statistical tests for the return of the actual optimal
portfolio.

We show the portfolio generated by our algorithm indeed has similar weight structure to the optimal
one.  For each of the 5250 month and specification combinations (105 months $\times$ 25 weight
limits $\times$ 2 rating group), we calculate the average numbers of CDS with weight $u$ or $l$ from
the 100 simulated portfolios.  We find they are very close to the numbers from the optimal
portfolio, with the average (largest) absolute difference being 0.028 (2.37) for the 5250
month-specification combinations.

We also verify there is enough randomness in the simulated portfolio weights.  For a given month and
specification, we calculate the average correlation of the weight vectors from the 100 simulated
portfolios.  We find the average correlation is low across the simulated portfolios. Specifically,
the absolute average correlation is smaller than 0.1 for 73\% of the 5250 cases, and smaller than
0.3 for 90\% of the cases.  These observations support the effectiveness of our simulation method.

\subsubsection{Dealing with Empty  $S_o$ Sets} \label{ap:weight_gen}

In the below, we describe how we deal with the scenario where the set $S_o$ in (\ref{eq:So}) is
empty in the simulation. 
For a given month, if we cannot generate a feasible portfolio weight by the above algorithm in 100
attempts ({each attempt being characterize by the choice of sets $S_u$ and $S_l$, and then
  checking whether $S_o$ is empty}), we take the following steps.  Recall the sets $S_u$ and $S_l$
denote the CDS selected with weight $u$ and $l$ respectively.  If the total weighted PVLGD from the
CDS in $S_u$ and $S_l$ is positive:
\begin{equation}
 u\sum_{j\in S_u} PV_{j,t} + l \sum_{j\in S_l}  PV_{j,t} >0,\label{eq:pv_sum}
\end{equation}
{we move the CDS that has the largest PVLGD in $S_u$ into $S_l$ and then move the CDS with the
  smallest PVLGD in $S_l$ in $S_u$}. This decreases the total weighted PVLGD from the selected CDS
while keeping the numbers of positive and negative weights unchanged.  Similarly, if the quantity in
(\ref{eq:pv_sum}) is negative, we switch the CDS that has the smallest PVLGD in $S_u$ with the CDS
that has the largest PVLGD in $S_l$.  We recalculate the admissible set $S_o$ in (\ref{eq:So}) after
switching to see if we can find a CDS to satisfy the zero-PVLGD constraint. If so, we return the
sampled portfolio weights. Otherwise, we continue the above steps and switch the next most extreme
PVLGD CDS between the sets $S_u$ and $S_l$.

In very rare situations, the above switching steps do not lead to a feasible solution (the switching
steps fall into a loop without generating a non-empty $S_o$).  In these cases, we drop the most
extreme PVLGD CDS from the set $S_u$ or $S_o$ sequentially (depending on whether the sum in
(\ref{eq:So}) is positive or negative) until we can generate a feasible portfolio. This leads a
simulated portfolio that has different numbers of weights set at the bounds. However, such cases are
very rare.

\subsection{Test for Joint Significance of Strategy Performance}\label{s:joint_test}

For each rating class (IG or HY), we implement our trading strategy for 25 weight speficiations
$(l,u)$.  We now describe how we test the joint significance of their performance.  Since our
trading strategy outcomes are not independent, to test their statistical significance, we generate
25 correlated Bernoulli variables each of which takes a value of one with a 5\% probability.  To do
this, we first generate 26 standard normal variables by
\begin{equation*}
X_i =\sqrt{ c} F+\sqrt{1-c}S_i,
\end{equation*}
where $F$ is a common standard normal shock and $S_i$ for $i=1,2,...,25$ are independent standard
normal variables.  Then, each Bernoulli variable is defined as $\mathbf{1}\{X_i\geq 1.645\}$, which
assigns a value of one 5\% of the time ({for tests at the 10\% level, we use $1.282$}).  The
single parameter $c\in [0,1]$ controls for the level of correlation.  For each $c$, we run $10^6$
simulated draws and calculate the fraction of trials $\phi_c$ with a number of successful draws
equal to the number of significantly positive returns for IG and HY.%
\footnote{We use equal to, and not greater than, because, for example, 12/25 and 24/25
    successful draws reflect two quite different correlation structures.}
We then vary the value of $c$ with an increment of 0.02 to obtain the largest $\phi_c$ across all
$c$'s.  This is the p-value that is least favorable for our test (since it is the highest one
possible).  For the IG tests shown in Table \ref{t:porttest}, the maximal p-value we obtain for
observing 24 out of 25 significant results is 0.44\%. For HY, the maximal p-values associated
with either the 12/25 (at the 5\% level) or the 14/25 (at the 10\% level) results are 0.42\% and
0.72\% respectively.

\section*{References}

\begin{list}{}{\topsep 0pt \leftmargin .2in \listparindent -0.2in \itemindent
    -0.2in \parsep \parskip}

\item Bali, T., A. Goyal, D. Huang, F.  Jiang, and Q. Wen, 2022, ``Predicting corporate bond
  returns: Merton meets machine learning,'' working paper.

\item Banz, R. W, 1981, ``The Relationship between Return and Market Value of Common Stocks,'' {\it
    Journal of Financial Economics}, 9 (1), 3-–18.

\item Bao, J., J. Pan, and J. Wang, 2011, ``The illiquidity of corporate bonds,'' {\it Journal of
    Finance}, 66 (3), 911--946.

\item Bartram, S., M. Grinblatt, and Y. Nozawa, 2020, ``Book-to-market, mispricing, and the
  cross-section of corporate bond returns,'' working paper.

\item Bernard, V. and J. Thomas, 1989, ``Post-earnings-announcement drift: Delayed price response or
  risk premium?'' {\it Journal of Accounting Research}, 27, 1--36.

\item Bharath, ST. and T. Shumway, 2008, ``Forecasting default with the Merton distance to default
  model,'' {\it Review of Financial Studies}, 21(3), 1339--69.

\item Chordia, T., A. Goyal, Y. Nozawa, A. Subrahmanyam, and Q. Tong, 2017, ``Are capital market
  anomalies common to equity and corporate bond markets? An empirical investigation,'' {\it Journal
    of Financial and Quantitative Analysis}, 52 (4), 1301--1342.

\item Collin-Dufresne, P., R. Goldstein, and J.S. Martin, 2001, ``The determinants of credit spread
  changes,'' {\it Journal of Finance}, 56 (6), 2177--2207.

\item Cremers, M., J. Driessen,  P. Maenhout, and D. Weinbaum, 2008, ``Individual stock-option
  prices and credit spreads,'' {\it Journal of Banking \& Finance}, 32(12), 2706--2715.

\item Ericsson, J., K. Jacobs, and R. Oviedo, 2009, ``The determinants of credit default swap
  premia,'' {\it Journal of Financial and Quantitative Analysis}, 44 (1), 109--132.

\item Fama, E. F., and K. R. French, 1992, ``The Cross-Section of Expected Stock Returns,'' {\it
    Journal of Finance}, 47 (2), 427–-465.

\item Fama, E. F., and K. R. French, 2008, ``Dissecting Anomalies,'' {\it Journal of Finance}, 63
  (4), 1653–-1678.

\item Gompers, P., J. Ishii, and A. Metrick, 2003, ``Corporate governance and equity prices,'' {\it
    Quarterly Journal of Economics}, 118 (1), 107--155.
  
\item Guo, X., H. Lin, C. Wu, and G. Zhou, 2021, ``Investor sentiment and the cross-section of
  corporate bond returns,'' working paper.

\item Jegadeesh, N., 1990, ``Evidence of predictable behavior of security returns,'' {\it The
    Journal of Finance}, 45 (3), 881--898.

\item Jegadeesh, N., and S. Titman, 1993, ``Returns to buying winners and selling losers:
  implications for stock market efficiency,'' {\it Journal of Finance}, 48 (1), 65–-91.

\item Jostova, G., S. Nikolova, A. Philipov, and C. W. Stahel, 2013, ``Momentum in corporate bond
  returns,'' {\it Review of Financial Studies}, 26 (7), 1649–-1693

\item Lehmann, B., 1990, ``Fads, martingales, and market efficiency,'' {\it The Quarterly Journal of
    Economics}, 105 (1), 1--28.

\item Lee, M., P. Meyer-Brauns, S. Rizova, and S. Wang, 2020, ``The cross-section of corporate bond
  returns,'' {\it Dimensional Fund Advisors working paper}.

\item Merton, R., 1974, ``On the pricing of corporate debt: The risk structure of interest rates,''
  {\it Journal of Finance}, 29, 449--470.

\item Penman, S., F. Reggiani, S. A. Richardson, I. Tuna, 2014, ``An accounting-based characteristic
  model for asset pricing'', working paper.

\item Vassalou, M. and Y. Xing, 2004, ``Default risk in equity returns,'' {\it Journal of Finance},
  59 (2), 831--868.

\end{list}
  

\newcommand\incbbg[1]{
  \includegraphics[width=0.94\textwidth,trim={0 0 0 0.625cm},clip]{figures/#1}
}

\begin{figure}[H]
  \centering
  \mbox{\incbbg{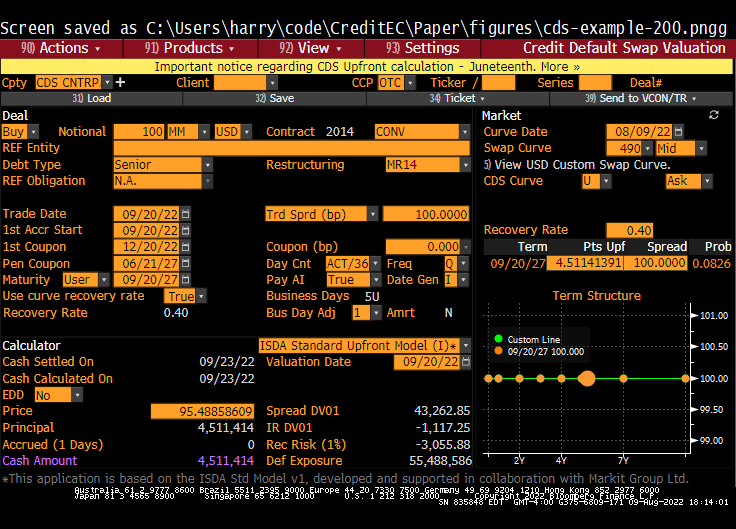}} \\[20pt]
  \mbox{\incbbg{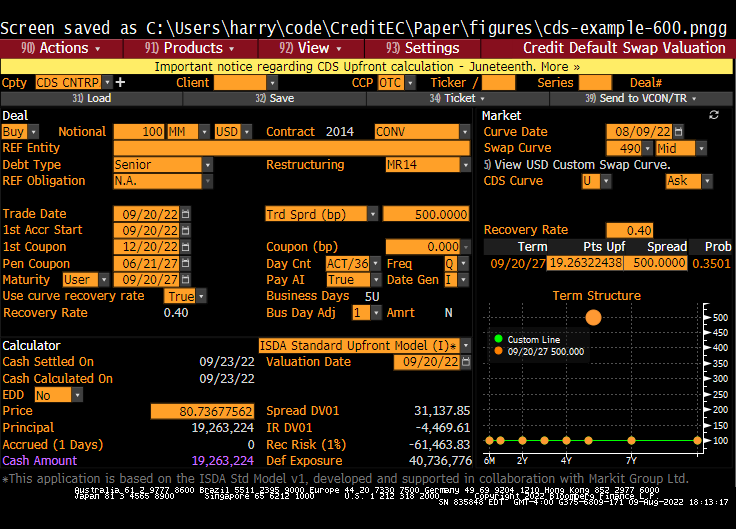}} 
  \caption{Sample Bloomberg pricing screens for CDS PVLGDs with $S=100$ or 500 (accessed via
    \texttt{CDSW} command), using the ISDA Standard Upfront Model. The {\it Pts Upf} amount should
    be compared to our PVLGD calculations in Table \ref{t:pvlgds} of the main
    paper. } \label{f:bbg-cdsw}
\end{figure}

\contempRegTable{tables/table2_6mth_InS_pvlgd}{full sample}{6}{PVLGD}
\contempRegTable{tables/table2_12mth_InS_logCDS}{full sample}{12}{logCDS}

\begin{table}
    \centering
    {\bf Dependence of 12-month PVLGD Changes on Forecasting Variables:\\ Full Sample
      Text Model with Firm and Time Fixed Effect} \\[5pt]
    \footnotesize \begin{tabular}{@{}lccccccccc@{}c}
\toprule
        &      (1) &     (2) &      (3) &      (4) &      (5) \\
\midrule
         $R^f$ &   $-$2.029 &  $-$0.456 &    0.377 &    0.519 &          \\
             &  ($-$0.86) & ($-$0.22) &   (0.22) &   (0.31) &          \\
     IV &    5.545 &   5.506 &    4.345 &    4.369 &          \\
             &   (6.72) &  (7.42) &   (5.58) &   (5.65) &          \\
      LEV &   11.068 &  10.024 &    9.856 &    9.830 &          \\
             &  (12.46) & (10.56) &  (10.86) &  (10.87) &          \\
       PVLGD &          &         &   $-$0.533 &   $-$0.491 &   $-$0.526 \\
\bottomrule
\end{tabular}
    \caption{This table reports regressions results of 12-month PVLGD changes on the contemporaneous
      changes in company market leverage, risk-free rate, and option implied volatility, adding
      PVLGD level, implied credit score, control factors, entity fixed effects, and year-month fixed
      effects. {The specification includes all variables in (\ref{eq:contemp-reg}), as well as
        time (year-month) and firm fixed effects.} The table reports results based on the full
      sample implied credit score. Column (1) presents results with contemporaneous changes, column
      (2) adds control factors, column (3) adds PVLGD levels, column (4) add Credit Scores, and
      column (5) drops contemporaneous changes. We cluster standard errors by entity and month, the
      t-statistics are reported in parenthesis. Only control variables that are statistically
      significant in at least one specifications are included.  All statistics
      are based on full sample series winsorized at 1\% and 99\% percentiles.}
    \label{t:contemp-full sample-12-PVLGD-FE}
\end{table}

\contempRegTableLasso{tables/table2_12mth_OOS_pvlgd}{rolling sample}{12}{PVLGD}
\contempRegTableLasso{tables/table2_6mth_OOS_pvlgd}{rolling sample}{6}{PVLGD}
\makeatletter\@input{xx.tex}\makeatother